\documentclass[aps,prl,nofootinbib,twocolumn,superscriptaddress,preprintnumbers,balancelastpage,longbibliography]{revtex4-2}

\usepackage{graphicx}
\usepackage{subfigure}
\usepackage{bm}
\usepackage{amssymb}
\usepackage{amsmath}
\usepackage{epsfig}
\usepackage{cancel}

\usepackage{color}
\usepackage{mathtools}
\usepackage{cases}
\usepackage{booktabs}
\usepackage[utf8]{inputenc}
\definecolor{blue}{rgb}{0,0.3,0.8}
\usepackage{physics}
\usepackage{placeins}
\usepackage{graphicx}
\graphicspath{{Figure_v3/}}

\usepackage[
colorlinks=true,
filecolor=black,
anchorcolor=blue,
linkcolor=blue,
citecolor=cyan, 
urlcolor=blue,
linktocpage=true,
plainpages=false,
breaklinks=true,
pdfstartview=FitH
]{hyperref}
\usepackage{bm}
\usepackage{braket}
\usepackage{placeins}
\usepackage{titletoc}  

\begin{document}

\title{Rapid post-merger gravitational-wave signals from magnetic black hole superradiance: novel approach to detect magnetic monopole and ultralight boson}

\author{Zhong-Hao Luo}
\email{luozhh58@mail2.sysu.edu.cn}
\author{Fa Peng Huang}
\email{Corresponding author. huangfp8@sysu.edu.cn}
\affiliation{MOE Key Laboratory of TianQin Mission, TianQin Research Center for Gravitational Physics \& School of Physics and Astronomy, Frontiers Science Center for TianQin, Gravitational Wave Research Center of CNSA, Sun Yat-sen University (Zhuhai Campus), Zhuhai 519082, China}
\author{Pengming Zhang}
\email{zhangpm5@mail.sysu.edu.cn}
\affiliation{School of Physics and Astronomy, Sun Yat-sen University, Zhuhai 519082, China}
\author{Chen Zhang}
\affiliation{Liaoning Key Laboratory of Cosmology and Astrophysics, College of Sciences, Northeastern University, Shenyang 110819, China}

\raggedbottom

\begin{abstract}
We present an analytic framework demonstrating that a spinning black hole endowed with a net magnetic charge exhibits a dramatically amplified superradiant instability against charged scalar fields---enhanced by several orders of magnitude compared with the neutral Kerr case. The amplification arises from a monopole-induced reduction of the centrifugal barrier,
\(\ell(\ell+1)\!\to\!\ell(\ell+1)-q^{2}\), encoded by an effective quantum number $\ell_q$, with
$q=eP=N/2$ fixed by Dirac quantization. This shift deepens the gravitational bound-state potential well and produces a parametrically larger instability growth rate.
The resulting rapid growth can build a macroscopic charged scalar cloud that acts as a coherent source of near-monochromatic continuous gravitational waves (GWs).
{ For a charged complex scalar, however, the stress tensor of a single monopole-harmonic sector is stationary and does not contain a $2\omega$ annihilation source. The leading oscillatory GW source instead comes from coherent north--south, equivalently charge-conjugate, cross terms in the stress tensor.
These cross terms select the channels
$(\tilde\ell,\tilde m,\tilde\omega)=(2q,+2q,+2\omega)$ and its conjugate $(2q,-2q,-2\omega)$.
We find an enhanced GW power,
$P_{\rm GW}\propto (E_N E_S/M^2)\,\alpha^{4\ell_q+8}$, or equivalently $P_{\rm GW}\propto (E_c/M)^2\alpha^{4\ell_q+8}$ for a balanced two-sector cloud, instead of the neutral-Kerr scaling $\alpha^{4\ell+10}$.
This motivates a potential smoking-gun search strategy for magnetic monopoles and ultralight bosons: prompt post-merger follow-up of binary-black-hole merger remnants with ground-based and space-based GW experiments.
In contrast to the Kerr case, where the signal turn-on can be delayed to years--centuries, a magnetic remnant can form a cloud and emit a stronger continuous signal within weeks to months. For magnetic supermassive remnants ($M\sim10^6 M_\odot$), the follow-up GW signal can appear within weeks of a binary-black-hole merger, with strain $h\sim10^{-20}$ in the mHz band.}
\end{abstract}

\maketitle

{\bf Introduction}---Black-hole superradiance extracts rotational energy from spinning black holes (BHs) and, in the presence of light bosonic fields, can trigger superradiant instabilities that build macroscopic bosonic condensates around the BH \cite{Starobinskii:1973vzb,Zouros:1979iw,Brito:2015oca,Cardoso:2004nk,PhysRevD.22.2323,Dolan:2007mj,Dolan:2018dqv,East:2017mrj,Xie:2022uvp,Alexander:2009tp,Cornwall:1997ms,PhysRevLett.95.151301,RoperPol:2019wvy,Seto:2007tn,Vachaspati:2001nb}.
These clouds generically emit long-lived, nearly monochromatic gravitational waves (GWs), making them well-motivated targets for ground- and space-based interferometers \cite{Arvanitaki:2014wva,Yoshino:2013ofa,Brito:2017zvb,Isi:2018pzk,Yang:2023vwm,Abbott:2022ScalarCloudsO3}.
Recent developments have sharpened waveform modeling, including cloud depletion, frequency evolution, and level-transition features \cite{PhysRevD.107.104003,Baumann:2019eav,Baumann:2019ztm,Xie:2025npy,Yang:2023aak}.

{ Most studies focus on neutral Kerr BHs, but astrophysical BHs may acquire a
small monopole-induced magnetic charge, for which even an order-one Dirac
number can qualitatively modify charged-boson superradiance.} Among
hypothetical objects in particle physics and cosmology, magnetic monopoles
hold a particularly compelling status.
{
Their discovery would not merely add a new particle to the spectrum, but
would provide direct evidence for nontrivial gauge structures underlying
fundamental interactions and offer a physical explanation for the
quantization of electric charge.
}
They are especially well motivated because they arise generically in many
quantum field theories, including grand or partially unified gauge theories
\cite{PhysRevLett.45.1,tHooft:1974kcl,Polyakov:1974ek,Langacker:1980kd,
Murayama:2009nj,Zou:2025jlp,Cho:1996qd}; their existence is consistent with
known principles of nature and would provide an elegant explanation for the
quantization of gauge charge through the Dirac condition
\cite{Dirac:1931kp,Preskill:1984gd}. Their classical description is captured
by the Wu--Yang construction
\cite{Wu:1976ge,PhysRevD.12.3845}. Yet monopoles are expected to be extremely
heavy and exceedingly rare in the late Universe
\cite{PhysRevLett.43.1365,Longo:1981jp,Stojkovic:2004hz,
Stojkovic:2005zh,Preskill:1984gd}, making them notoriously difficult to
detect: despite decades of direct searches
\cite{Giacomelli:2003yu,ATLAS:2023esy,ParticleDataGroup:2010dbb,
MACRO:2002jdv,IceCube:2021eye} and astrophysical bounds
\cite{Parker:1970xv,Turner:1982ag,Kobayashi:2023ryr,Zhang:2024mze,
Graesser:2021vkr,Gould:2017zwi,Hook:2017vyc}, their abundance remains only
weakly constrained.
{
The sharp contrast between their fundamental theoretical significance and
persistent experimental elusiveness makes qualitatively new, indirect
detection strategies particularly valuable.
}
In particular, strong-gravity environments provide a natural arena
\cite{Maldacena:2020skw,Bai:2020spd,Diamond:2021scl,Zhang:2023tfv}, where BHs
can act as amplifiers and even tiny magnetic charges may leave observable
imprints. In this context, Pere\~niguez \textit{et al.}
\cite{Pereniguez:2024fkn} numerically studied charged-scalar superradiance
around rotating magnetic BHs and found dramatically shorter instability
timescales even for an order-one monopole number, together with
``north''/``south'' monopole-harmonic modes whose most unstable clouds align
with the rotation axis and localize in opposite hemispheres rather than
forming an equatorial torus as in the Kerr case.

Despite this progress, several key issues remain open.
First, the \emph{physical origin} and analytic structure of the instability enhancement in rotating magnetic BHs remain unclear.
Here we present an analytic derivation for a gauge-charged massive scalar on a magnetically charged background and treat the Dirac-quantized parameter $q\equiv eP=N/2$ as a free parameter~\cite{Wu:1976ge}.
We show that the magnetic charge effectively reduces the centrifugal barrier via
$\ell(\ell+1)/r^2\rightarrow [\ell(\ell+1)-q^2]/r^2\equiv \ell_q(\ell_q+1)/r^2$,
which implies the scaling $\omega_I\propto \alpha^{4\ell_q+5}$ with $\alpha\equiv \mu M$ and explains the large growth-rate enhancement relative to Kerr BH.

Second, GW emission from hemispherically localized monopole-harmonic clouds has not been computed systematically.
We identify the dominant bound-state profiles as the north/south monopole-harmonic modes $Y^{\pm}_{qq,\pm q}$~\cite{Pereniguez:2024fkn}. { For a charged complex scalar, the stress tensor of either sector alone is stationary, so the leading $2\omega$ GW source is not a single-sector self-annihilation term. Instead, it arises from coherent north--south, or charge-conjugate, cross terms in the stress tensor.
For $\ell=q$ and $|m|=q$ we find $P_{\rm GW}\propto (E_NE_S/M^2)\,\alpha^{4\ell_q+8}$, together with a cross-channel selection rule
$(\tilde\ell,\tilde m,\tilde\omega)=(2q,+2q,+2\omega)$ and its conjugate $(2q,-2q,-2\omega)$.}

Third, the enhanced growth motivates a self-consistent treatment of backreaction and GW depletion~\cite{Siemonsen:2022yyf,Guo:2022mpr,Brito:2014wla,East:2018glu}.
We therefore integrate the coupled BH--cloud evolution and translate the modified growth history and cross-source emission into observable continuous-wave waveforms. { Neither fundamental magnetic monopoles nor astrophysical BHs
carrying a net magnetic charge have been observationally confirmed. We
therefore treat the present analysis conditionally: the predicted GW signal
applies if a rapidly rotating magnetically charged remnant and a sufficiently
long-lived gauge-charged bosonic field are realized, while the source
abundance and event rate are not addressed here. Ordinary plasma cannot
directly neutralize the magnetic charge of the BH; magnetic discharge instead
requires oppositely magnetically charged states and may be suppressed for
heavy monopoles~\cite{Pereniguez:2024fkn}. A dark-$U(1)$ or sufficiently
feebly coupled realization may provide an alternative long-lived regime.}

Finally, within this conditional scenario, these effects can be
observationally relevant: rapidly spinning post-merger remnants provide the
high spin required for superradiance, and the monopole-induced enhancement
can shorten the cloud build-up time from the neutral-Kerr scale
(years--centuries, depending on $\alpha$) to days--weeks
(or $\sim$year), bringing the onset of monochromatic GWs into the same
observing window.

This work provides an analytic explanation for the magnetic-BH enhancement
of superradiance and identifies distinctive GW power, growth-time, and
angular-selection signatures that provide concrete targets for future
searches, provided that the required source conditions are realized.

{\bf Setup and monopole-harmonic clouds---}
We consider a rotating BH of mass $M$ and spin $a=J/M$ that carries a magnetic charge $P$ after capturing magnetic monopoles~\cite{Mouland:2024zgk,Myung:2022dpp,Zou:2021mwa}. {
A $U(1)$-charged massive scalar $\Psi$ with mass $\mu$ and gauge charge $e$
obeys
\begin{equation}
(D_\rho D^\rho-\mu^2)\Psi=0,
\qquad
D_\rho\equiv \nabla_\rho-ieA_\rho .
\end{equation}
Here, $\nabla_\rho$ denotes the Levi-Civita covariant derivative associated with the background spacetime metric. The gauge field is taken to be the Dirac-type monopole potential~\cite{Dirac:1931kp}
\begin{equation}
A_\mu dx^\mu=-P\cos\theta\,d\phi,
\quad
q\equiv eP=\frac{N}{2},
\quad
N\in\mathbb{Z}_{\ge0}.
\end{equation}}
We separate variables as $\Psi=e^{-i\omega t+im\phi}S_{q\ell m}(\theta)R(r)$, where, in the ultralight-scalar-mass limit, the angular dependence is described by the monopole harmonics $Y_{q\ell m}(\theta,\phi)=S_{q\ell m}(\theta)e^{im\phi}$~\cite{Wu:1976ge,Pereniguez:2024fkn}. Moreover, the north-south monopole modes $Y^{\pm}_{qq,\pm q}$ are the most superradiant unstable modes, which take the form 
\begin{equation}
Y^+_{qq,q}
=
A_q e^{iq\phi}
\cos^{2q}\frac{\theta}{2},
\ 
Y^-_{qq,-q}
=
A_q e^{-iq\phi}
\sin^{2q}\frac{\theta}{2},
\end{equation}
with $A_{\pm q}=\sqrt{(2q+1)/(4\pi)}$.
{ Here $Y^{\pm}_{qq,\pm q}$ denote the regular representatives of the
lowest monopole harmonics on the north/south gauge patches for the two
charge-conjugate cloud sectors. These sectors carry opposite gauge
charges $\pm e$ and hence $\pm q$, where
$q\equiv |eP|$. The apparent phase singularities
$e^{\pm iq\phi}$ at the poles are gauge artifacts and can be removed by
an intermediate gauge choice \cite{Pereniguez:2024fkn}, so the harmonics
are regular on the $z$ axis ($\cos\theta=\pm1$) and behave as
\begin{align}
    Y^{\pm}_{qq,\pm q}\sim(1\pm\cos\theta)^q .
\end{align}
As a result, the maximally superradiant modes form polar clouds rather
than sharply separated hemispheres: the north mode peaks at $\theta=0$
and smoothly vanishes at $\theta=\pi$, while the south mode behaves
oppositely, leaving \textit{a finite overlap} near the equator. This
north--south two-lobe morphology is illustrated in
Fig.~\ref{Fig:Schematic monopole-cloud morphology} of the SM,
Sec.~\ref{Monopole harmonics and the north/south cloud morphology}.}

{\bf Result I: Analytic origin of the superradiance rate enhancement in the presence of a magnetic monopole---}
In the far region ($r\gg M$) the radial equation reduces to 
\begin{equation}
\frac{d^2(rR)}{dr^2}+\left[\omega^2-\mu^2\Bigl(1-\frac{2M}{r}\Bigr)-\frac{\ell(\ell+1)-q^2}{r^2}\right](rR)=0.
\label{eq:far_radial}
\end{equation}
Thus the centrifugal barrier is weakened by the monopole charge. We define the reduced orbital index $\ell_q$ via
\begin{equation}
\ell(\ell+1)-q^2\equiv \ell_q(\ell_q+1),~ 
\ell_q=\sqrt{\Bigl(\ell+\frac12\Bigr)^2-q^2}-\frac12,
\label{eq:ellqdef}
\end{equation}
which is precisely the Wu--Yang structure for charged motion around a monopole \cite{Wu:1976ge}.
A smaller $\ell_q$ deepens the effective potential well and increases the number of superradiant amplification cycles (Fig.~\ref{Fig:potential well}).

Matched asymptotics in the overlap region (derivation in SM, Sec.~\ref{Matched-asymptotic derivation of the growth rate}) give the imaginary part of the bound-state frequency,
\begin{equation}
\omega_I=\frac{\delta\nu_I}{M}\left(\frac{\alpha}{n_r+\ell_q+1}\right)^3\left[1-\left(\frac{\alpha}{n_r+\ell_q+1}\right)^2\right]^{-1/2},
\label{eq:omegaI_general}
\end{equation}
with $\alpha\equiv \mu M$ and $\delta\nu$ determined by Gamma-function matching (see Eq.~\eqref{Growth rate alpha} in SM, Sec.~\ref{Scaling estimate from the reduced angular barrier}).
In the small-coupling limit, the scaling is
\begin{equation}
\omega_I \propto \alpha^{\,4\ell_q+5},
\label{eq:omegaI_scaling}
\end{equation}
to be contrasted with the Kerr scaling $\omega_I^{\rm (Kerr)}\propto \alpha^{4\ell+5}$. {
Since $\ell_q<\ell$ for $q\neq0$ and $\alpha<1$ in the small-coupling regime, the exponent is reduced and the growth can be enhanced by orders of magnitude.} 
For the dominant magnetic mode $(\ell,m)=(q,q)$ with $N=3$ ($q=3/2$), $\ell_q\simeq0.82$, implying $\omega_I/\omega_{I\rm (Kerr)}\sim \mathcal{O}(10^2)$ near the peak (for the BH spin $a/M=0.9$, see Fig.~\ref{Fig:Analytical growth rate} in SM, Sec.~\ref{Superradiant growth rate: scaling estimate and matched-asymptotic derivation}); this agrees at the order-of-magnitude level with the numerical results of \cite{Pereniguez:2024fkn} and with our analytic curves in Fig.~\ref{Fig:Analytical growth rate}.

{\bf Result II: Power enhancement from the north--south cross source and cross-source emission---}

\begin{figure}[h]
    \centering
    \includegraphics[width=0.95\linewidth]{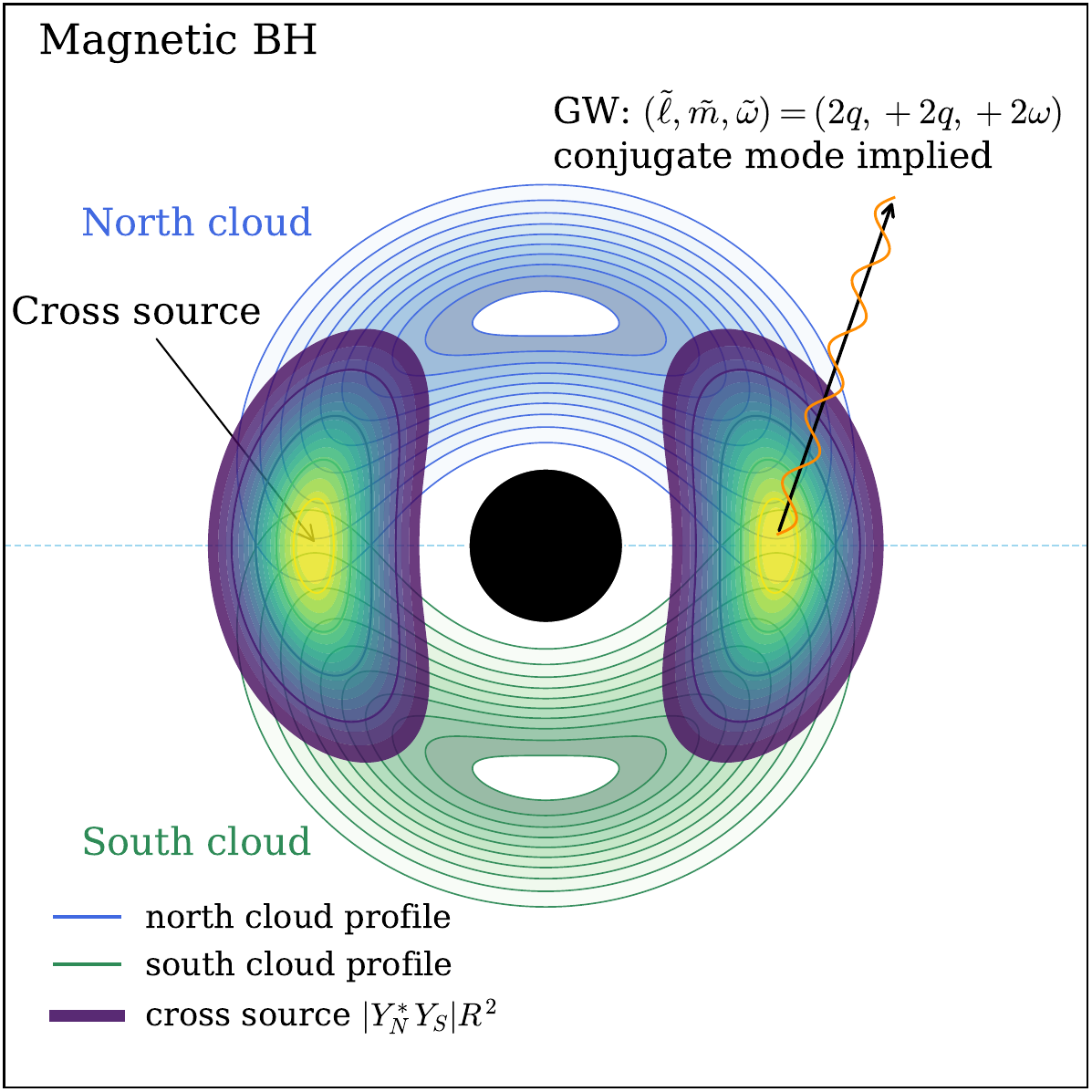}
    \caption{Schematic illustration of the cross-source GW source from a two-sector charged scalar cloud around a magnetically charged BH. The north and south monopole-harmonic profiles form the coherent configuration in Eq.~\eqref{eq:NS_complex_config_main}. A single sector is stationary and does not radiate at $2\omega$; the oscillatory source is the north--south cross term. The channel $T^{SN,-}_{\mu\nu}\propto e^{-2i\omega t+2iq\phi}$ selects $(\tilde\ell,\tilde m,\tilde\omega)=(2q,+2q,+2\omega)$, while the conjugate channel $T^{NS,+}_{\mu\nu}\propto e^{+2i\omega t-2iq\phi}$ supplies $(2q,-2q,-2\omega)$.}
\label{Fig:Cross-source GW}
\end{figure}

For the GW calculation, one must distinguish the spatial cloud profile
from the oscillatory stress-tensor source. A single charged complex sector,
e.g. $\psi_N=e^{-i\omega t}Y^+_{qq,q}R(r)$, gives
$T^{NN}_{\mu\nu}\sim D_\mu^c\psi_N^*D_\nu\psi_N
\propto e^{+i\omega t}e^{-i\omega t}$, where
$D^c_\mu\psi^*\equiv(D_\mu\psi)^*$. The fast phase therefore cancels,
so an isolated sector does not generate an annihilation-type
$2\omega$ GW source. The radiating source instead requires a coherent north--south, or
charge-conjugate, two-sector configuration
$\Psi=c_N\psi_N+c_S\psi_S$, with
\begin{equation}
\psi_N=e^{-i\omega t}Y^+_{qq,q}R(r),
\qquad
\psi_S=e^{+i\omega t}Y^-_{qq,-q}R(r).
\label{eq:NS_complex_config_main}
\end{equation}
The oscillatory stress tensor comes from the cross terms,
\begin{equation}
T^{SN,-}_{\mu\nu}\propto e^{-2i\omega t+2iq\phi},
\qquad
T^{NS,+}_{\mu\nu}\propto e^{+2i\omega t-2iq\phi},
\label{eq:cross_source_main}
\end{equation}
so the GW frequency remains $|\tilde\omega|\simeq2\omega\simeq2\mu$,
but the microscopic source is a cross term rather than the
self-annihilation of an isolated sector. 

In the flat-spacetime approximation for the sourced wave equation
~\cite{PhysRevD.83.044026}
(see SM, Sec.~\ref{North/south selection rules and cross-source emission}),
projecting the oscillatory stress tensor onto the radiative angular
harmonics shows that the two cross-source components excite the following
GW modes:
\begin{equation}
\begin{aligned}
T_{\mu\nu}^{SN,-}
&\ \longrightarrow\ h^{SN,-}:
\quad
(\tilde\ell,\tilde m,\tilde\omega)
=(2q,+2q,+2\omega),\\
T_{\mu\nu}^{NS,+}
&\ \longrightarrow\ h^{NS,+}:
\quad
(\tilde\ell,\tilde m,\tilde\omega)
=(2q,-2q,-2\omega).
\end{aligned}
\end{equation}
Here $(\tilde\ell,\tilde m)$ denote the angular multipole indices of the
radiative GW mode, while $\tilde\omega$ is its Fourier frequency. The second channel is the complex-conjugate, negative-frequency partner
of the first one; equivalently, the physical real waveform contains one
selected positive-frequency angular channel together with its conjugate,
rather than two independent radiative modes. 

\begin{figure}[h]
    \centering
    \includegraphics[width=1\linewidth]{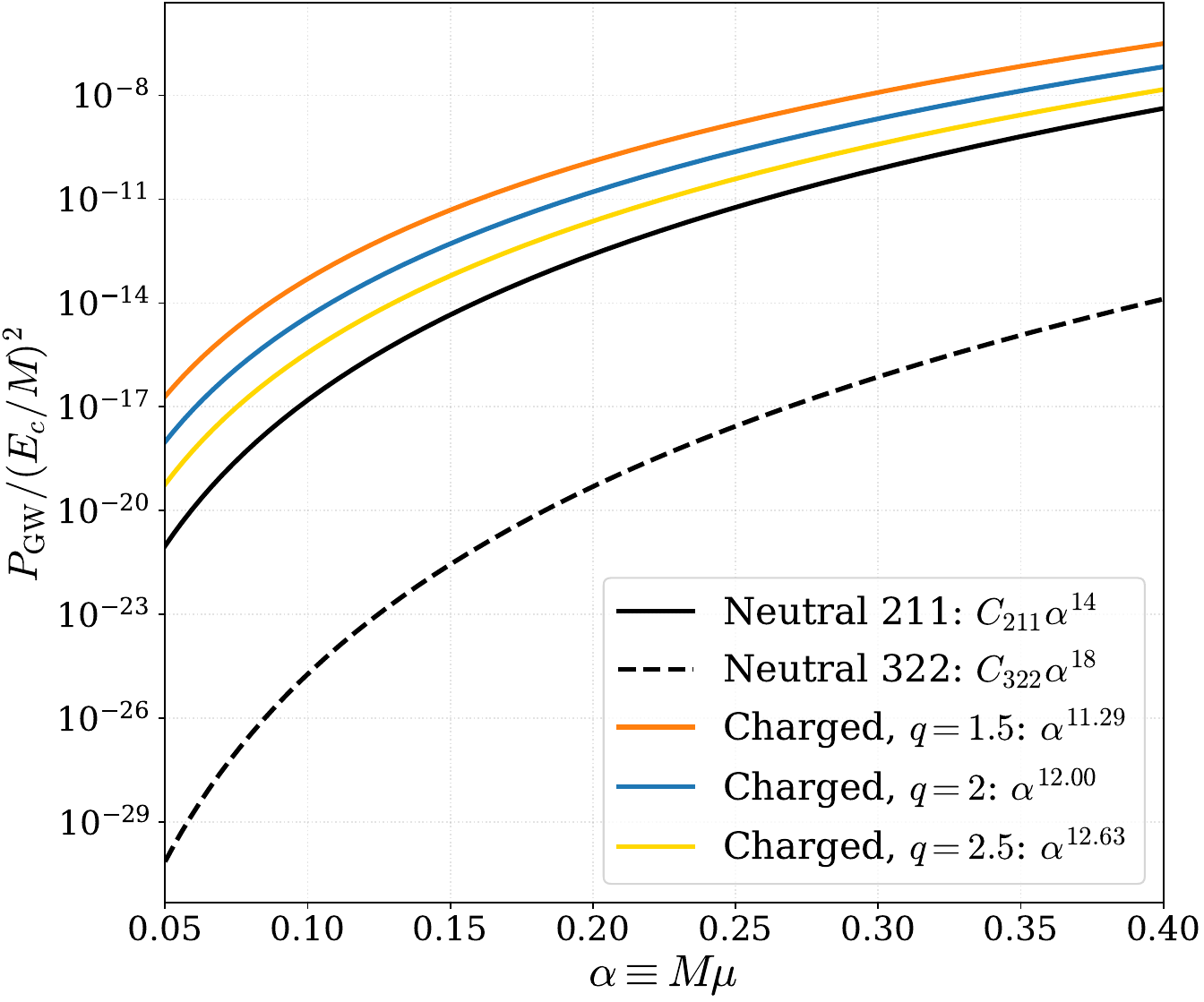}
    \caption{{ Normalized GW power $P_{\rm GW}(\alpha)/(E_{c}/M)^2$ for the dominant cross-source channel, with the balanced-cloud convention $E_N=E_S=E_c/2$. Top: neutral Kerr (211, 322) versus charged clouds with $\ell=q$ for representative $q$.}}

    \label{Fig:GW power}
\end{figure}

For the mirror-symmetric configuration, we explicitly evaluate both
frequency components and find that their GW mode amplitudes are related
by complex conjugation, consistently with
$\tilde h(-\Omega)=\tilde h(\Omega)^*$
(see SM, Sec.~\ref{GW emission in the flat-spacetime approximation}).
They therefore carry the same power, while the negative-frequency
component contains no independent radiative information and only the
independent positive-frequency mode is retained in the definition of
$P_{\rm GW}$. The resulting GW power takes the form
\begin{equation}
P_{\rm GW}
=
C_{n\ell\ell_q}
\left(\frac{E_N E_S}{M^2}\right)
\alpha^{\,4\ell_q+8},
\label{eq:PGW_main_product}
\end{equation}
where $E_N$ and $E_S$ are the energies stored in the two coherent
sectors, and $C_{n\ell\ell_q}$ is a dimensionless coefficient fixed
by the angular integrals and radial normalization
(see SM, Sec.~\ref{GW emission in the flat-spacetime approximation}).
For a balanced configuration, $E_N=E_S=E_s$ and
$E_c=E_N+E_S=2E_s$, so that Eq.~\eqref{eq:PGW_main_product} can be
rewritten as
\begin{equation}
P_{\rm GW}
=
\bar C_{n\ell\ell_q}
\left(\frac{E_c}{M}\right)^2
\alpha^{\,4\ell_q+8},
\label{eq:PGW_main}
\end{equation}
with $E_NE_S=E_c^2/4$ absorbed into
$\bar C_{n\ell\ell_q}$. This flat-spacetime estimate captures the
leading parametric $\alpha$ scaling and the monopole-harmonic selection
rule, while curvature corrections are expected to modify primarily the
numerical coefficient. In contrast to Kerr, where
$P_{\rm GW}\propto\alpha^{4\ell+10}$ \cite{Yoshino:2013ofa}, the
enhancement here has two origins: the reduced index $\ell_q$ weakens the
small-$r$ suppression of the bound-state wavefunction and increases its
support in the effective GW-emitting region, while the leading source
projection $\langle u,T\rangle$ onto the outgoing GW mode remains
nonvanishing, avoiding the additional $\alpha^2$ suppression caused by
the leading-order cancellations in the Kerr case. Accordingly,
Fig.~\ref{Fig:GW power} shows that, at fixed $\alpha$ and cloud energy,
charged clouds can radiate much more efficiently than neutral clouds in
the Kerr case.
\color{black}

{\bf Result III: Novel GW signatures---rapid post-merger continuous-wave signal---}
\begin{figure*}[t]
    \centering
    \includegraphics[width=1\linewidth]{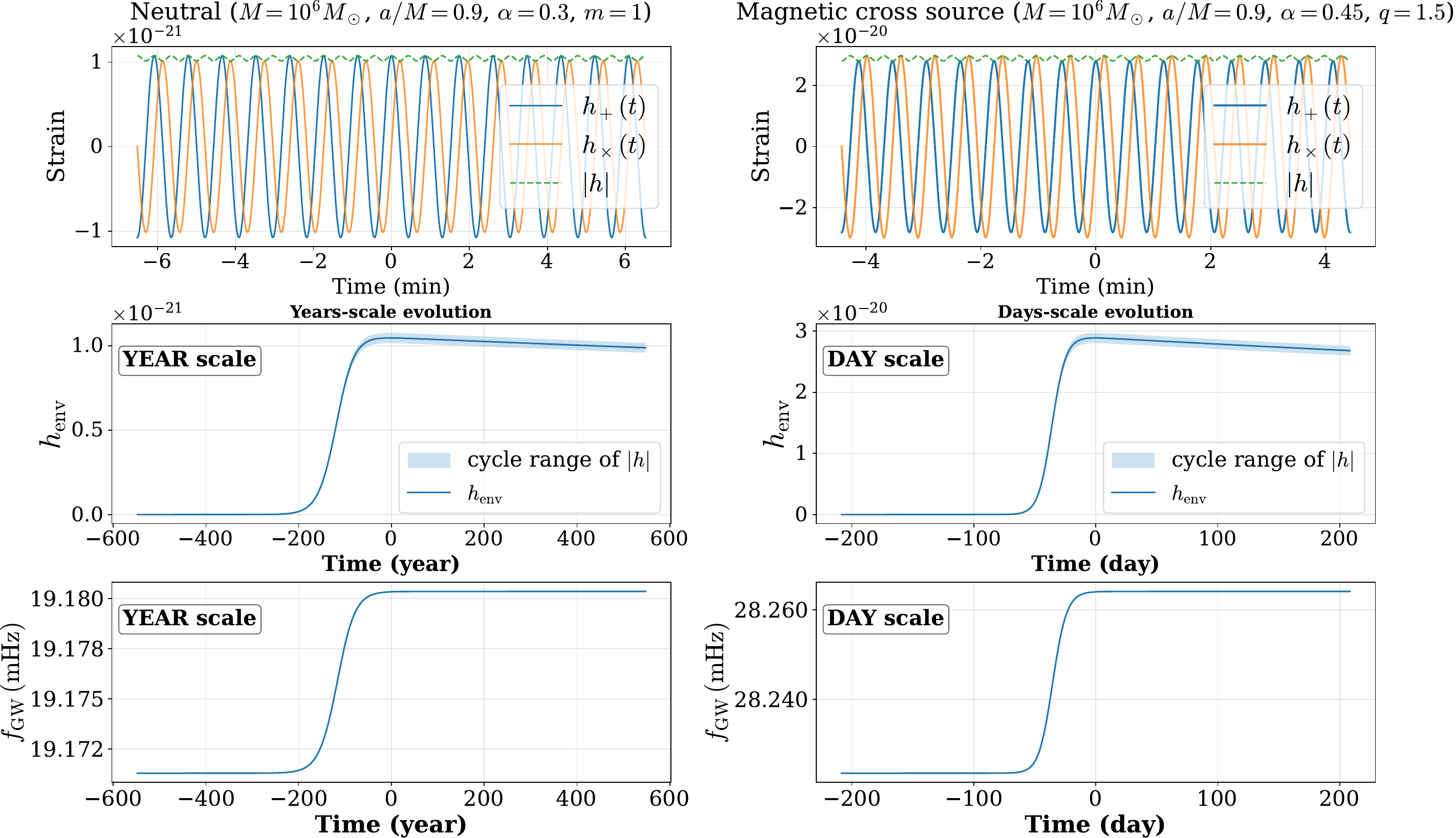}
\caption{{
GW strain from boson clouds around a rotating BH with
$M=10^{6}M_{\odot}$ and $a/M=0.9$.
Left: neutral cloud with $\alpha=0.3$, $m=1$, and
$\mu\simeq4.0\times10^{-17}\,\mathrm{eV}$.
Right: magnetic two-sector cloud with $\alpha=0.45$, $q=1.5$, and
$\mu\simeq6.0\times10^{-17}\,\mathrm{eV}$, whose coherent north--south
cross term sources the GW.
Top: short-window waveforms $h_{+}=\operatorname{Re}h$,
$h_{\times}=-\operatorname{Im}h$, and $|h|$.
Middle: secular envelope $h_{\rm env}$, with the shaded band showing the
cycle range of $|h|$.
Bottom: instantaneous frequency $f_{\rm GW}(t)$.
The magnetic waveform combines the selected
$(\tilde{\ell},\tilde m,\tilde\omega)=(2q,+2q,+2\omega)$ mode with its
negative-frequency conjugate, which does
not constitute an additional independent contribution to the GW
luminosity.
}}
\label{Fig:GWwaveformcomparison}
\end{figure*}
A rapid post-merger onset follows from the enhanced growth found in Result~I: in the linear regime $M_c(t)\simeq M_{c,0}e^{2\omega_I t}$, so a useful idealized scale is the e-folding time $\tau_e\equiv(2\omega_I)^{-1}$, which using $\omega_I\sim\mu\,\alpha^{4\ell_q+5}$ implies
\begin{equation}
\frac{\tau_e}{M}\sim \alpha^{-(4\ell_q+6)},
\label{eq:tau_scaling}
\end{equation}
and a rough build-up time $t_{\rm grow}\sim \tau_e\ln(M_c/M_{c,0})$; adopting peak growth rates $M\omega_I\sim10^{-6}$ for the magnetic $N=3$ mode and $M\omega_I\sim2\times10^{-8}$ for the neutral 211 mode gives the illustrative estimates in Tab.~\ref{tab:efolding}.

However, $\tau_e$ is only a reference scale, since it neglects GW backreaction. In the Kerr case one typically has $\tau_{\rm SR}\ll\tau_{\rm GW}$ during growth, so the cloud approaches the threshold $m\Omega_H\simeq\omega_R$ before GW losses become important. For magnetically charged BHs, however, the enhancement in Eq.~\eqref{eq:PGW_main} can make GW losses relevant already during growth, quenching the instability and shifting the effective turn-on time. A self-consistent evolution therefore requires integrating \cite{Brito:2015oca,Siemonsen:2022yyf}
\begin{align}
    \frac{dM}{dt}&=-2\omega_{I}M_c,\label{eq:BHmassevo}\\
    \frac{dJ}{dt}&=-\frac{2m\omega_I}{\omega_R}M_c,\label{eq:BHJevo}\\
    \frac{dM_c}{dt}&=2\omega_{I}M_c-P_{\text{GW}}. \label{eq:cloudmassevo}
\end{align}
This evolution also determines the spin history, $a(t)/M(t)=J(t)/M(t)^2$: GW quenching turns on earlier in the magnetic case, leaving only a modest $\Delta(a/M)$ in our fiducial examples (Fig.~\ref{Fig:Cloud evolution}).

\begin{table}[h]
\centering
\setlength{\tabcolsep}{3.5pt}
\renewcommand{\arraystretch}{1.05}
\begin{ruledtabular}
\begin{tabular}{lcc}
BH mass $M$ &
\begin{tabular}[c]{@{}c@{}}Magnetic $N=3$\\ $M\omega_I\!\sim\!10^{-6}$\end{tabular} &
\begin{tabular}[c]{@{}c@{}}Neutral 211\\ $M\omega_{I\rm (Kerr)}\!\sim\!2\times10^{-8}$\end{tabular} \\
\hline
$10\,M_{\odot}$     & $2.47\times10^{1}\,$s ($0.41$ min) & $1.23\times10^{3}\,$s ($20.5$ min) \\
$10^{4}\,M_{\odot}$ & $2.47\times10^{4}\,$s ($6.86$ hr)  & $1.23\times10^{6}\,$s ($14.3$ d) \\
$10^{6}\,M_{\odot}$ & $2.47\times10^{6}\,$s ($28.6$ d)   & $1.23\times10^{8}\,$s ($3.9$ yr) \\
\end{tabular}
\end{ruledtabular}
\caption{Using representative peak growth rates for a BH spin $a/M=0.9$, we estimate the e-folding time $\tau_e=(2\omega_I)^{-1}$.}
\label{tab:efolding}
\end{table}

The resulting $M_c(t)$ sets the GW luminosity and strain envelope via $P_{\rm GW}[M_c(t)]$, while evolving BH parameters shift $\omega_R(t)$ and hence the instantaneous frequency $f_{\rm GW}(t)$ (Fig.~\ref{Fig:GWwaveformcomparison}). 
In the nonrelativistic regime $f_{\rm GW}\simeq \tilde\omega/(2\pi)\simeq \mu/\pi$; equivalently, at fixed $\alpha=\mu M$ one has $f_{\rm GW}\propto 1/M$, so supermassive, intermediate-mass, and stellar-mass remnants populate the space-based~\cite{LISA:2017pwj,Robson2019LISASens,Cutler1998,TianQin2016,TaijiNSR2017,TaijiSources}, decihertz~\cite{Seto:2001qf,Yagi:2011wg,Crowder:2005nr}, and ground-based bands~\cite{Abbott:2016xvh,VIRGO:2014yos,KAGRA:2020tym,Reitze:2019iox,ET:2025xjr}, respectively. 
For Fig.~\ref{Fig:GWwaveformcomparison} we find $f_{\rm GW}\simeq 19.18\,{\rm mHz}$ (neutral, $\mu\simeq 4.0\times10^{-17}\,{\rm eV}$) and $f_{\rm GW}\simeq 28.26\,{\rm mHz}$ (magnetic, $\mu\simeq 6.0\times10^{-17}\,{\rm eV}$), in the LISA/TianQin/Taiji band~\cite{LISA:2017pwj,Robson2019LISASens,TianQin2016,TaijiNSR2017}. 
For a quasi-monochromatic signal, a coherent integration over $T$ gives $\rho\sim h_0\sqrt{T/S_n(f_{\rm GW})}$~\cite{LISA:2017pwj,Robson2019LISASens}. 
In our fiducial examples $h_0\sim10^{-21}$ (neutral) and can reach $h_0\sim10^{-20}$ (magnetic), with the charged signal turning on and saturating within $\mathcal{O}(10)$ days, motivating prompt post-merger follow-ups: a $\mathcal{O}(10)$--day baseline can already yield substantial signal-to-noise ratio, with longer baselines further improving sensitivity.

The observable message of the magnetic configuration is therefore the rapid growth, enhanced GW power, and the fixed cross-source selection rule. The polarization degree is not used as a primary discriminator in this work.

{\bf Conclusion and discussion}---We have shown that magnetic charge sourced by monopoles can qualitatively reshape the superradiance. 
Through the replacement $\ell(\ell+1)\!\to\!\ell(\ell+1)-q^2$ (encoded in $\ell_q$), the centrifugal barrier is reduced and the bound states deepen, yielding a parametrically faster instability than in the neutral Kerr case and enabling cloud growth and a near-monochromatic continuous-wave signal on observationally relevant timescales (days--years). 
For a charged complex scalar, a single monopole-harmonic sector has a stationary stress tensor and does not by itself produce a $2\omega$ annihilation source. The radiating component instead arises from coherent north--south, or charge-conjugate, cross terms, whose azimuthal phases enforce the selected channels $(\tilde\ell,\tilde m,\tilde\omega)=(2q,+2q,+2\omega)$ and its conjugate $(2q,-2q,-2\omega)$.
Together with the enhanced GW power relative to Kerr at comparable parameters, these features motivate targeted continuous-wave searches and prompt post-merger follow-up of rapidly spinning remnants. 
A practical caveat is the fate of the cloud's gauge charge if the boson is electromagnetically charged, since ambient plasma could neutralize or screen the cloud and weaken charged-cloud imprints; nevertheless, in the supermassive-BH regime the cloud can form within $\mathcal{O}(10)$ days, plausibly faster than environmental neutralization, with dark $U(1)$ (or sufficiently feebly coupled) charges providing an additional long-lived possibility. { In this work, we focus exclusively on the distinctive superradiant and GW phenomenology of rotating magnetically charged BHs, while neglecting additional gauge radiation channels, such as possible dark photon emission, which might further deplete the cloud and quantitatively reduce the signal strength. A detailed study of concrete particle-physics realizations in which such losses are sufficiently suppressed is left for future work. Within this viable parameter window, the prompt emergence of a rapidly growing continuous signal would provide a smoking gun of magnetic monopoles as well as ultralight bosons through GW observations.}

\let\oldaddcontentsline\addcontentsline
\renewcommand{\addcontentsline}[3]{}

{\bf Acknowledgments}---
This work is supported by the National Natural Science Foundation of China (NNSFC) under Grant No.12475111, No.12205387, No.12375084
and the Fundamental Research Funds for the Central Universities, Sun Yat-sen University.

\bibliography{Reference}

\begin{thebibliography}{84}%
\makeatletter
\providecommand \@ifxundefined [1]{%
 \@ifx{#1\undefined}
}%
\providecommand \@ifnum [1]{%
 \ifnum #1\expandafter \@firstoftwo
 \else \expandafter \@secondoftwo
 \fi
}%
\providecommand \@ifx [1]{%
 \ifx #1\expandafter \@firstoftwo
 \else \expandafter \@secondoftwo
 \fi
}%
\providecommand \natexlab [1]{#1}%
\providecommand \enquote  [1]{``#1''}%
\providecommand \bibnamefont  [1]{#1}%
\providecommand \bibfnamefont [1]{#1}%
\providecommand \citenamefont [1]{#1}%
\providecommand \href@noop [0]{\@secondoftwo}%
\providecommand \href [0]{\begingroup \@sanitize@url \@href}%
\providecommand \@href[1]{\@@startlink{#1}\@@href}%
\providecommand \@@href[1]{\endgroup#1\@@endlink}%
\providecommand \@sanitize@url [0]{\catcode `\\12\catcode `\$12\catcode `\&12\catcode `\#12\catcode `\^12\catcode `\_12\catcode `\%12\relax}%
\providecommand \@@startlink[1]{}%
\providecommand \@@endlink[0]{}%
\providecommand \url  [0]{\begingroup\@sanitize@url \@url }%
\providecommand \@url [1]{\endgroup\@href {#1}{\urlprefix }}%
\providecommand \urlprefix  [0]{URL }%
\providecommand \Eprint [0]{\href }%
\providecommand \doibase [0]{https://doi.org/}%
\providecommand \selectlanguage [0]{\@gobble}%
\providecommand \bibinfo  [0]{\@secondoftwo}%
\providecommand \bibfield  [0]{\@secondoftwo}%
\providecommand \translation [1]{[#1]}%
\providecommand \BibitemOpen [0]{}%
\providecommand \bibitemStop [0]{}%
\providecommand \bibitemNoStop [0]{.\EOS\space}%
\providecommand \EOS [0]{\spacefactor3000\relax}%
\providecommand \BibitemShut  [1]{\csname bibitem#1\endcsname}%
\let\auto@bib@innerbib\@empty
\bibitem [{\citenamefont {Starobinskii}(1973)}]{Starobinskii:1973vzb}%
  \BibitemOpen
  \bibfield  {author} {\bibinfo {author} {\bibfnamefont {A.~A.}\ \bibnamefont {Starobinskii}},\ }\bibfield  {title} {\bibinfo {title} {{Amplification of waves during reflection from a rotating ''black hole''}},\ }\href@noop {} {\bibfield  {journal} {\bibinfo  {journal} {Sov. Phys. JETP}\ }\textbf {\bibinfo {volume} {37}},\ \bibinfo {pages} {28} (\bibinfo {year} {1973})}\BibitemShut {NoStop}%
\bibitem [{\citenamefont {Zouros}\ and\ \citenamefont {Eardley}(1979)}]{Zouros:1979iw}%
  \BibitemOpen
  \bibfield  {author} {\bibinfo {author} {\bibfnamefont {T.~J.~M.}\ \bibnamefont {Zouros}}\ and\ \bibinfo {author} {\bibfnamefont {D.~M.}\ \bibnamefont {Eardley}},\ }\bibfield  {title} {\bibinfo {title} {{INSTABILITIES OF MASSIVE SCALAR PERTURBATIONS OF A ROTATING BLACK HOLE}},\ }\href {https://doi.org/10.1016/0003-4916(79)90237-9} {\bibfield  {journal} {\bibinfo  {journal} {Annals Phys.}\ }\textbf {\bibinfo {volume} {118}},\ \bibinfo {pages} {139} (\bibinfo {year} {1979})}\BibitemShut {NoStop}%
\bibitem [{\citenamefont {Brito}\ \emph {et~al.}(2015{\natexlab{a}})\citenamefont {Brito}, \citenamefont {Cardoso},\ and\ \citenamefont {Pani}}]{Brito:2015oca}%
  \BibitemOpen
  \bibfield  {author} {\bibinfo {author} {\bibfnamefont {R.}~\bibnamefont {Brito}}, \bibinfo {author} {\bibfnamefont {V.}~\bibnamefont {Cardoso}},\ and\ \bibinfo {author} {\bibfnamefont {P.}~\bibnamefont {Pani}},\ }\bibfield  {title} {\bibinfo {title} {{Superradiance}: {New Frontiers in Black Hole Physics}},\ }\href {https://doi.org/10.1007/978-3-319-19000-6} {\bibfield  {journal} {\bibinfo  {journal} {Lect. Notes Phys.}\ }\textbf {\bibinfo {volume} {906}},\ \bibinfo {pages} {pp.1} (\bibinfo {year} {2015}{\natexlab{a}})},\ \Eprint {https://arxiv.org/abs/1501.06570} {arXiv:1501.06570 [gr-qc]} \BibitemShut {NoStop}%
\bibitem [{\citenamefont {Cardoso}\ \emph {et~al.}(2004)\citenamefont {Cardoso}, \citenamefont {Dias}, \citenamefont {Lemos},\ and\ \citenamefont {Yoshida}}]{Cardoso:2004nk}%
  \BibitemOpen
  \bibfield  {author} {\bibinfo {author} {\bibfnamefont {V.}~\bibnamefont {Cardoso}}, \bibinfo {author} {\bibfnamefont {O.~J.~C.}\ \bibnamefont {Dias}}, \bibinfo {author} {\bibfnamefont {J.~P.~S.}\ \bibnamefont {Lemos}},\ and\ \bibinfo {author} {\bibfnamefont {S.}~\bibnamefont {Yoshida}},\ }\bibfield  {title} {\bibinfo {title} {{The Black hole bomb and superradiant instabilities}},\ }\href {https://doi.org/10.1103/PhysRevD.70.049903} {\bibfield  {journal} {\bibinfo  {journal} {Phys. Rev. D}\ }\textbf {\bibinfo {volume} {70}},\ \bibinfo {pages} {044039} (\bibinfo {year} {2004})},\ \bibinfo {note} {[Erratum: Phys.Rev.D 70, 049903 (2004)]},\ \Eprint {https://arxiv.org/abs/hep-th/0404096} {arXiv:hep-th/0404096} \BibitemShut {NoStop}%
\bibitem [{\citenamefont {Detweiler}(1980)}]{PhysRevD.22.2323}%
  \BibitemOpen
  \bibfield  {author} {\bibinfo {author} {\bibfnamefont {S.}~\bibnamefont {Detweiler}},\ }\bibfield  {title} {\bibinfo {title} {Klein-gordon equation and rotating black holes},\ }\href {https://doi.org/10.1103/PhysRevD.22.2323} {\bibfield  {journal} {\bibinfo  {journal} {Phys. Rev. D}\ }\textbf {\bibinfo {volume} {22}},\ \bibinfo {pages} {2323} (\bibinfo {year} {1980})}\BibitemShut {NoStop}%
\bibitem [{\citenamefont {Dolan}(2007)}]{Dolan:2007mj}%
  \BibitemOpen
  \bibfield  {author} {\bibinfo {author} {\bibfnamefont {S.~R.}\ \bibnamefont {Dolan}},\ }\bibfield  {title} {\bibinfo {title} {{Instability of the massive Klein-Gordon field on the Kerr spacetime}},\ }\href {https://doi.org/10.1103/PhysRevD.76.084001} {\bibfield  {journal} {\bibinfo  {journal} {Phys. Rev. D}\ }\textbf {\bibinfo {volume} {76}},\ \bibinfo {pages} {084001} (\bibinfo {year} {2007})},\ \Eprint {https://arxiv.org/abs/0705.2880} {arXiv:0705.2880 [gr-qc]} \BibitemShut {NoStop}%
\bibitem [{\citenamefont {Dolan}(2018)}]{Dolan:2018dqv}%
  \BibitemOpen
  \bibfield  {author} {\bibinfo {author} {\bibfnamefont {S.~R.}\ \bibnamefont {Dolan}},\ }\bibfield  {title} {\bibinfo {title} {{Instability of the Proca field on Kerr spacetime}},\ }\href {https://doi.org/10.1103/PhysRevD.98.104006} {\bibfield  {journal} {\bibinfo  {journal} {Phys. Rev. D}\ }\textbf {\bibinfo {volume} {98}},\ \bibinfo {pages} {104006} (\bibinfo {year} {2018})},\ \Eprint {https://arxiv.org/abs/1806.01604} {arXiv:1806.01604 [gr-qc]} \BibitemShut {NoStop}%
\bibitem [{\citenamefont {East}(2017)}]{East:2017mrj}%
  \BibitemOpen
  \bibfield  {author} {\bibinfo {author} {\bibfnamefont {W.~E.}\ \bibnamefont {East}},\ }\bibfield  {title} {\bibinfo {title} {{Superradiant instability of massive vector fields around spinning black holes in the relativistic regime}},\ }\href {https://doi.org/10.1103/PhysRevD.96.024004} {\bibfield  {journal} {\bibinfo  {journal} {Phys. Rev. D}\ }\textbf {\bibinfo {volume} {96}},\ \bibinfo {pages} {024004} (\bibinfo {year} {2017})},\ \Eprint {https://arxiv.org/abs/1705.01544} {arXiv:1705.01544 [gr-qc]} \BibitemShut {NoStop}%
\bibitem [{\citenamefont {Xie}\ and\ \citenamefont {Huang}(2024)}]{Xie:2022uvp}%
  \BibitemOpen
  \bibfield  {author} {\bibinfo {author} {\bibfnamefont {N.}~\bibnamefont {Xie}}\ and\ \bibinfo {author} {\bibfnamefont {F.~P.}\ \bibnamefont {Huang}},\ }\bibfield  {title} {\bibinfo {title} {{Imprints of ultralight axions on the gravitational wave and pulsar timing measurement}},\ }\href {https://doi.org/10.1007/s11433-023-2172-7} {\bibfield  {journal} {\bibinfo  {journal} {Sci. China Phys. Mech. Astron.}\ }\textbf {\bibinfo {volume} {67}},\ \bibinfo {pages} {210411} (\bibinfo {year} {2024})},\ \Eprint {https://arxiv.org/abs/2207.11145} {arXiv:2207.11145 [hep-ph]} \BibitemShut {NoStop}%
\bibitem [{\citenamefont {Alexander}\ and\ \citenamefont {Yunes}(2009)}]{Alexander:2009tp}%
  \BibitemOpen
  \bibfield  {author} {\bibinfo {author} {\bibfnamefont {S.}~\bibnamefont {Alexander}}\ and\ \bibinfo {author} {\bibfnamefont {N.}~\bibnamefont {Yunes}},\ }\bibfield  {title} {\bibinfo {title} {{Chern-Simons Modified General Relativity}},\ }\href {https://doi.org/10.1016/j.physrep.2009.07.002} {\bibfield  {journal} {\bibinfo  {journal} {Phys. Rept.}\ }\textbf {\bibinfo {volume} {480}},\ \bibinfo {pages} {1} (\bibinfo {year} {2009})},\ \Eprint {https://arxiv.org/abs/0907.2562} {arXiv:0907.2562 [hep-th]} \BibitemShut {NoStop}%
\bibitem [{\citenamefont {Cornwall}(1997)}]{Cornwall:1997ms}%
  \BibitemOpen
  \bibfield  {author} {\bibinfo {author} {\bibfnamefont {J.~M.}\ \bibnamefont {Cornwall}},\ }\bibfield  {title} {\bibinfo {title} {{Speculations on primordial magnetic helicity}},\ }\href {https://doi.org/10.1103/PhysRevD.56.6146} {\bibfield  {journal} {\bibinfo  {journal} {Phys. Rev. D}\ }\textbf {\bibinfo {volume} {56}},\ \bibinfo {pages} {6146} (\bibinfo {year} {1997})},\ \Eprint {https://arxiv.org/abs/hep-th/9704022} {arXiv:hep-th/9704022} \BibitemShut {NoStop}%
\bibitem [{\citenamefont {Kahniashvili}\ \emph {et~al.}(2005)\citenamefont {Kahniashvili}, \citenamefont {Gogoberidze},\ and\ \citenamefont {Ratra}}]{PhysRevLett.95.151301}%
  \BibitemOpen
  \bibfield  {author} {\bibinfo {author} {\bibfnamefont {T.}~\bibnamefont {Kahniashvili}}, \bibinfo {author} {\bibfnamefont {G.}~\bibnamefont {Gogoberidze}},\ and\ \bibinfo {author} {\bibfnamefont {B.}~\bibnamefont {Ratra}},\ }\bibfield  {title} {\bibinfo {title} {Polarized cosmological gravitational waves from primordial helical turbulence},\ }\href {https://doi.org/10.1103/PhysRevLett.95.151301} {\bibfield  {journal} {\bibinfo  {journal} {Phys. Rev. Lett.}\ }\textbf {\bibinfo {volume} {95}},\ \bibinfo {pages} {151301} (\bibinfo {year} {2005})}\BibitemShut {NoStop}%
\bibitem [{\citenamefont {Roper~Pol}\ \emph {et~al.}(2020)\citenamefont {Roper~Pol}, \citenamefont {Mandal}, \citenamefont {Brandenburg}, \citenamefont {Kahniashvili},\ and\ \citenamefont {Kosowsky}}]{RoperPol:2019wvy}%
  \BibitemOpen
  \bibfield  {author} {\bibinfo {author} {\bibfnamefont {A.}~\bibnamefont {Roper~Pol}}, \bibinfo {author} {\bibfnamefont {S.}~\bibnamefont {Mandal}}, \bibinfo {author} {\bibfnamefont {A.}~\bibnamefont {Brandenburg}}, \bibinfo {author} {\bibfnamefont {T.}~\bibnamefont {Kahniashvili}},\ and\ \bibinfo {author} {\bibfnamefont {A.}~\bibnamefont {Kosowsky}},\ }\bibfield  {title} {\bibinfo {title} {{Numerical simulations of gravitational waves from early-universe turbulence}},\ }\href {https://doi.org/10.1103/PhysRevD.102.083512} {\bibfield  {journal} {\bibinfo  {journal} {Phys. Rev. D}\ }\textbf {\bibinfo {volume} {102}},\ \bibinfo {pages} {083512} (\bibinfo {year} {2020})},\ \Eprint {https://arxiv.org/abs/1903.08585} {arXiv:1903.08585 [astro-ph.CO]} \BibitemShut {NoStop}%
\bibitem [{\citenamefont {Seto}\ and\ \citenamefont {Taruya}(2007)}]{Seto:2007tn}%
  \BibitemOpen
  \bibfield  {author} {\bibinfo {author} {\bibfnamefont {N.}~\bibnamefont {Seto}}\ and\ \bibinfo {author} {\bibfnamefont {A.}~\bibnamefont {Taruya}},\ }\bibfield  {title} {\bibinfo {title} {{Measuring a Parity Violation Signature in the Early Universe via Ground-based Laser Interferometers}},\ }\href {https://doi.org/10.1103/PhysRevLett.99.121101} {\bibfield  {journal} {\bibinfo  {journal} {Phys. Rev. Lett.}\ }\textbf {\bibinfo {volume} {99}},\ \bibinfo {pages} {121101} (\bibinfo {year} {2007})},\ \Eprint {https://arxiv.org/abs/0707.0535} {arXiv:0707.0535 [astro-ph]} \BibitemShut {NoStop}%
\bibitem [{\citenamefont {Vachaspati}(2001)}]{Vachaspati:2001nb}%
  \BibitemOpen
  \bibfield  {author} {\bibinfo {author} {\bibfnamefont {T.}~\bibnamefont {Vachaspati}},\ }\bibfield  {title} {\bibinfo {title} {{Estimate of the primordial magnetic field helicity}},\ }\href {https://doi.org/10.1103/PhysRevLett.87.251302} {\bibfield  {journal} {\bibinfo  {journal} {Phys. Rev. Lett.}\ }\textbf {\bibinfo {volume} {87}},\ \bibinfo {pages} {251302} (\bibinfo {year} {2001})},\ \Eprint {https://arxiv.org/abs/astro-ph/0101261} {arXiv:astro-ph/0101261} \BibitemShut {NoStop}%
\bibitem [{\citenamefont {Arvanitaki}\ \emph {et~al.}(2015)\citenamefont {Arvanitaki}, \citenamefont {Baryakhtar},\ and\ \citenamefont {Huang}}]{Arvanitaki:2014wva}%
  \BibitemOpen
  \bibfield  {author} {\bibinfo {author} {\bibfnamefont {A.}~\bibnamefont {Arvanitaki}}, \bibinfo {author} {\bibfnamefont {M.}~\bibnamefont {Baryakhtar}},\ and\ \bibinfo {author} {\bibfnamefont {X.}~\bibnamefont {Huang}},\ }\bibfield  {title} {\bibinfo {title} {{Discovering the QCD Axion with Black Holes and Gravitational Waves}},\ }\href {https://doi.org/10.1103/PhysRevD.91.084011} {\bibfield  {journal} {\bibinfo  {journal} {Phys. Rev. D}\ }\textbf {\bibinfo {volume} {91}},\ \bibinfo {pages} {084011} (\bibinfo {year} {2015})},\ \Eprint {https://arxiv.org/abs/1411.2263} {arXiv:1411.2263 [hep-ph]} \BibitemShut {NoStop}%
\bibitem [{\citenamefont {Yoshino}\ and\ \citenamefont {Kodama}(2014)}]{Yoshino:2013ofa}%
  \BibitemOpen
  \bibfield  {author} {\bibinfo {author} {\bibfnamefont {H.}~\bibnamefont {Yoshino}}\ and\ \bibinfo {author} {\bibfnamefont {H.}~\bibnamefont {Kodama}},\ }\bibfield  {title} {\bibinfo {title} {{Gravitational radiation from an axion cloud around a black hole: Superradiant phase}},\ }\href {https://doi.org/10.1093/ptep/ptu029} {\bibfield  {journal} {\bibinfo  {journal} {PTEP}\ }\textbf {\bibinfo {volume} {2014}},\ \bibinfo {pages} {043E02} (\bibinfo {year} {2014})},\ \Eprint {https://arxiv.org/abs/1312.2326} {arXiv:1312.2326 [gr-qc]} \BibitemShut {NoStop}%
\bibitem [{\citenamefont {Brito}\ \emph {et~al.}(2017)\citenamefont {Brito}, \citenamefont {Ghosh}, \citenamefont {Barausse}, \citenamefont {Berti}, \citenamefont {Cardoso}, \citenamefont {Dvorkin}, \citenamefont {Klein},\ and\ \citenamefont {Pani}}]{Brito:2017zvb}%
  \BibitemOpen
  \bibfield  {author} {\bibinfo {author} {\bibfnamefont {R.}~\bibnamefont {Brito}}, \bibinfo {author} {\bibfnamefont {S.}~\bibnamefont {Ghosh}}, \bibinfo {author} {\bibfnamefont {E.}~\bibnamefont {Barausse}}, \bibinfo {author} {\bibfnamefont {E.}~\bibnamefont {Berti}}, \bibinfo {author} {\bibfnamefont {V.}~\bibnamefont {Cardoso}}, \bibinfo {author} {\bibfnamefont {I.}~\bibnamefont {Dvorkin}}, \bibinfo {author} {\bibfnamefont {A.}~\bibnamefont {Klein}},\ and\ \bibinfo {author} {\bibfnamefont {P.}~\bibnamefont {Pani}},\ }\bibfield  {title} {\bibinfo {title} {{Gravitational wave searches for ultralight bosons with LIGO and LISA}},\ }\href {https://doi.org/10.1103/PhysRevD.96.064050} {\bibfield  {journal} {\bibinfo  {journal} {Phys. Rev. D}\ }\textbf {\bibinfo {volume} {96}},\ \bibinfo {pages} {064050} (\bibinfo {year} {2017})},\ \Eprint {https://arxiv.org/abs/1706.06311} {arXiv:1706.06311 [gr-qc]} \BibitemShut {NoStop}%
\bibitem [{\citenamefont {Isi}\ \emph {et~al.}(2019)\citenamefont {Isi}, \citenamefont {Sun}, \citenamefont {Brito},\ and\ \citenamefont {Melatos}}]{Isi:2018pzk}%
  \BibitemOpen
  \bibfield  {author} {\bibinfo {author} {\bibfnamefont {M.}~\bibnamefont {Isi}}, \bibinfo {author} {\bibfnamefont {L.}~\bibnamefont {Sun}}, \bibinfo {author} {\bibfnamefont {R.}~\bibnamefont {Brito}},\ and\ \bibinfo {author} {\bibfnamefont {A.}~\bibnamefont {Melatos}},\ }\bibfield  {title} {\bibinfo {title} {{Directed searches for gravitational waves from ultralight bosons}},\ }\href {https://doi.org/10.1103/PhysRevD.99.084042} {\bibfield  {journal} {\bibinfo  {journal} {Phys. Rev. D}\ }\textbf {\bibinfo {volume} {99}},\ \bibinfo {pages} {084042} (\bibinfo {year} {2019})},\ \bibinfo {note} {[Erratum: Phys.Rev.D 102, 049901 (2020)]},\ \Eprint {https://arxiv.org/abs/1810.03812} {arXiv:1810.03812 [gr-qc]} \BibitemShut {NoStop}%
\bibitem [{\citenamefont {Yang}\ and\ \citenamefont {Huang}(2023)}]{Yang:2023vwm}%
  \BibitemOpen
  \bibfield  {author} {\bibinfo {author} {\bibfnamefont {J.}~\bibnamefont {Yang}}\ and\ \bibinfo {author} {\bibfnamefont {F.~P.}\ \bibnamefont {Huang}},\ }\bibfield  {title} {\bibinfo {title} {{Gravitational waves from axions annihilation through quantum field theory}},\ }\href {https://doi.org/10.1103/PhysRevD.108.103002} {\bibfield  {journal} {\bibinfo  {journal} {Phys. Rev. D}\ }\textbf {\bibinfo {volume} {108}},\ \bibinfo {pages} {103002} (\bibinfo {year} {2023})},\ \Eprint {https://arxiv.org/abs/2306.12375} {arXiv:2306.12375 [hep-ph]} \BibitemShut {NoStop}%
\bibitem [{\citenamefont {Abbott}\ \emph {et~al.}(2022)\citenamefont {Abbott}, \citenamefont {others (LIGO Scientific~Collaboration}, \citenamefont {Collaboration},\ and\ \citenamefont {Collaboration)}}]{Abbott:2022ScalarCloudsO3}%
  \BibitemOpen
  \bibfield  {author} {\bibinfo {author} {\bibfnamefont {R.}~\bibnamefont {Abbott}}, \bibinfo {author} {\bibnamefont {others (LIGO Scientific~Collaboration}}, \bibinfo {author} {\bibfnamefont {V.}~\bibnamefont {Collaboration}},\ and\ \bibinfo {author} {\bibfnamefont {K.}~\bibnamefont {Collaboration)}},\ }\bibfield  {title} {\bibinfo {title} {All-sky search for gravitational wave emission from scalar boson clouds around spinning black holes in ligo o3 data},\ }\href {https://doi.org/10.1103/PhysRevD.105.102001} {\bibfield  {journal} {\bibinfo  {journal} {Phys. Rev. D}\ }\textbf {\bibinfo {volume} {105}},\ \bibinfo {pages} {102001} (\bibinfo {year} {2022})},\ \Eprint {https://arxiv.org/abs/2111.15507} {arXiv:2111.15507 [gr-qc]} \BibitemShut {NoStop}%
\bibitem [{\citenamefont {Siemonsen}\ \emph {et~al.}(2023{\natexlab{a}})\citenamefont {Siemonsen}, \citenamefont {May},\ and\ \citenamefont {East}}]{PhysRevD.107.104003}%
  \BibitemOpen
  \bibfield  {author} {\bibinfo {author} {\bibfnamefont {N.}~\bibnamefont {Siemonsen}}, \bibinfo {author} {\bibfnamefont {T.}~\bibnamefont {May}},\ and\ \bibinfo {author} {\bibfnamefont {W.~E.}\ \bibnamefont {East}},\ }\bibfield  {title} {\bibinfo {title} {Modeling the black hole superradiance gravitational waveform},\ }\href {https://doi.org/10.1103/PhysRevD.107.104003} {\bibfield  {journal} {\bibinfo  {journal} {Phys. Rev. D}\ }\textbf {\bibinfo {volume} {107}},\ \bibinfo {pages} {104003} (\bibinfo {year} {2023}{\natexlab{a}})}\BibitemShut {NoStop}%
\bibitem [{\citenamefont {Baumann}\ \emph {et~al.}(2019)\citenamefont {Baumann}, \citenamefont {Chia}, \citenamefont {Stout},\ and\ \citenamefont {ter Haar}}]{Baumann:2019eav}%
  \BibitemOpen
  \bibfield  {author} {\bibinfo {author} {\bibfnamefont {D.}~\bibnamefont {Baumann}}, \bibinfo {author} {\bibfnamefont {H.~S.}\ \bibnamefont {Chia}}, \bibinfo {author} {\bibfnamefont {J.}~\bibnamefont {Stout}},\ and\ \bibinfo {author} {\bibfnamefont {L.}~\bibnamefont {ter Haar}},\ }\bibfield  {title} {\bibinfo {title} {{The Spectra of Gravitational Atoms}},\ }\href {https://doi.org/10.1088/1475-7516/2019/12/006} {\bibfield  {journal} {\bibinfo  {journal} {JCAP}\ }\textbf {\bibinfo {volume} {12}},\ \bibinfo {pages} {006}},\ \Eprint {https://arxiv.org/abs/1908.10370} {arXiv:1908.10370 [gr-qc]} \BibitemShut {NoStop}%
\bibitem [{\citenamefont {Baumann}\ \emph {et~al.}(2020)\citenamefont {Baumann}, \citenamefont {Chia}, \citenamefont {Porto},\ and\ \citenamefont {Stout}}]{Baumann:2019ztm}%
  \BibitemOpen
  \bibfield  {author} {\bibinfo {author} {\bibfnamefont {D.}~\bibnamefont {Baumann}}, \bibinfo {author} {\bibfnamefont {H.~S.}\ \bibnamefont {Chia}}, \bibinfo {author} {\bibfnamefont {R.~A.}\ \bibnamefont {Porto}},\ and\ \bibinfo {author} {\bibfnamefont {J.}~\bibnamefont {Stout}},\ }\bibfield  {title} {\bibinfo {title} {{Gravitational Collider Physics}},\ }\href {https://doi.org/10.1103/PhysRevD.101.083019} {\bibfield  {journal} {\bibinfo  {journal} {Phys. Rev. D}\ }\textbf {\bibinfo {volume} {101}},\ \bibinfo {pages} {083019} (\bibinfo {year} {2020})},\ \Eprint {https://arxiv.org/abs/1912.04932} {arXiv:1912.04932 [gr-qc]} \BibitemShut {NoStop}%
\bibitem [{\citenamefont {Xie}\ and\ \citenamefont {Huang}(2025)}]{Xie:2025npy}%
  \BibitemOpen
  \bibfield  {author} {\bibinfo {author} {\bibfnamefont {N.}~\bibnamefont {Xie}}\ and\ \bibinfo {author} {\bibfnamefont {F.~P.}\ \bibnamefont {Huang}},\ }\bibfield  {title} {\bibinfo {title} {{Self-interaction effects on the Kerr black hole superradiance and their observational implications}},\ }\href {https://doi.org/10.1103/xmhn-cpv4} {\bibfield  {journal} {\bibinfo  {journal} {Phys. Rev. D}\ }\textbf {\bibinfo {volume} {112}},\ \bibinfo {pages} {055028} (\bibinfo {year} {2025})},\ \Eprint {https://arxiv.org/abs/2503.10347} {arXiv:2503.10347 [hep-ph]} \BibitemShut {NoStop}%
\bibitem [{\citenamefont {Yang}\ \emph {et~al.}(2024)\citenamefont {Yang}, \citenamefont {Xie},\ and\ \citenamefont {Huang}}]{Yang:2023aak}%
  \BibitemOpen
  \bibfield  {author} {\bibinfo {author} {\bibfnamefont {J.}~\bibnamefont {Yang}}, \bibinfo {author} {\bibfnamefont {N.}~\bibnamefont {Xie}},\ and\ \bibinfo {author} {\bibfnamefont {F.~P.}\ \bibnamefont {Huang}},\ }\bibfield  {title} {\bibinfo {title} {{Implication of nano-Hertz stochastic gravitational wave background on ultralight axion particles}},\ }\href {https://doi.org/10.1088/1475-7516/2024/11/045} {\bibfield  {journal} {\bibinfo  {journal} {JCAP}\ }\textbf {\bibinfo {volume} {11}},\ \bibinfo {pages} {045}},\ \Eprint {https://arxiv.org/abs/2306.17113} {arXiv:2306.17113 [hep-ph]} \BibitemShut {NoStop}%
\bibitem [{\citenamefont {Langacker}\ and\ \citenamefont {Pi}(1980{\natexlab{a}})}]{PhysRevLett.45.1}%
  \BibitemOpen
  \bibfield  {author} {\bibinfo {author} {\bibfnamefont {P.}~\bibnamefont {Langacker}}\ and\ \bibinfo {author} {\bibfnamefont {S.-Y.}\ \bibnamefont {Pi}},\ }\bibfield  {title} {\bibinfo {title} {Magnetic monopoles in grand unified theories},\ }\href {https://doi.org/10.1103/PhysRevLett.45.1} {\bibfield  {journal} {\bibinfo  {journal} {Phys. Rev. Lett.}\ }\textbf {\bibinfo {volume} {45}},\ \bibinfo {pages} {1} (\bibinfo {year} {1980}{\natexlab{a}})}\BibitemShut {NoStop}%
\bibitem [{\citenamefont {'t~Hooft}(1974)}]{tHooft:1974kcl}%
  \BibitemOpen
  \bibfield  {author} {\bibinfo {author} {\bibfnamefont {G.}~\bibnamefont {'t~Hooft}},\ }\bibfield  {title} {\bibinfo {title} {{Magnetic Monopoles in Unified Gauge Theories}},\ }\href {https://doi.org/10.1016/0550-3213(74)90486-6} {\bibfield  {journal} {\bibinfo  {journal} {Nucl. Phys. B}\ }\textbf {\bibinfo {volume} {79}},\ \bibinfo {pages} {276} (\bibinfo {year} {1974})}\BibitemShut {NoStop}%
\bibitem [{\citenamefont {Polyakov}(1974)}]{Polyakov:1974ek}%
  \BibitemOpen
  \bibfield  {author} {\bibinfo {author} {\bibfnamefont {A.~M.}\ \bibnamefont {Polyakov}},\ }\bibfield  {title} {\bibinfo {title} {{Particle Spectrum in Quantum Field Theory}},\ }\href@noop {} {\bibfield  {journal} {\bibinfo  {journal} {JETP Lett.}\ }\textbf {\bibinfo {volume} {20}},\ \bibinfo {pages} {194} (\bibinfo {year} {1974})}\BibitemShut {NoStop}%
\bibitem [{\citenamefont {Langacker}\ and\ \citenamefont {Pi}(1980{\natexlab{b}})}]{Langacker:1980kd}%
  \BibitemOpen
  \bibfield  {author} {\bibinfo {author} {\bibfnamefont {P.}~\bibnamefont {Langacker}}\ and\ \bibinfo {author} {\bibfnamefont {S.-Y.}\ \bibnamefont {Pi}},\ }\bibfield  {title} {\bibinfo {title} {{Magnetic Monopoles in Grand Unified Theories}},\ }\href {https://doi.org/10.1103/PhysRevLett.45.1} {\bibfield  {journal} {\bibinfo  {journal} {Phys. Rev. Lett.}\ }\textbf {\bibinfo {volume} {45}},\ \bibinfo {pages} {1} (\bibinfo {year} {1980}{\natexlab{b}})}\BibitemShut {NoStop}%
\bibitem [{\citenamefont {Murayama}\ and\ \citenamefont {Shu}(2010)}]{Murayama:2009nj}%
  \BibitemOpen
  \bibfield  {author} {\bibinfo {author} {\bibfnamefont {H.}~\bibnamefont {Murayama}}\ and\ \bibinfo {author} {\bibfnamefont {J.}~\bibnamefont {Shu}},\ }\bibfield  {title} {\bibinfo {title} {{Topological Dark Matter}},\ }\href {https://doi.org/10.1016/j.physletb.2010.02.037} {\bibfield  {journal} {\bibinfo  {journal} {Phys. Lett. B}\ }\textbf {\bibinfo {volume} {686}},\ \bibinfo {pages} {162} (\bibinfo {year} {2010})},\ \Eprint {https://arxiv.org/abs/0905.1720} {arXiv:0905.1720 [hep-ph]} \BibitemShut {NoStop}%
\bibitem [{\citenamefont {Zou}\ \emph {et~al.}(2025)\citenamefont {Zou}, \citenamefont {Zhang},\ and\ \citenamefont {Cho}}]{Zou:2025jlp}%
  \BibitemOpen
  \bibfield  {author} {\bibinfo {author} {\bibfnamefont {L.}~\bibnamefont {Zou}}, \bibinfo {author} {\bibfnamefont {P.}~\bibnamefont {Zhang}},\ and\ \bibinfo {author} {\bibfnamefont {Y.~M.}\ \bibnamefont {Cho}},\ }\href@noop {} {\bibinfo {title} {{New Monopoles in Non-Abelian Gauge Theories}}} (\bibinfo {year} {2025}),\ \Eprint {https://arxiv.org/abs/2512.08234} {arXiv:2512.08234 [hep-ph]} \BibitemShut {NoStop}%
\bibitem [{\citenamefont {Cho}\ and\ \citenamefont {Maison}(1997)}]{Cho:1996qd}%
  \BibitemOpen
  \bibfield  {author} {\bibinfo {author} {\bibfnamefont {Y.~M.}\ \bibnamefont {Cho}}\ and\ \bibinfo {author} {\bibfnamefont {D.}~\bibnamefont {Maison}},\ }\bibfield  {title} {\bibinfo {title} {{Monopoles in Weinberg-Salam model}},\ }\href {https://doi.org/10.1016/S0370-2693(96)01492-X} {\bibfield  {journal} {\bibinfo  {journal} {Phys. Lett. B}\ }\textbf {\bibinfo {volume} {391}},\ \bibinfo {pages} {360} (\bibinfo {year} {1997})},\ \Eprint {https://arxiv.org/abs/hep-th/9601028} {arXiv:hep-th/9601028} \BibitemShut {NoStop}%
\bibitem [{\citenamefont {Dirac}(1931)}]{Dirac:1931kp}%
  \BibitemOpen
  \bibfield  {author} {\bibinfo {author} {\bibfnamefont {P.~A.~M.}\ \bibnamefont {Dirac}},\ }\bibfield  {title} {\bibinfo {title} {{Quantised singularities in the electromagnetic field,}},\ }\href {https://doi.org/10.1098/rspa.1931.0130} {\bibfield  {journal} {\bibinfo  {journal} {Proc. Roy. Soc. Lond. A}\ }\textbf {\bibinfo {volume} {133}},\ \bibinfo {pages} {60} (\bibinfo {year} {1931})}\BibitemShut {NoStop}%
\bibitem [{\citenamefont {Preskill}(1984)}]{Preskill:1984gd}%
  \BibitemOpen
  \bibfield  {author} {\bibinfo {author} {\bibfnamefont {J.}~\bibnamefont {Preskill}},\ }\bibfield  {title} {\bibinfo {title} {{MAGNETIC MONOPOLES}},\ }\href {https://doi.org/10.1146/annurev.ns.34.120184.002333} {\bibfield  {journal} {\bibinfo  {journal} {Ann. Rev. Nucl. Part. Sci.}\ }\textbf {\bibinfo {volume} {34}},\ \bibinfo {pages} {461} (\bibinfo {year} {1984})}\BibitemShut {NoStop}%
\bibitem [{\citenamefont {Wu}\ and\ \citenamefont {Yang}(1976)}]{Wu:1976ge}%
  \BibitemOpen
  \bibfield  {author} {\bibinfo {author} {\bibfnamefont {T.~T.}\ \bibnamefont {Wu}}\ and\ \bibinfo {author} {\bibfnamefont {C.~N.}\ \bibnamefont {Yang}},\ }\bibfield  {title} {\bibinfo {title} {{Dirac Monopole Without Strings: Monopole Harmonics}},\ }\href {https://doi.org/10.1016/0550-3213(76)90143-7} {\bibfield  {journal} {\bibinfo  {journal} {Nucl. Phys. B}\ }\textbf {\bibinfo {volume} {107}},\ \bibinfo {pages} {365} (\bibinfo {year} {1976})}\BibitemShut {NoStop}%
\bibitem [{\citenamefont {Wu}\ and\ \citenamefont {Yang}(1975)}]{PhysRevD.12.3845}%
  \BibitemOpen
  \bibfield  {author} {\bibinfo {author} {\bibfnamefont {T.~T.}\ \bibnamefont {Wu}}\ and\ \bibinfo {author} {\bibfnamefont {C.~N.}\ \bibnamefont {Yang}},\ }\bibfield  {title} {\bibinfo {title} {Concept of nonintegrable phase factors and global formulation of gauge fields},\ }\href {https://doi.org/10.1103/PhysRevD.12.3845} {\bibfield  {journal} {\bibinfo  {journal} {Phys. Rev. D}\ }\textbf {\bibinfo {volume} {12}},\ \bibinfo {pages} {3845} (\bibinfo {year} {1975})}\BibitemShut {NoStop}%
\bibitem [{\citenamefont {Preskill}(1979)}]{PhysRevLett.43.1365}%
  \BibitemOpen
  \bibfield  {author} {\bibinfo {author} {\bibfnamefont {J.~P.}\ \bibnamefont {Preskill}},\ }\bibfield  {title} {\bibinfo {title} {Cosmological production of superheavy magnetic monopoles},\ }\href {https://doi.org/10.1103/PhysRevLett.43.1365} {\bibfield  {journal} {\bibinfo  {journal} {Phys. Rev. Lett.}\ }\textbf {\bibinfo {volume} {43}},\ \bibinfo {pages} {1365} (\bibinfo {year} {1979})}\BibitemShut {NoStop}%
\bibitem [{\citenamefont {Longo}(1982)}]{Longo:1981jp}%
  \BibitemOpen
  \bibfield  {author} {\bibinfo {author} {\bibfnamefont {M.~J.}\ \bibnamefont {Longo}},\ }\bibfield  {title} {\bibinfo {title} {{Massive Magnetic Monopoles: Indirect and Direct Limits on Their Number Density and Flux}},\ }\href {https://doi.org/10.1103/PhysRevD.25.2399} {\bibfield  {journal} {\bibinfo  {journal} {Phys. Rev. D}\ }\textbf {\bibinfo {volume} {25}},\ \bibinfo {pages} {2399} (\bibinfo {year} {1982})}\BibitemShut {NoStop}%
\bibitem [{\citenamefont {Stojkovic}\ and\ \citenamefont {Freese}(2005)}]{Stojkovic:2004hz}%
  \BibitemOpen
  \bibfield  {author} {\bibinfo {author} {\bibfnamefont {D.}~\bibnamefont {Stojkovic}}\ and\ \bibinfo {author} {\bibfnamefont {K.}~\bibnamefont {Freese}},\ }\bibfield  {title} {\bibinfo {title} {{A Black hole solution to the cosmological monopole problem}},\ }\href {https://doi.org/10.1016/j.physletb.2004.12.019} {\bibfield  {journal} {\bibinfo  {journal} {Phys. Lett. B}\ }\textbf {\bibinfo {volume} {606}},\ \bibinfo {pages} {251} (\bibinfo {year} {2005})},\ \Eprint {https://arxiv.org/abs/hep-ph/0403248} {arXiv:hep-ph/0403248} \BibitemShut {NoStop}%
\bibitem [{\citenamefont {Stojkovic}\ \emph {et~al.}(2005)\citenamefont {Stojkovic}, \citenamefont {Freese},\ and\ \citenamefont {Starkman}}]{Stojkovic:2005zh}%
  \BibitemOpen
  \bibfield  {author} {\bibinfo {author} {\bibfnamefont {D.}~\bibnamefont {Stojkovic}}, \bibinfo {author} {\bibfnamefont {K.}~\bibnamefont {Freese}},\ and\ \bibinfo {author} {\bibfnamefont {G.~D.}\ \bibnamefont {Starkman}},\ }\bibfield  {title} {\bibinfo {title} {{Holes in the walls: Primordial black holes as a solution to the cosmological domain wall problem}},\ }\href {https://doi.org/10.1103/PhysRevD.72.045012} {\bibfield  {journal} {\bibinfo  {journal} {Phys. Rev. D}\ }\textbf {\bibinfo {volume} {72}},\ \bibinfo {pages} {045012} (\bibinfo {year} {2005})},\ \Eprint {https://arxiv.org/abs/hep-ph/0505026} {arXiv:hep-ph/0505026} \BibitemShut {NoStop}%
\bibitem [{\citenamefont {Giacomelli}\ and\ \citenamefont {Patrizii}(2003)}]{Giacomelli:2003yu}%
  \BibitemOpen
  \bibfield  {author} {\bibinfo {author} {\bibfnamefont {G.}~\bibnamefont {Giacomelli}}\ and\ \bibinfo {author} {\bibfnamefont {L.}~\bibnamefont {Patrizii}},\ }\bibfield  {title} {\bibinfo {title} {{Magnetic monopole searches}},\ }\href@noop {} {\bibfield  {journal} {\bibinfo  {journal} {ICTP Lect. Notes Ser.}\ }\textbf {\bibinfo {volume} {14}},\ \bibinfo {pages} {121} (\bibinfo {year} {2003})},\ \Eprint {https://arxiv.org/abs/hep-ex/0302011} {arXiv:hep-ex/0302011} \BibitemShut {NoStop}%
\bibitem [{\citenamefont {Aad}\ \emph {et~al.}(2023)\citenamefont {Aad} \emph {et~al.}}]{ATLAS:2023esy}%
  \BibitemOpen
  \bibfield  {author} {\bibinfo {author} {\bibfnamefont {G.}~\bibnamefont {Aad}} \emph {et~al.} (\bibinfo {collaboration} {ATLAS}),\ }\bibfield  {title} {\bibinfo {title} {{Search for magnetic monopoles and stable particles with high electric charges in $ \sqrt{s} $ = 13 TeV pp collisions with the ATLAS detector}},\ }\href {https://doi.org/10.1007/JHEP11(2023)112} {\bibfield  {journal} {\bibinfo  {journal} {JHEP}\ }\textbf {\bibinfo {volume} {11}},\ \bibinfo {pages} {112}},\ \Eprint {https://arxiv.org/abs/2308.04835} {arXiv:2308.04835 [hep-ex]} \BibitemShut {NoStop}%
\bibitem [{\citenamefont {Nakamura}\ \emph {et~al.}(2010)\citenamefont {Nakamura} \emph {et~al.}}]{ParticleDataGroup:2010dbb}%
  \BibitemOpen
  \bibfield  {author} {\bibinfo {author} {\bibfnamefont {K.}~\bibnamefont {Nakamura}} \emph {et~al.} (\bibinfo {collaboration} {Particle Data Group}),\ }\bibfield  {title} {\bibinfo {title} {{Review of particle physics}},\ }\href {https://doi.org/10.1088/0954-3899/37/7A/075021} {\bibfield  {journal} {\bibinfo  {journal} {J. Phys. G}\ }\textbf {\bibinfo {volume} {37}},\ \bibinfo {pages} {075021} (\bibinfo {year} {2010})}\BibitemShut {NoStop}%
\bibitem [{\citenamefont {Ambrosio}\ \emph {et~al.}(2002)\citenamefont {Ambrosio} \emph {et~al.}}]{MACRO:2002jdv}%
  \BibitemOpen
  \bibfield  {author} {\bibinfo {author} {\bibfnamefont {M.}~\bibnamefont {Ambrosio}} \emph {et~al.} (\bibinfo {collaboration} {MACRO}),\ }\bibfield  {title} {\bibinfo {title} {{Final results of magnetic monopole searches with the MACRO experiment}},\ }\href {https://doi.org/10.1140/epjc/s2002-01046-9} {\bibfield  {journal} {\bibinfo  {journal} {Eur. Phys. J. C}\ }\textbf {\bibinfo {volume} {25}},\ \bibinfo {pages} {511} (\bibinfo {year} {2002})},\ \Eprint {https://arxiv.org/abs/hep-ex/0207020} {arXiv:hep-ex/0207020} \BibitemShut {NoStop}%
\bibitem [{\citenamefont {Abbasi}\ \emph {et~al.}(2022)\citenamefont {Abbasi} \emph {et~al.}}]{IceCube:2021eye}%
  \BibitemOpen
  \bibfield  {author} {\bibinfo {author} {\bibfnamefont {R.}~\bibnamefont {Abbasi}} \emph {et~al.} (\bibinfo {collaboration} {IceCube}),\ }\bibfield  {title} {\bibinfo {title} {{Search for Relativistic Magnetic Monopoles with Eight Years of IceCube Data}},\ }\href {https://doi.org/10.1103/PhysRevLett.128.051101} {\bibfield  {journal} {\bibinfo  {journal} {Phys. Rev. Lett.}\ }\textbf {\bibinfo {volume} {128}},\ \bibinfo {pages} {051101} (\bibinfo {year} {2022})},\ \Eprint {https://arxiv.org/abs/2109.13719} {arXiv:2109.13719 [astro-ph.HE]} \BibitemShut {NoStop}%
\bibitem [{\citenamefont {Parker}(1970)}]{Parker:1970xv}%
  \BibitemOpen
  \bibfield  {author} {\bibinfo {author} {\bibfnamefont {E.~N.}\ \bibnamefont {Parker}},\ }\bibfield  {title} {\bibinfo {title} {{The Origin of Magnetic Fields}},\ }\href {https://doi.org/10.1086/150442} {\bibfield  {journal} {\bibinfo  {journal} {Astrophys. J.}\ }\textbf {\bibinfo {volume} {160}},\ \bibinfo {pages} {383} (\bibinfo {year} {1970})}\BibitemShut {NoStop}%
\bibitem [{\citenamefont {Turner}\ \emph {et~al.}(1982)\citenamefont {Turner}, \citenamefont {Parker},\ and\ \citenamefont {Bogdan}}]{Turner:1982ag}%
  \BibitemOpen
  \bibfield  {author} {\bibinfo {author} {\bibfnamefont {M.~S.}\ \bibnamefont {Turner}}, \bibinfo {author} {\bibfnamefont {E.~N.}\ \bibnamefont {Parker}},\ and\ \bibinfo {author} {\bibfnamefont {T.~J.}\ \bibnamefont {Bogdan}},\ }\bibfield  {title} {\bibinfo {title} {{Magnetic Monopoles and the Survival of Galactic Magnetic Fields}},\ }\href {https://doi.org/10.1103/PhysRevD.26.1296} {\bibfield  {journal} {\bibinfo  {journal} {Phys. Rev. D}\ }\textbf {\bibinfo {volume} {26}},\ \bibinfo {pages} {1296} (\bibinfo {year} {1982})}\BibitemShut {NoStop}%
\bibitem [{\citenamefont {Kobayashi}\ and\ \citenamefont {Perri}(2023)}]{Kobayashi:2023ryr}%
  \BibitemOpen
  \bibfield  {author} {\bibinfo {author} {\bibfnamefont {T.}~\bibnamefont {Kobayashi}}\ and\ \bibinfo {author} {\bibfnamefont {D.}~\bibnamefont {Perri}},\ }\bibfield  {title} {\bibinfo {title} {{Parker bounds on monopoles with arbitrary charge from galactic and primordial magnetic fields}},\ }\href {https://doi.org/10.1103/PhysRevD.108.083005} {\bibfield  {journal} {\bibinfo  {journal} {Phys. Rev. D}\ }\textbf {\bibinfo {volume} {108}},\ \bibinfo {pages} {083005} (\bibinfo {year} {2023})},\ \Eprint {https://arxiv.org/abs/2307.07553} {arXiv:2307.07553 [hep-ph]} \BibitemShut {NoStop}%
\bibitem [{\citenamefont {Zhang}\ \emph {et~al.}(2024)\citenamefont {Zhang}, \citenamefont {Zhang}, \citenamefont {Fu}, \citenamefont {Zhang},\ and\ \citenamefont {Zhang}}]{Zhang:2024mze}%
  \BibitemOpen
  \bibfield  {author} {\bibinfo {author} {\bibfnamefont {C.}~\bibnamefont {Zhang}}, \bibinfo {author} {\bibfnamefont {S.-H.}\ \bibnamefont {Zhang}}, \bibinfo {author} {\bibfnamefont {B.}~\bibnamefont {Fu}}, \bibinfo {author} {\bibfnamefont {J.-F.}\ \bibnamefont {Zhang}},\ and\ \bibinfo {author} {\bibfnamefont {X.}~\bibnamefont {Zhang}},\ }\bibfield  {title} {\bibinfo {title} {{On the cosmological abundance of magnetic monopoles}},\ }\href {https://doi.org/10.1007/JHEP08(2024)220} {\bibfield  {journal} {\bibinfo  {journal} {JHEP}\ }\textbf {\bibinfo {volume} {08}},\ \bibinfo {pages} {220}},\ \Eprint {https://arxiv.org/abs/2404.04926} {arXiv:2404.04926 [hep-ph]} \BibitemShut {NoStop}%
\bibitem [{\citenamefont {Graesser}\ \emph {et~al.}(2022)\citenamefont {Graesser}, \citenamefont {Shoemaker},\ and\ \citenamefont {Arellano}}]{Graesser:2021vkr}%
  \BibitemOpen
  \bibfield  {author} {\bibinfo {author} {\bibfnamefont {M.~L.}\ \bibnamefont {Graesser}}, \bibinfo {author} {\bibfnamefont {I.~M.}\ \bibnamefont {Shoemaker}},\ and\ \bibinfo {author} {\bibfnamefont {N.~T.}\ \bibnamefont {Arellano}},\ }\bibfield  {title} {\bibinfo {title} {{Milli-magnetic monopole dark matter and the survival of galactic magnetic fields}},\ }\href {https://doi.org/10.1007/JHEP03(2022)105} {\bibfield  {journal} {\bibinfo  {journal} {JHEP}\ }\textbf {\bibinfo {volume} {03}},\ \bibinfo {pages} {105}},\ \Eprint {https://arxiv.org/abs/2105.05769} {arXiv:2105.05769 [hep-ph]} \BibitemShut {NoStop}%
\bibitem [{\citenamefont {Gould}\ and\ \citenamefont {Rajantie}(2017)}]{Gould:2017zwi}%
  \BibitemOpen
  \bibfield  {author} {\bibinfo {author} {\bibfnamefont {O.}~\bibnamefont {Gould}}\ and\ \bibinfo {author} {\bibfnamefont {A.}~\bibnamefont {Rajantie}},\ }\bibfield  {title} {\bibinfo {title} {{Magnetic monopole mass bounds from heavy ion collisions and neutron stars}},\ }\href {https://doi.org/10.1103/PhysRevLett.119.241601} {\bibfield  {journal} {\bibinfo  {journal} {Phys. Rev. Lett.}\ }\textbf {\bibinfo {volume} {119}},\ \bibinfo {pages} {241601} (\bibinfo {year} {2017})},\ \Eprint {https://arxiv.org/abs/1705.07052} {arXiv:1705.07052 [hep-ph]} \BibitemShut {NoStop}%
\bibitem [{\citenamefont {Hook}\ and\ \citenamefont {Huang}(2017)}]{Hook:2017vyc}%
  \BibitemOpen
  \bibfield  {author} {\bibinfo {author} {\bibfnamefont {A.}~\bibnamefont {Hook}}\ and\ \bibinfo {author} {\bibfnamefont {J.}~\bibnamefont {Huang}},\ }\bibfield  {title} {\bibinfo {title} {{Bounding millimagnetically charged particles with magnetars}},\ }\href {https://doi.org/10.1103/PhysRevD.96.055010} {\bibfield  {journal} {\bibinfo  {journal} {Phys. Rev. D}\ }\textbf {\bibinfo {volume} {96}},\ \bibinfo {pages} {055010} (\bibinfo {year} {2017})},\ \Eprint {https://arxiv.org/abs/1705.01107} {arXiv:1705.01107 [hep-ph]} \BibitemShut {NoStop}%
\bibitem [{\citenamefont {Maldacena}(2021)}]{Maldacena:2020skw}%
  \BibitemOpen
  \bibfield  {author} {\bibinfo {author} {\bibfnamefont {J.}~\bibnamefont {Maldacena}},\ }\bibfield  {title} {\bibinfo {title} {{Comments on magnetic black holes}},\ }\href {https://doi.org/10.1007/JHEP04(2021)079} {\bibfield  {journal} {\bibinfo  {journal} {JHEP}\ }\textbf {\bibinfo {volume} {04}},\ \bibinfo {pages} {079}},\ \Eprint {https://arxiv.org/abs/2004.06084} {arXiv:2004.06084 [hep-th]} \BibitemShut {NoStop}%
\bibitem [{\citenamefont {Bai}\ \emph {et~al.}(2020)\citenamefont {Bai}, \citenamefont {Berger}, \citenamefont {Korwar},\ and\ \citenamefont {Orlofsky}}]{Bai:2020spd}%
  \BibitemOpen
  \bibfield  {author} {\bibinfo {author} {\bibfnamefont {Y.}~\bibnamefont {Bai}}, \bibinfo {author} {\bibfnamefont {J.}~\bibnamefont {Berger}}, \bibinfo {author} {\bibfnamefont {M.}~\bibnamefont {Korwar}},\ and\ \bibinfo {author} {\bibfnamefont {N.}~\bibnamefont {Orlofsky}},\ }\bibfield  {title} {\bibinfo {title} {{Phenomenology of magnetic black holes with electroweak-symmetric coronas}},\ }\href {https://doi.org/10.1007/JHEP10(2020)210} {\bibfield  {journal} {\bibinfo  {journal} {JHEP}\ }\textbf {\bibinfo {volume} {10}},\ \bibinfo {pages} {210}},\ \Eprint {https://arxiv.org/abs/2007.03703} {arXiv:2007.03703 [hep-ph]} \BibitemShut {NoStop}%
\bibitem [{\citenamefont {Diamond}\ and\ \citenamefont {Kaplan}(2022)}]{Diamond:2021scl}%
  \BibitemOpen
  \bibfield  {author} {\bibinfo {author} {\bibfnamefont {M.~D.}\ \bibnamefont {Diamond}}\ and\ \bibinfo {author} {\bibfnamefont {D.~E.}\ \bibnamefont {Kaplan}},\ }\bibfield  {title} {\bibinfo {title} {{Constraints on relic magnetic black holes}},\ }\href {https://doi.org/10.1007/JHEP03(2022)157} {\bibfield  {journal} {\bibinfo  {journal} {JHEP}\ }\textbf {\bibinfo {volume} {03}},\ \bibinfo {pages} {157}},\ \Eprint {https://arxiv.org/abs/2103.01850} {arXiv:2103.01850 [hep-ph]} \BibitemShut {NoStop}%
\bibitem [{\citenamefont {Zhang}\ and\ \citenamefont {Zhang}(2023)}]{Zhang:2023tfv}%
  \BibitemOpen
  \bibfield  {author} {\bibinfo {author} {\bibfnamefont {C.}~\bibnamefont {Zhang}}\ and\ \bibinfo {author} {\bibfnamefont {X.}~\bibnamefont {Zhang}},\ }\bibfield  {title} {\bibinfo {title} {{Gravitational capture of magnetic monopoles by primordial black holes in the early universe}},\ }\href {https://doi.org/10.1007/JHEP10(2023)037} {\bibfield  {journal} {\bibinfo  {journal} {JHEP}\ }\textbf {\bibinfo {volume} {10}},\ \bibinfo {pages} {037}},\ \Eprint {https://arxiv.org/abs/2302.07002} {arXiv:2302.07002 [hep-ph]} \BibitemShut {NoStop}%
\bibitem [{\citenamefont {Pere{\~n}iguez}\ \emph {et~al.}(2024)\citenamefont {Pere{\~n}iguez}, \citenamefont {de~Amicis}, \citenamefont {Brito},\ and\ \citenamefont {Panosso~Macedo}}]{Pereniguez:2024fkn}%
  \BibitemOpen
  \bibfield  {author} {\bibinfo {author} {\bibfnamefont {D.}~\bibnamefont {Pere{\~n}iguez}}, \bibinfo {author} {\bibfnamefont {M.}~\bibnamefont {de~Amicis}}, \bibinfo {author} {\bibfnamefont {R.}~\bibnamefont {Brito}},\ and\ \bibinfo {author} {\bibfnamefont {R.}~\bibnamefont {Panosso~Macedo}},\ }\bibfield  {title} {\bibinfo {title} {{Superradiant Instability of Magnetic Black Holes}},\ }\href {https://doi.org/10.1103/PhysRevD.110.104001} {\bibfield  {journal} {\bibinfo  {journal} {Phys. Rev. D}\ }\textbf {\bibinfo {volume} {110}},\ \bibinfo {pages} {104001} (\bibinfo {year} {2024})},\ \Eprint {https://arxiv.org/abs/2402.05178} {arXiv:2402.05178 [gr-qc]} \BibitemShut {NoStop}%
\bibitem [{\citenamefont {Siemonsen}\ \emph {et~al.}(2023{\natexlab{b}})\citenamefont {Siemonsen}, \citenamefont {May},\ and\ \citenamefont {East}}]{Siemonsen:2022yyf}%
  \BibitemOpen
  \bibfield  {author} {\bibinfo {author} {\bibfnamefont {N.}~\bibnamefont {Siemonsen}}, \bibinfo {author} {\bibfnamefont {T.}~\bibnamefont {May}},\ and\ \bibinfo {author} {\bibfnamefont {W.~E.}\ \bibnamefont {East}},\ }\bibfield  {title} {\bibinfo {title} {{Modeling the black hole superradiance gravitational waveform}},\ }\href {https://doi.org/10.1103/PhysRevD.107.104003} {\bibfield  {journal} {\bibinfo  {journal} {Phys. Rev. D}\ }\textbf {\bibinfo {volume} {107}},\ \bibinfo {pages} {104003} (\bibinfo {year} {2023}{\natexlab{b}})},\ \Eprint {https://arxiv.org/abs/2211.03845} {arXiv:2211.03845 [gr-qc]} \BibitemShut {NoStop}%
\bibitem [{\citenamefont {Guo}\ \emph {et~al.}(2023)\citenamefont {Guo}, \citenamefont {Bao},\ and\ \citenamefont {Zhang}}]{Guo:2022mpr}%
  \BibitemOpen
  \bibfield  {author} {\bibinfo {author} {\bibfnamefont {Y.-d.}\ \bibnamefont {Guo}}, \bibinfo {author} {\bibfnamefont {S.-s.}\ \bibnamefont {Bao}},\ and\ \bibinfo {author} {\bibfnamefont {H.}~\bibnamefont {Zhang}},\ }\bibfield  {title} {\bibinfo {title} {{Subdominant modes of the scalar superradiant instability and gravitational wave beats}},\ }\href {https://doi.org/10.1103/PhysRevD.107.075009} {\bibfield  {journal} {\bibinfo  {journal} {Phys. Rev. D}\ }\textbf {\bibinfo {volume} {107}},\ \bibinfo {pages} {075009} (\bibinfo {year} {2023})},\ \Eprint {https://arxiv.org/abs/2212.07186} {arXiv:2212.07186 [gr-qc]} \BibitemShut {NoStop}%
\bibitem [{\citenamefont {Brito}\ \emph {et~al.}(2015{\natexlab{b}})\citenamefont {Brito}, \citenamefont {Cardoso},\ and\ \citenamefont {Pani}}]{Brito:2014wla}%
  \BibitemOpen
  \bibfield  {author} {\bibinfo {author} {\bibfnamefont {R.}~\bibnamefont {Brito}}, \bibinfo {author} {\bibfnamefont {V.}~\bibnamefont {Cardoso}},\ and\ \bibinfo {author} {\bibfnamefont {P.}~\bibnamefont {Pani}},\ }\bibfield  {title} {\bibinfo {title} {{Black holes as particle detectors: evolution of superradiant instabilities}},\ }\href {https://doi.org/10.1088/0264-9381/32/13/134001} {\bibfield  {journal} {\bibinfo  {journal} {Class. Quant. Grav.}\ }\textbf {\bibinfo {volume} {32}},\ \bibinfo {pages} {134001} (\bibinfo {year} {2015}{\natexlab{b}})},\ \Eprint {https://arxiv.org/abs/1411.0686} {arXiv:1411.0686 [gr-qc]} \BibitemShut {NoStop}%
\bibitem [{\citenamefont {East}(2018)}]{East:2018glu}%
  \BibitemOpen
  \bibfield  {author} {\bibinfo {author} {\bibfnamefont {W.~E.}\ \bibnamefont {East}},\ }\bibfield  {title} {\bibinfo {title} {{Massive Boson Superradiant Instability of Black Holes: Nonlinear Growth, Saturation, and Gravitational Radiation}},\ }\href {https://doi.org/10.1103/PhysRevLett.121.131104} {\bibfield  {journal} {\bibinfo  {journal} {Phys. Rev. Lett.}\ }\textbf {\bibinfo {volume} {121}},\ \bibinfo {pages} {131104} (\bibinfo {year} {2018})},\ \Eprint {https://arxiv.org/abs/1807.00043} {arXiv:1807.00043 [gr-qc]} \BibitemShut {NoStop}%
\bibitem [{\citenamefont {Mouland}\ and\ \citenamefont {Tong}(2024)}]{Mouland:2024zgk}%
  \BibitemOpen
  \bibfield  {author} {\bibinfo {author} {\bibfnamefont {R.}~\bibnamefont {Mouland}}\ and\ \bibinfo {author} {\bibfnamefont {D.}~\bibnamefont {Tong}},\ }\bibfield  {title} {\bibinfo {title} {{On the Hilbert Space of Dyons}},\ }\href {https://doi.org/10.1103/PhysRevD.110.085014} {\bibfield  {journal} {\bibinfo  {journal} {Phys. Rev. D}\ }\textbf {\bibinfo {volume} {110}},\ \bibinfo {pages} {085014} (\bibinfo {year} {2024})},\ \Eprint {https://arxiv.org/abs/2401.01924} {arXiv:2401.01924 [hep-th]} \BibitemShut {NoStop}%
\bibitem [{\citenamefont {Myung}(2022)}]{Myung:2022dpp}%
  \BibitemOpen
  \bibfield  {author} {\bibinfo {author} {\bibfnamefont {Y.~S.}\ \bibnamefont {Myung}},\ }\bibfield  {title} {\bibinfo {title} {{Superradiant instability of dyonic Reissner{\textendash}Nordstr{\"o}m black holes}},\ }\href {https://doi.org/10.1140/epjc/s10052-022-10911-y} {\bibfield  {journal} {\bibinfo  {journal} {Eur. Phys. J. C}\ }\textbf {\bibinfo {volume} {82}},\ \bibinfo {pages} {933} (\bibinfo {year} {2022})},\ \Eprint {https://arxiv.org/abs/2206.01390} {arXiv:2206.01390 [gr-qc]} \BibitemShut {NoStop}%
\bibitem [{\citenamefont {Zou}\ \emph {et~al.}(2021)\citenamefont {Zou}, \citenamefont {Xu}, \citenamefont {Mai},\ and\ \citenamefont {Huang}}]{Zou:2021mwa}%
  \BibitemOpen
  \bibfield  {author} {\bibinfo {author} {\bibfnamefont {Y.-F.}\ \bibnamefont {Zou}}, \bibinfo {author} {\bibfnamefont {J.-H.}\ \bibnamefont {Xu}}, \bibinfo {author} {\bibfnamefont {Z.-F.}\ \bibnamefont {Mai}},\ and\ \bibinfo {author} {\bibfnamefont {J.-H.}\ \bibnamefont {Huang}},\ }\bibfield  {title} {\bibinfo {title} {{Dyonic Reissner{\textendash}Nordstrom black holes and superradiant stability}},\ }\href {https://doi.org/10.1140/epjc/s10052-021-09642-3} {\bibfield  {journal} {\bibinfo  {journal} {Eur. Phys. J. C}\ }\textbf {\bibinfo {volume} {81}},\ \bibinfo {pages} {855} (\bibinfo {year} {2021})},\ \Eprint {https://arxiv.org/abs/2105.14702} {arXiv:2105.14702 [gr-qc]} \BibitemShut {NoStop}%
\bibitem [{\citenamefont {Arvanitaki}\ and\ \citenamefont {Dubovsky}(2011)}]{PhysRevD.83.044026}%
  \BibitemOpen
  \bibfield  {author} {\bibinfo {author} {\bibfnamefont {A.}~\bibnamefont {Arvanitaki}}\ and\ \bibinfo {author} {\bibfnamefont {S.}~\bibnamefont {Dubovsky}},\ }\bibfield  {title} {\bibinfo {title} {Exploring the string axiverse with precision black hole physics},\ }\href {https://doi.org/10.1103/PhysRevD.83.044026} {\bibfield  {journal} {\bibinfo  {journal} {Phys. Rev. D}\ }\textbf {\bibinfo {volume} {83}},\ \bibinfo {pages} {044026} (\bibinfo {year} {2011})}\BibitemShut {NoStop}%
\bibitem [{\citenamefont {Amaro-Seoane}\ \emph {et~al.}(2017)\citenamefont {Amaro-Seoane} \emph {et~al.}}]{LISA:2017pwj}%
  \BibitemOpen
  \bibfield  {author} {\bibinfo {author} {\bibfnamefont {P.}~\bibnamefont {Amaro-Seoane}} \emph {et~al.} (\bibinfo {collaboration} {LISA}),\ }\href@noop {} {\bibinfo {title} {Laser interferometer space antenna}} (\bibinfo {year} {2017}),\ \Eprint {https://arxiv.org/abs/1702.00786} {arXiv:1702.00786 [astro-ph.IM]} \BibitemShut {NoStop}%
\bibitem [{\citenamefont {Robson}\ \emph {et~al.}(2019)\citenamefont {Robson}, \citenamefont {Cornish},\ and\ \citenamefont {Liu}}]{Robson2019LISASens}%
  \BibitemOpen
  \bibfield  {author} {\bibinfo {author} {\bibfnamefont {T.}~\bibnamefont {Robson}}, \bibinfo {author} {\bibfnamefont {N.~J.}\ \bibnamefont {Cornish}},\ and\ \bibinfo {author} {\bibfnamefont {C.}~\bibnamefont {Liu}},\ }\bibfield  {title} {\bibinfo {title} {The construction and use of {LISA} sensitivity curves},\ }\href {https://doi.org/10.1088/1361-6382/ab1101} {\bibfield  {journal} {\bibinfo  {journal} {Classical and Quantum Gravity}\ }\textbf {\bibinfo {volume} {36}},\ \bibinfo {pages} {105011} (\bibinfo {year} {2019})},\ \Eprint {https://arxiv.org/abs/1803.01944} {arXiv:1803.01944 [astro-ph.HE]} \BibitemShut {NoStop}%
\bibitem [{\citenamefont {Cutler}(1998)}]{Cutler1998}%
  \BibitemOpen
  \bibfield  {author} {\bibinfo {author} {\bibfnamefont {C.}~\bibnamefont {Cutler}},\ }\bibfield  {title} {\bibinfo {title} {Angular resolution of the {LISA} gravitational wave detector},\ }\href {https://doi.org/10.1103/PhysRevD.57.7089} {\bibfield  {journal} {\bibinfo  {journal} {Physical Review D}\ }\textbf {\bibinfo {volume} {57}},\ \bibinfo {pages} {7089} (\bibinfo {year} {1998})}\BibitemShut {NoStop}%
\bibitem [{\citenamefont {Luo}\ \emph {et~al.}(2016)\citenamefont {Luo} \emph {et~al.}}]{TianQin2016}%
  \BibitemOpen
  \bibfield  {author} {\bibinfo {author} {\bibfnamefont {J.}~\bibnamefont {Luo}} \emph {et~al.},\ }\bibfield  {title} {\bibinfo {title} {Tianqin: a space-borne gravitational wave detector},\ }\href {https://doi.org/10.1088/0264-9381/33/3/035010} {\bibfield  {journal} {\bibinfo  {journal} {Classical and Quantum Gravity}\ }\textbf {\bibinfo {volume} {33}},\ \bibinfo {pages} {035010} (\bibinfo {year} {2016})},\ \Eprint {https://arxiv.org/abs/1512.02076} {arXiv:1512.02076 [astro-ph.IM]} \BibitemShut {NoStop}%
\bibitem [{\citenamefont {Hu}\ and\ \citenamefont {Wu}(2017)}]{TaijiNSR2017}%
  \BibitemOpen
  \bibfield  {author} {\bibinfo {author} {\bibfnamefont {W.-R.}\ \bibnamefont {Hu}}\ and\ \bibinfo {author} {\bibfnamefont {Y.-L.}\ \bibnamefont {Wu}},\ }\bibfield  {title} {\bibinfo {title} {The {Taiji} program in space for gravitational wave physics and the nature of gravity},\ }\href {https://doi.org/10.1093/nsr/nwx116} {\bibfield  {journal} {\bibinfo  {journal} {National Science Review}\ }\textbf {\bibinfo {volume} {4}},\ \bibinfo {pages} {685} (\bibinfo {year} {2017})}\BibitemShut {NoStop}%
\bibitem [{\citenamefont {Ruan}\ \emph {et~al.}(2020)\citenamefont {Ruan}, \citenamefont {Guo}, \citenamefont {Cai},\ and\ \citenamefont {Zhang}}]{TaijiSources}%
  \BibitemOpen
  \bibfield  {author} {\bibinfo {author} {\bibfnamefont {W.-H.}\ \bibnamefont {Ruan}}, \bibinfo {author} {\bibfnamefont {Z.-K.}\ \bibnamefont {Guo}}, \bibinfo {author} {\bibfnamefont {R.-G.}\ \bibnamefont {Cai}},\ and\ \bibinfo {author} {\bibfnamefont {Y.-Z.}\ \bibnamefont {Zhang}},\ }\href@noop {} {\bibinfo {title} {{Taiji} program: Gravitational-wave sources}} (\bibinfo {year} {2020}),\ \Eprint {https://arxiv.org/abs/1807.09495} {arXiv:1807.09495 [gr-qc]} \BibitemShut {NoStop}%
\bibitem [{\citenamefont {Seto}\ \emph {et~al.}(2001)\citenamefont {Seto}, \citenamefont {Kawamura},\ and\ \citenamefont {Nakamura}}]{Seto:2001qf}%
  \BibitemOpen
  \bibfield  {author} {\bibinfo {author} {\bibfnamefont {N.}~\bibnamefont {Seto}}, \bibinfo {author} {\bibfnamefont {S.}~\bibnamefont {Kawamura}},\ and\ \bibinfo {author} {\bibfnamefont {T.}~\bibnamefont {Nakamura}},\ }\bibfield  {title} {\bibinfo {title} {{Possibility of direct measurement of the acceleration of the universe using 0.1-Hz band laser interferometer gravitational wave antenna in space}},\ }\href {https://doi.org/10.1103/PhysRevLett.87.221103} {\bibfield  {journal} {\bibinfo  {journal} {Phys. Rev. Lett.}\ }\textbf {\bibinfo {volume} {87}},\ \bibinfo {pages} {221103} (\bibinfo {year} {2001})},\ \Eprint {https://arxiv.org/abs/astro-ph/0108011} {arXiv:astro-ph/0108011} \BibitemShut {NoStop}%
\bibitem [{\citenamefont {Yagi}\ and\ \citenamefont {Seto}(2011)}]{Yagi:2011wg}%
  \BibitemOpen
  \bibfield  {author} {\bibinfo {author} {\bibfnamefont {K.}~\bibnamefont {Yagi}}\ and\ \bibinfo {author} {\bibfnamefont {N.}~\bibnamefont {Seto}},\ }\bibfield  {title} {\bibinfo {title} {{Detector configuration of DECIGO/BBO and identification of cosmological neutron-star binaries}},\ }\href {https://doi.org/10.1103/PhysRevD.83.044011} {\bibfield  {journal} {\bibinfo  {journal} {Phys. Rev. D}\ }\textbf {\bibinfo {volume} {83}},\ \bibinfo {pages} {044011} (\bibinfo {year} {2011})},\ \bibinfo {note} {[Erratum: Phys.Rev.D 95, 109901 (2017)]},\ \Eprint {https://arxiv.org/abs/1101.3940} {arXiv:1101.3940 [astro-ph.CO]} \BibitemShut {NoStop}%
\bibitem [{\citenamefont {Crowder}\ and\ \citenamefont {Cornish}(2005)}]{Crowder:2005nr}%
  \BibitemOpen
  \bibfield  {author} {\bibinfo {author} {\bibfnamefont {J.}~\bibnamefont {Crowder}}\ and\ \bibinfo {author} {\bibfnamefont {N.~J.}\ \bibnamefont {Cornish}},\ }\bibfield  {title} {\bibinfo {title} {{Beyond LISA: Exploring future gravitational wave missions}},\ }\href {https://doi.org/10.1103/PhysRevD.72.083005} {\bibfield  {journal} {\bibinfo  {journal} {Phys. Rev. D}\ }\textbf {\bibinfo {volume} {72}},\ \bibinfo {pages} {083005} (\bibinfo {year} {2005})},\ \Eprint {https://arxiv.org/abs/gr-qc/0506015} {arXiv:gr-qc/0506015} \BibitemShut {NoStop}%
\bibitem [{\citenamefont {Abbott}\ \emph {et~al.}(2016)\citenamefont {Abbott} \emph {et~al.}}]{Abbott:2016xvh}%
  \BibitemOpen
  \bibfield  {author} {\bibinfo {author} {\bibfnamefont {B.~P.}\ \bibnamefont {Abbott}} \emph {et~al.},\ }\bibfield  {title} {\bibinfo {title} {{Sensitivity of the Advanced LIGO detectors at the beginning of gravitational wave astronomy}},\ }\href {https://doi.org/10.1103/PhysRevD.93.112004} {\bibfield  {journal} {\bibinfo  {journal} {Phys. Rev. D}\ }\textbf {\bibinfo {volume} {93}},\ \bibinfo {pages} {112004} (\bibinfo {year} {2016})},\ \bibinfo {note} {[Addendum: Phys.Rev.D 97, 059901 (2018)]},\ \Eprint {https://arxiv.org/abs/1604.00439} {arXiv:1604.00439 [astro-ph.IM]} \BibitemShut {NoStop}%
\bibitem [{\citenamefont {Acernese}\ \emph {et~al.}(2015)\citenamefont {Acernese} \emph {et~al.}}]{VIRGO:2014yos}%
  \BibitemOpen
  \bibfield  {author} {\bibinfo {author} {\bibfnamefont {F.}~\bibnamefont {Acernese}} \emph {et~al.} (\bibinfo {collaboration} {VIRGO}),\ }\bibfield  {title} {\bibinfo {title} {{Advanced Virgo: a second-generation interferometric gravitational wave detector}},\ }\href {https://doi.org/10.1088/0264-9381/32/2/024001} {\bibfield  {journal} {\bibinfo  {journal} {Class. Quant. Grav.}\ }\textbf {\bibinfo {volume} {32}},\ \bibinfo {pages} {024001} (\bibinfo {year} {2015})},\ \Eprint {https://arxiv.org/abs/1408.3978} {arXiv:1408.3978 [gr-qc]} \BibitemShut {NoStop}%
\bibitem [{\citenamefont {Akutsu}\ \emph {et~al.}(2021)\citenamefont {Akutsu} \emph {et~al.}}]{KAGRA:2020tym}%
  \BibitemOpen
  \bibfield  {author} {\bibinfo {author} {\bibfnamefont {T.}~\bibnamefont {Akutsu}} \emph {et~al.} (\bibinfo {collaboration} {KAGRA}),\ }\bibfield  {title} {\bibinfo {title} {{Overview of KAGRA: Detector design and construction history}},\ }\href {https://doi.org/10.1093/ptep/ptaa125} {\bibfield  {journal} {\bibinfo  {journal} {PTEP}\ }\textbf {\bibinfo {volume} {2021}},\ \bibinfo {pages} {05A101} (\bibinfo {year} {2021})},\ \Eprint {https://arxiv.org/abs/2005.05574} {arXiv:2005.05574 [physics.ins-det]} \BibitemShut {NoStop}%
\bibitem [{\citenamefont {Reitze}\ \emph {et~al.}(2019)\citenamefont {Reitze} \emph {et~al.}}]{Reitze:2019iox}%
  \BibitemOpen
  \bibfield  {author} {\bibinfo {author} {\bibfnamefont {D.}~\bibnamefont {Reitze}} \emph {et~al.},\ }\bibfield  {title} {\bibinfo {title} {{Cosmic Explorer: The U.S. Contribution to Gravitational-Wave Astronomy beyond LIGO}},\ }\href@noop {} {\bibfield  {journal} {\bibinfo  {journal} {Bull. Am. Astron. Soc.}\ }\textbf {\bibinfo {volume} {51}},\ \bibinfo {pages} {035} (\bibinfo {year} {2019})},\ \Eprint {https://arxiv.org/abs/1907.04833} {arXiv:1907.04833 [astro-ph.IM]} \BibitemShut {NoStop}%
\bibitem [{\citenamefont {Abac}\ \emph {et~al.}(2025)\citenamefont {Abac} \emph {et~al.}}]{ET:2025xjr}%
  \BibitemOpen
  \bibfield  {author} {\bibinfo {author} {\bibfnamefont {A.}~\bibnamefont {Abac}} \emph {et~al.} (\bibinfo {collaboration} {ET}),\ }\href@noop {} {\bibinfo {title} {The science of the einstein telescope}} (\bibinfo {year} {2025}),\ \Eprint {https://arxiv.org/abs/2503.12263} {arXiv:2503.12263 [gr-qc]} \BibitemShut {NoStop}%
\bibitem [{\citenamefont {MacRobert}(1955)}]{MacRobert1955HigherTF}%
  \BibitemOpen
  \bibfield  {author} {\bibinfo {author} {\bibfnamefont {T.~M.}\ \bibnamefont {MacRobert}},\ }\bibfield  {title} {\bibinfo {title} {Higher transcendental functions},\ }\href {https://api.semanticscholar.org/CorpusID:21329182} {\bibfield  {journal} {\bibinfo  {journal} {Nature}\ }\textbf {\bibinfo {volume} {175}},\ \bibinfo {pages} {317} (\bibinfo {year} {1955})}\BibitemShut {NoStop}%
\bibitem [{\citenamefont {Leaver}(1985)}]{Leaver:1985ax}%
  \BibitemOpen
  \bibfield  {author} {\bibinfo {author} {\bibfnamefont {E.~W.}\ \bibnamefont {Leaver}},\ }\bibfield  {title} {\bibinfo {title} {{An Analytic representation for the quasi normal modes of Kerr black holes}},\ }\href {https://doi.org/10.1098/rspa.1985.0119} {\bibfield  {journal} {\bibinfo  {journal} {Proc. Roy. Soc. Lond. A}\ }\textbf {\bibinfo {volume} {402}},\ \bibinfo {pages} {285} (\bibinfo {year} {1985})}\BibitemShut {NoStop}%
\bibitem [{\citenamefont {Griffiths}\ and\ \citenamefont {Schroeter}(2018)}]{griffiths_schroeter_2018_iqm}%
  \BibitemOpen
  \bibfield  {author} {\bibinfo {author} {\bibfnamefont {D.~J.}\ \bibnamefont {Griffiths}}\ and\ \bibinfo {author} {\bibfnamefont {D.~F.}\ \bibnamefont {Schroeter}},\ }\href@noop {} {\emph {\bibinfo {title} {Introduction to Quantum Mechanics}}},\ \bibinfo {edition} {3rd}\ ed.\ (\bibinfo  {publisher} {Cambridge University Press},\ \bibinfo {year} {2018})\BibitemShut {NoStop}%
\bibitem [{\citenamefont {Whittaker}\ and\ \citenamefont {Watson}(1996)}]{Whittaker_Watson_1996}%
  \BibitemOpen
  \bibfield  {author} {\bibinfo {author} {\bibfnamefont {E.~T.}\ \bibnamefont {Whittaker}}\ and\ \bibinfo {author} {\bibfnamefont {G.~N.}\ \bibnamefont {Watson}},\ }\href@noop {} {\emph {\bibinfo {title} {A Course of Modern Analysis}}},\ \bibinfo {edition} {4th}\ ed.,\ Cambridge Mathematical Library\ (\bibinfo  {publisher} {Cambridge University Press},\ \bibinfo {year} {1996})\BibitemShut {NoStop}%
\end{thebibliography}%

\let\addcontentsline\oldaddcontentsline%
\clearpage

\bigskip

\onecolumngrid
\begin{center}
  \textbf{\large Supplemental Material
  }\\[.2cm]
  \vspace{0.05in}
  {Zhong-Hao Luo, Fa Peng Huang, Pengming Zhang, and Chen Zhang}
\end{center}

\twocolumngrid

\setcounter{equation}{0}
\setcounter{figure}{0}
\setcounter{table}{0}
\setcounter{section}{0}
\setcounter{page}{1}
\makeatletter

\setcounter{secnumdepth}{2}
\renewcommand{\thesection}{\Roman{section}}
\renewcommand{\thesubsection}{\alph{subsection}}

\onecolumngrid

\startcontents[sections]
\tableofcontents


\renewcommand{\theequation}{S\arabic{equation}}
\renewcommand{\thefigure}{S\arabic{figure}}
\renewcommand{\thetable}{S\arabic{table}}

\renewcommand{\theHequation}{S\arabic{equation}}
\renewcommand{\theHfigure}{S\arabic{figure}}
\renewcommand{\theHtable}{S\arabic{table}}

\makeatother

\section{Magnetic Kerr background and monopole-harmonic clouds}
\label{app:background}

This Supplemental Material collects technical steps that underlie the results quoted in the main text.
We work in units $G=c=\hbar=1$ and follow the sign conventions of the main paper.

\subsection{Metric, horizons, and electromagnetic potential}

A rotating \emph{magnetically} charged BH is described by the Kerr--Newman family with vanishing gauge charge and magnetic charge $P$.
In Boyer--Lindquist coordinates $(t,r,\theta,\phi)$ the metric is
\begin{align}
\begin{aligned}
ds^{2}=&-\left(1-\frac{2Mr-P^{2}}{\Sigma}\right)dt^{2}
+\frac{\Sigma}{\Delta}\,dr^{2}
+\Sigma\,d\theta^{2}\\
&+\left[(r^{2}+a^{2})+\frac{(2Mr-P^{2})a^{2}\sin^{2}\theta}{\Sigma}\right]\sin^{2}\theta\,d\phi^{2}
-\frac{2a(2Mr-P^{2})\sin^{2}\theta}{\Sigma}\,dt\,d\phi,
\end{aligned}
\label{eq:SM_metric}
\end{align}
where
\begin{equation}
\Sigma\equiv r^{2}+a^{2}\cos^{2}\theta,\qquad 
\Delta\equiv r^{2}-2Mr+a^{2}+P^{2}=(r-r_{+})(r-r_{-}).
\end{equation}
The outer/inner horizons are
\begin{equation}
r_{\pm}=M\pm\sqrt{M^{2}-a^{2}-P^{2}},
\end{equation}
so regular horizons require $a^{2}+P^{2}\le M^{2}$.
The horizon angular velocity and surface gravity are
\begin{equation}
\Omega_{H}=\frac{a}{r_{+}^{2}+a^{2}},\qquad 
\kappa=\frac{r_{+}-r_{-}}{2(r_{+}^{2}+a^{2})}.
\label{eq:SM_Omega_kappa}
\end{equation}

A magnetic monopole cannot be described globally by a single smooth vector potential; one uses Wu--Yang north/south patches \cite{Wu:1976ge}. A convenient choice is the Dirac form of the gauge potential of magnetic charges $A_\rho=(0,0,0,-P\cos{\theta})$ \cite{Pereniguez:2024fkn,Mouland:2024zgk,Dirac:1931kp}. While the Dirac gauge potential needs to introduce the Dirac quantization 
\begin{equation}
q\equiv eP=\frac{N}{2}\ge 0,\qquad N\in\mathbb{Z}.
\end{equation}
with $N=0,1,2,...$, this condition eliminates the global singularity. Notably, we define the magnetic parameter $q\geq0$ to be non-negative, which differs by a sign from the convention used in \cite{Pereniguez:2024fkn}.

\subsection{Angular separation equation of the charged Klein--Gordon equation}

The gauge-charged massive scalar perturbation $\Psi$ in the magnetic BH background can be described by the covariant Klein-Gordon equation \cite{Pereniguez:2024fkn,Zou:2021mwa,Myung:2022dpp}
\begin{equation}
\left[(\nabla_{\mu}-ieA_{\mu})(\nabla^{\mu}-ieA^{\mu})-\mu^{2}\right]\Psi=0.
\label{Klein--Gordon equation}
\end{equation}
Because the background is stationary and axisymmetric, we take the ansatz
\begin{equation}
\Psi(t,r,\theta,\phi)=e^{-i\omega t}R(r)\,S_{q\ell m}(\theta)\,e^{im\phi},
\label{eq:SM_sep_ansatz}
\end{equation}
with $(\omega,m)$ conserved quantum numbers. Separation leads to an angular equation of the angular function $S_{q\ell m}(\theta)$
\begin{align}
&\frac{1}{\sin\theta}\frac{d}{d\theta}\!\left(\sin\theta\,\frac{dS}{d\theta}\right)
+\biggl[
a^{2}(\omega^{2}-\mu^{2})\cos^{2}\theta
-\frac{(m+q\cos\theta)^{2}}{\sin^{2}\theta}
-2a\omega q\cos\theta-q^{2}
+\Lambda_{q\ell m}
\biggr]S_{q\ell m}(\theta)=0.
\label{eq:SM_ang_full}
\end{align}
Here, \(\Lambda_{q \ell m}\) represents the separation constant. For bound states of scalar fields around magnetic BHs, the frequency-mass relationship implies $\omega\approx\mu$. In the ultralight limit  $\mu M\ll1$ (and hence $\omega M\ll1$), the above angular equation Eq.~\eqref{eq:SM_ang_full} reduces to 
\begin{align}
\begin{aligned}
        -\frac{1}{\sin\theta}\frac{d}{{d}\theta}\left(\sin{\theta}\frac{{d}S_{q\ell m}(\theta)}{{d}\theta}\right) +\frac{(m+q\cos{\theta})^2}{\sin^2{\theta}}S_{q\ell m}(\theta)=(\Lambda_{q \ell m}-q^2)S_{q\ell m}(\theta),
    \label{angulareq}
\end{aligned}
\end{align}
which is identical to Eq.~(22) in \cite{Wu:1976ge}, except for the form of separation constant $\Lambda_{q \ell m}\approx \ell(\ell+1)+O(a\omega)$. The approximate expansion of $\Lambda_{q \ell m}$ stems from neglecting the terms $a^2(\omega^2-\mu^2)\cos^2{\theta}$ and $2a\omega q\cos{\theta}$ in Eq.~\eqref{eq:SM_ang_full}. Hence, the separation constant $\Lambda_{q \ell m}$ is approximated as $\Lambda_{q \ell m} \simeq  \ell(\ell+1)$. More importantly, the spheroidal harmonics \(S_{q\ell m}(\theta) e^{im\phi}\) reduce to the monopole harmonics \(Y_{q \ell m}(\theta,\phi)\) \cite{Wu:1976ge,Pereniguez:2024fkn,Zou:2021mwa}, rather than the usual spherical harmonics $Y_{\ell m}(\theta,\phi)$. 

\subsection{Monopole harmonics and the north/south cloud morphology}
\label{Monopole harmonics and the north/south cloud morphology}

\begin{figure}[h]
    \centering
    \includegraphics[width=0.45\linewidth]{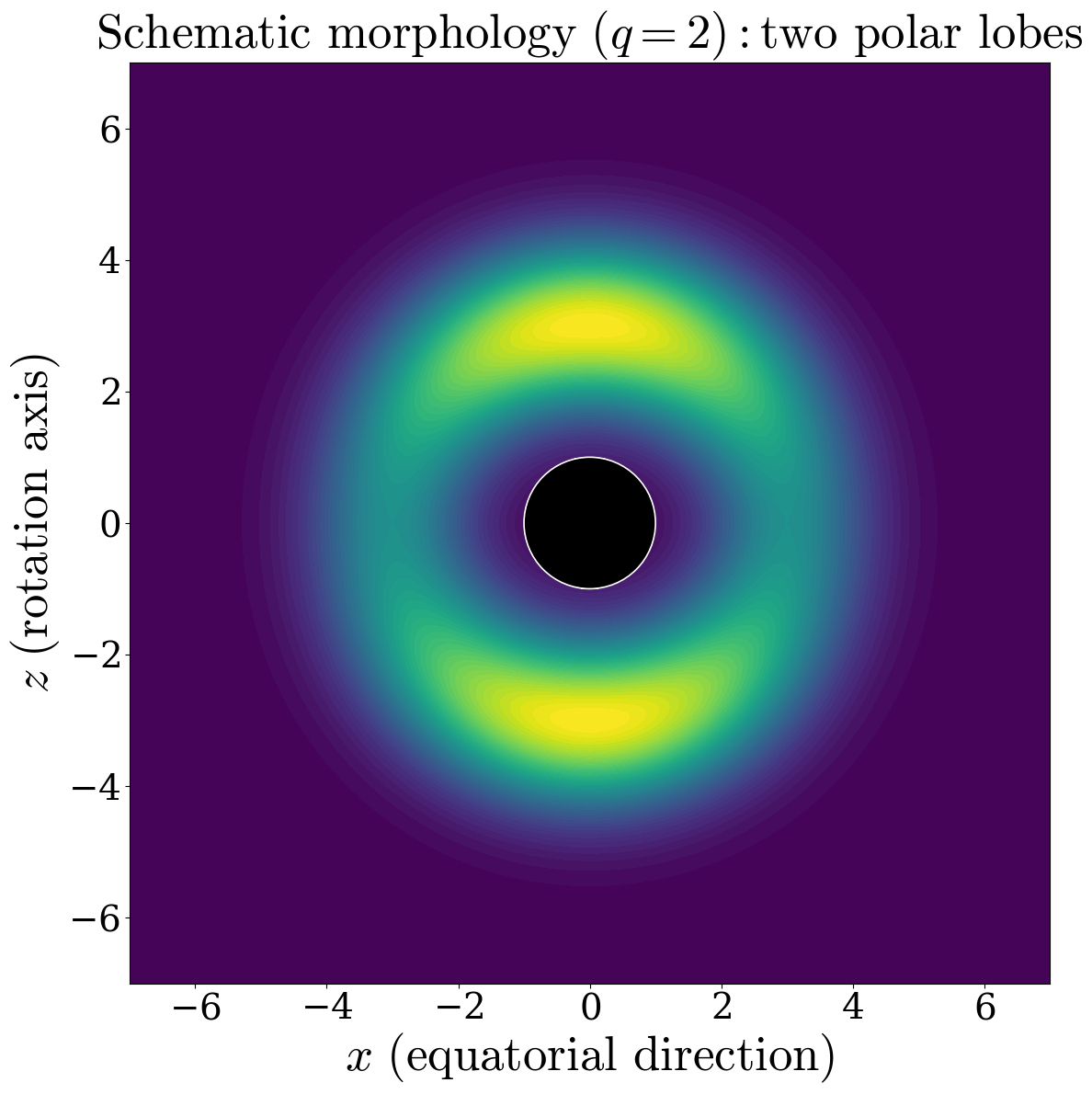}
    \caption{Schematic meridional ($x$--$z$) slice of the cloud profile built from the maximally unstable monopole-harmonic modes $Y^{+}_{qq,q}$ (north) and $Y^{-}_{qq,-q}$ (south).}
    \label{Fig:Schematic monopole-cloud morphology}
\end{figure}

{A convenient explicit representation of monopole harmonics, in the convention used below, is \cite{Wu:1976ge}
\begin{align}
    \begin{aligned}
    &Y_{q\ell m}=S_{q\ell m}(\theta)e^{im\phi}=N_{q \ell m}(1-\cos{\theta})^{\alpha_J/2}(1+\cos{\theta})^{\beta_J/2}P_{\zeta}^{\alpha_J,\beta_J}(\cos{\theta})e^{im\phi},\\
    &\ell=|q|,|q|+1,...,\quad m=-\ell,-\ell+1,...,\ell,
    \end{aligned}
\end{align}
where $\alpha_J=m-q$, $\beta_J=m+q$, and $\zeta=\ell-m$. The quantity $N_{q \ell m}$ is normalization constant, and  $P_{\zeta}^{\alpha_J,\beta_J}(\cos{\theta})$ represents the Jacobi polynomial \cite{MacRobert1955HigherTF}.}
The mode that maximizes the superradiant growth rate is the \emph{lowest} allowed $\ell$ for a given $q$, namely $\ell=q$, with $|m|=q$. Moreover, the north-south monopole modes $Y^{\pm}_{qq,\pm q}$ are the most superradiant unstable modes, which take the form 
\begin{equation}
Y^{+}_{qq,q}=A_{q}\,e^{iq\phi}\left(\cos\frac{\theta}{2}\right)^{2q},
\qquad 
Y^{-}_{qq,-q}=A_{q}\,e^{-iq\phi}\left(\sin\frac{\theta}{2}\right)^{2q},
\label{eq:SM_Ypm}
\end{equation}
with $N_{qqq}\equiv A_q=N_{qq-q}\equiv A_{-q}=\sqrt{(2q+1)/(4\pi)}$. Here $ Y^{\pm}_{qq,\pm q}$ represent the two profiles peak at opposite poles but are not sharply confined
to separate hemispheres; both remain finite near the equatorial
region, producing a nonvanishing spatial overlap. It is noteworthy that the phase singularity $e^{\pm iq\phi}$ at the north and south poles can be removed by the intermediate gauge \cite{Pereniguez:2024fkn}. Consequently, the monopole harmonics are regular at both poles of the $z$-axis $\cos\theta=\pm1$
\begin{equation}
Y^{+}_{qq,q}\propto (1+\cos\theta)^{q}\quad (\theta\simeq 0),\qquad
Y^{-}_{qq,-q}\propto (1-\cos\theta)^{q}\quad (\theta\simeq \pi),
\end{equation}
so the scalar density is enhanced around the rotation axis and suppressed near the equator, yielding the characteristic north--south two-lobe morphology (Fig.~\ref{Fig:Schematic monopole-cloud morphology}).

\subsection{Why the centrifugal barrier is reduced: an intuitive angular-momentum picture}

The key analytic effect of the monopole is the weakening of the centrifugal barrier in the far region, which can be understood directly from gauge-covariant angular momentum.
For a charged particle in a monopole background, the canonical momentum is $\boldsymbol{\pi}=\mathbf{p}-e\mathbf{A}$ and the gauge-invariant total angular momentum is \cite{Wu:1976ge}
\begin{equation}
\mathbf{J}=\mathbf{r}\times\boldsymbol{\pi}-q\,\hat{\mathbf{r}},
\label{eq:SM_J_def}
\end{equation}
where $\hat{\mathbf{r}}$ is the radial unit vector.
Squaring gives
\begin{equation}
J^{2}=(\mathbf{r}\times\boldsymbol{\pi})^{2}+q^{2},
\qquad\Rightarrow\qquad
(\mathbf{r}\times\boldsymbol{\pi})^{2}=J^{2}-q^{2}.
\end{equation}
Upon separation one has $J^{2}\to \ell(\ell+1)$, so the centrifugal term becomes
\begin{equation}
\frac{\ell(\ell+1)}{r^{2}}
\ \longrightarrow\ 
\frac{\ell(\ell+1)-q^{2}}{r^{2}}
\equiv \frac{\ell_{q}(\ell_{q}+1)}{r^{2}},
\label{eq:SM_cent_reduction}
\end{equation}
which defines the reduced effective index $\ell_{q}$.
Equivalently,
\begin{equation}
\ell_{q}=\sqrt{\left(\ell+\tfrac12\right)^{2}-q^{2}}-\frac12,
\end{equation}
as quoted in Eq.~\eqref{eq:ellqdef} of the main text.
Since $\ell_{q}<\ell$ for $q\neq 0$, the effective potential barrier is reduced, leading to a deeper bound-state well and (as quantified in see SM, Sec.~\ref{Superradiant growth rate: scaling estimate and matched-asymptotic derivation}) a much larger superradiant growth rate.

\begin{figure*}[t]
    \centering
    \includegraphics[width=0.98\textwidth]{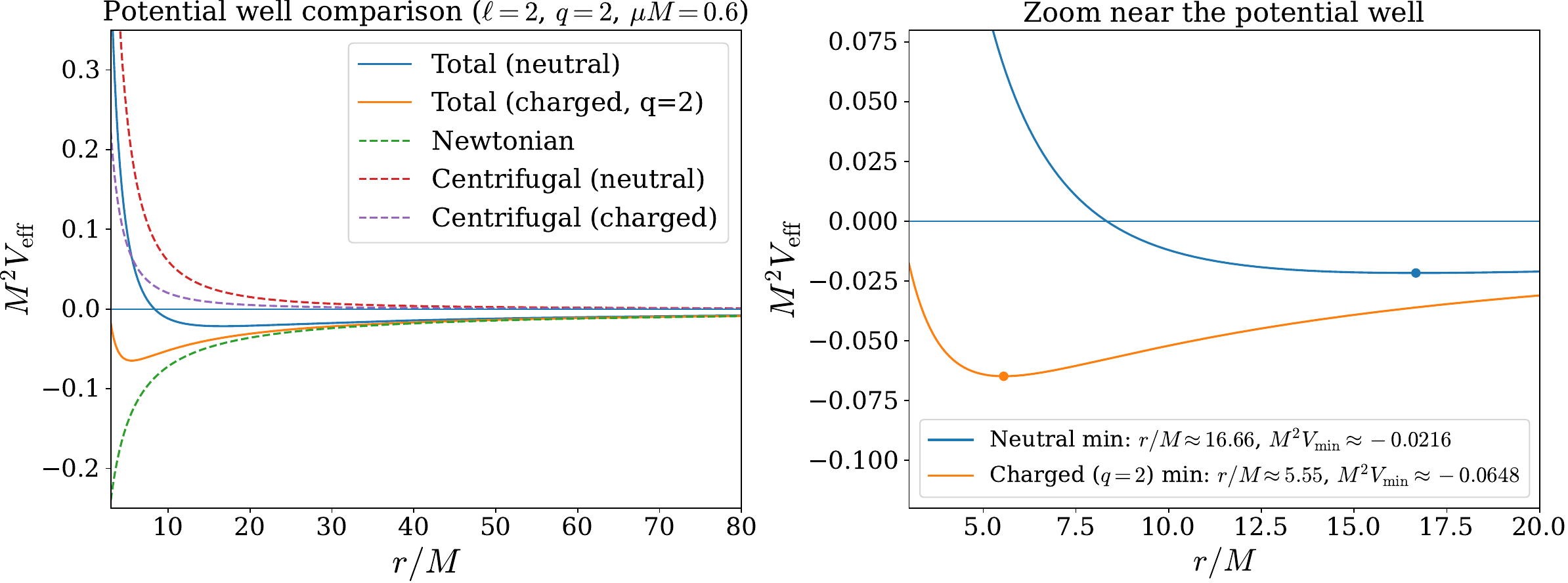}
    \caption{Comparison of the far-region effective potential for Kerr ($q=0$) and magnetic Kerr ($q\neq 0$). The reduced barrier for $q\neq 0$ deepens the potential well and increases the number of superradiant ``bounces''.}
    \label{Fig:potential well}
\end{figure*}

\section{Superradiant growth rate: scaling estimate and matched-asymptotic derivation}
\label{Superradiant growth rate: scaling estimate and matched-asymptotic derivation}

\subsection{Scaling estimate from the reduced angular barrier $\ell\to \ell_q$}
\label{Scaling estimate from the reduced angular barrier}

The radial separation equation of Eq.~\eqref{Klein--Gordon equation} determines the growth rate of charged scalar clouds
\begin{align}
 \frac{d}{d r}\left(\Delta(r)\frac{d}{d r}R(r)\right)+\left[\frac{\big(\left(r^{2}+a^{2}\right)\omega-ma\big)^{2}}{\Delta(r)}-\mu^{2}r^{2}-\Lambda_{q\ell m}+q^{2}+2am\omega-a^{2}\omega^{2}\right]R(r)=0.
 \label{radial equation}
\end{align}
In \cite{Pereniguez:2024fkn}, the numerical results of the growth rate have been computed via the Leaver method \cite{Dolan:2007mj,Leaver:1985ax}, revealing that the superradiant instability of magnetic BHs can be up to two orders of magnitude faster than that of neutral Kerr BHs. Although this dramatic enhancement is robustly supported by numerical computations, the underlying physical mechanism remains unclear. We therefore aim to elucidate the physical mechanism by solving the radial equation analytically \cite{Baumann:2019eav,Starobinskii:1973vzb}, through matching the far-region and near-region solutions of Eq.~\eqref{radial equation} in the overlap region.
 
In the far region (\(r\gg M\) and \(r \sim \ell/\omega\)), we find that the radial equation reduces to
\begin{align}
    \frac{d^2(rR)}{dr^2}+\left[\omega^2-\mu^2\big(1-\frac{2M}{r}\big)-\frac{\ell(\ell+1)-q^2}{r^2}\right](rR)=0,
    \label{farradialEq}
\end{align}
where the centrifugal potential term $\frac{\ell(\ell+1)-q^2}{r^2}$ is weakened compared to the neutral case. This potential structure also appears in the case of a charged scalar field around a magnetic monopole given by Wu-Yang in Eq.~(53) of \cite{Wu:1976ge}. Apart from the gravitational binding from the Newtonian potential $-\frac{2M\mu^2}{r}$, the charged scalar field is additionally bound by the magnetic charges of the BH.
Qualitatively, this extra binding creates a deeper effective potential well around the magnetic BH as shown in Fig.~\ref{Fig:potential well}. As a result, the bound state consequently undergoes more cycles of superradiant amplification, ultimately leading to an enhanced instability. Quantitatively, however, the precise growth rate requires matching the radial solutions across the overlap region.

After the calculation in Sec.~\ref{The far region solution}, we find that, significantly, although the $\ell(\ell+1)$ term in the centrifugal potential is modified to $\ell(\ell+1)-q^2\equiv\ell_q(\ell_q+1)$ in Eq.~\eqref{farradialEq}, the far-region radial solution can still be expressed in terms of confluent hypergeometric functions, simply by replacing the usual orbital quantum number $\ell$ with $\ell_q$ relative to the neutral scalar cloud case 
\begin{align}
    R(r)=(2kr)^{\ell_q} e^{-kr} U(\ell_q+1-\nu,2\ell_q+2;2kr),
\label{far region solution}
\end{align}
where 
\begin{align}
\begin{aligned}
\ell_q\equiv\sqrt{(\ell+1/2)^2-q^2}-1/2,\\
    \ell(\ell+1)\to\ell(\ell+1)-q^2\equiv\ell_q(\ell_q+1).
\end{aligned}
\end{align}
Here $\nu\equiv{M\mu^2}/{k}$ with $k^2\equiv\mu^2-\omega^2$, and \(U(\ell_q+1-\nu,2\ell_q+2; 2kr)\) is the confluent hypergeometric function. Analogous to the hydrogen atom \cite{griffiths_schroeter_2018_iqm}, we define $\nu_0\equiv n_r+\ell_q+1=n_r+\sqrt{(\ell+1/2)^2-q^2}+1/2
$. However, the presence of the horizon introduces a 
small complex correction $\nu=\nu_0+\delta\nu$. Consequently, the solution in Eq.~(\ref{far region solution}) can be rewritten as 
\begin{align}
    R(r)=(2kr)^{\ell_q} e^{-kr}U(-n_r-\delta \nu,2\ell_q+2;2kr).
    \label{far region radial solution}
\end{align}
Using the asymptotic properties of confluent hypergeometric functions and Gamma functions in Eq.~\eqref{property confluent hypergeometric functions} and Eq.~\eqref{property gamma functions}, the far-region solution can be expanded in the overlap region ($2kr\to 0$) as the centrifugal potential series 
\begin{align}
\begin{split}
    R(r)=&(2kr)^{\ell_q}e^{-kr}{U}(-n_r-\delta\nu,2\ell_q+2;2kr)
    \\
 \simeq &(-1)^{n_r}\frac{\Gamma(2\ell_q+n_r+2)}{\Gamma(2\ell_q+2)}(2kr)^{\ell_q}+(-1)^{n_r+1}\Gamma(n_r+1)\Gamma(2\ell_q+1)\delta\nu(2kr)^{-\ell_q-1}.
\end{split}
\label{farRpotential}
\end{align}
Notably, the non-integer nature of  $\ell_q$ precludes the use of the factorials as in \cite{Starobinskii:1973vzb} and necessitates the generalized Gamma function formalism.

In the near region, under the coordinate
transformation $ z\equiv \frac{r-r_+}{r_+-r_-}$, the radial equation in Eq.~(\ref{near radial eq}) can be approximated as
\begin{align}
    \begin{aligned}
& z(z+1)\frac{d}{dz}\left[z(z+1)\frac{dR}{dz}\right] 
+\big[{p}^2-(\ell(\ell+1)-q^2)z(z+1)\big]R=0,
    \end{aligned}
    \label{near radial eq}
\end{align}
where the sign of $p$ determines whether superradiance is active. This parameter is expressed 
in terms of the BH's angular velocity $\Omega_H=\frac{a}{r_+^2+a^2}$ and surface gravity $\kappa=\frac{r_+-r_-}{2(r_+^2+a^2)}$ 
as 
\begin{align}
p\equiv-\frac{\omega_*}{2\kappa}=-\frac{\omega-m\Omega_H}{2\kappa}.
\end{align}
The near-region radial equation~\eqref{near radial eq} possesses three regular singularities at \{\(0, -1, \infty\)\}. Its fundamental solutions $R(z)$ can be expressed as a linear combination of two linearly independent hypergeometric functions \cite{MacRobert1955HigherTF}, characterized by six indices $(x_0^\pm,x_{-1}^\pm,x_{\infty}^\pm)$ corresponding to these singularities. Solving the indicial equation~\eqref{indicial equation} of Eq.~\eqref{near radial eq}, we find that the indices at infinity $x_{\infty}^\pm$ precisely coincide with the $\ell_q$ factor
\begin{align}
&x_{\infty}^+=1/2+\sqrt{(\ell+1/2)^2-q^2}=\ell_q+1, \\&
x_{\infty}^-=1/2- \sqrt{(\ell+1/2)^2-q^2}=-\ell_q.
\end{align}
Physically, these indices $x_{\infty}^\pm$ correspond to the power-law behavior of the centrifugal potential series in the overlap region. Considering the asymptotic form ($r\gg M$) of the hypergeometric functions $w_3,w_4$ in Eq.~\eqref{centrifugal hypergeometry function in the overlap region}, the solution behaves as  
\begin{align}
\begin{aligned}
     R(r) &\simeq  A \left(\frac{-r}{r_+-r_-}\right)^{-x_{\infty}^-}+B \left(\frac{-r}{r_+-r_-}\right)^{-x_{\infty}^+}\\&=A \left(\frac{-r}{r_+-r_-}\right)^{\ell_q}+B \left(\frac{-r}{r_+-r_-}\right)^{-\ell_q-1}.
\end{aligned}
\label{nearRpotential}
\end{align}
where $A,B$ are the linear coefficients. 

Consequently, matching becomes possible in the overlap region between the far-region solution Eq.~\eqref{farRpotential} and near-region solution Eq.~\eqref{nearRpotential}, as both exhibit the same power-law dependence on $\ell_q$ in their centrifugal potential forms. The linear coefficients $A$ and $B$ are thereby determined by 
\begin{align}
    &A= (-1)^{n_r+\ell_q}[2k(r_+-r_-)]^{\ell_q}\frac{\Gamma(2\ell_q+n_r+2)}{\Gamma(2\ell_q+2)},
    \label{expression of A}
    \\&
    B= (-1)^{n_r+\ell_q}[2k(r_+-r_-)]^{-\ell_q-1}\Gamma(n_r+1)\Gamma(2\ell_q+1)\delta\nu.
    \label{expression of B}
\end{align}
Meanwhile, the near-region solution $R(z)$ can be expressed as the combination of the ingoing ($z^{-ip}$) and outgoing ($z^{ip}$) modes, which are associated with the hypergeometric function basis $(w_1,w_2)$ near the horizon. This basis can be transformed from the basis $(w_3,w_4)$ in Eq.~\eqref{R1,R2 transformed by R3,R4}, which corresponds to the centrifugal potential in the overlap region, 
\begin{align}
\begin{aligned}
R(z)&=(Af_{13}+Bf_{14})R_1+(Af_{23}+Bf_{24})R_2\\&\simeq(Af_{13}+Bf_{14})z^{ip}+(Af_{23}+Bf_{24})z^{-ip},
\end{aligned}
\end{align}
where $f_{13},f_{23},f_{14},f_{24}$ are the connection coefficients between different hypergeometric function bases, as given by Eq.~\eqref{the connection coefficients between different hypergeometric function bases}. Imposing the boundary condition at the event horizon (purely ingoing waves), the BH's reflectivity must vanish $\mathcal{R}\equiv\frac{Af_{13}+Bf_{14}}{Af_{23}+Bf_{24}}=0$. This immediately implies the relation between the coefficients \(A\) and \(B\)
\begin{align} f_{13}A+f_{14}B=0. \label{ABrelationship}
\end{align}
Substituting the expressions for $A$ and $B$ in Eqs.~\eqref{expression of A}--\eqref{expression of B} into the relation in Eq.~\eqref{ABrelationship} yields the horizon-corrected frequency parameter $\nu=\nu_0+\delta\nu$, connected to the angular frequency via $\omega=\sqrt{\mu^2-\frac{M\mu^2}{\nu}}$
\begin{align}
    \begin{aligned}
    \delta\nu=&2i{p} \left[2 k\left(r_{+}-r_{-}\right)\right]^{2 \ell_q+1} \times\left|\frac{\Gamma(1+\ell_q+2ip)}{\Gamma(1+2ip)}\right|^2   
    \\&\times\frac{\Gamma(2\ell_q+n_r+2)}{\Gamma(n_r+1)}\left[\frac{\Gamma(\ell_q+1)}{\Gamma(2\ell_q+1)\Gamma(2\ell_q+2)}\right]^2.
\end{aligned}
\label{Growth rate delta nu}
\end{align}
This result generalizes Detweiler's finding in \cite{PhysRevD.22.2323} by replacing factorials with Gamma functions and introducing $\ell_q$. Otherwise, if $\ell_q$ is a half-integer, continuing to use the factorial form  would lead to an underestimation of the result. The superradiant factor $2p\propto(m\Omega_H-\omega)$ governs the superradiance condition. The radial transmission factor $\left[2 k\left(r_{+}-r_{-}\right)\right]^{2 \ell_q+1}$ determines the amplitude scale,  while the phase coupling factor $\left|\frac{\Gamma(1+\ell_q+2ip)}{\Gamma(1+2ip)}\right|^2$ captures wave-horizon interaction phases. The remaining term $\frac{\Gamma(2\ell_q+n_r+2)}{\Gamma(n_r+1)}\left[\frac{\Gamma(\ell_q+1)}{\Gamma(2\ell_q+1)\Gamma(2\ell_q+2)}\right]^2 $ is the normalization coefficient. 

Given the definitions $k=\frac{M\mu^2}{\nu}$ and $k^2=\mu^2-\omega^2$, the parameter $\nu=\nu_0+\delta\nu$ induces expansions for both wavevector $k$ and angular frequency $\omega$. First, the wavevector expands as $k\simeq k_0+\delta k$, 
\begin{align} 
\begin{aligned}
&k=\frac{\alpha^2}{M(\nu_0+\delta\nu)}\simeq\frac{\alpha^2}{M\nu_0}-\frac{\alpha^2}{M\nu_0^2}\delta\nu=k_0+\delta k,
\\
&k_0=\frac{\alpha^2}{M\nu_0},\  \delta k= -\frac{\alpha^2}{M\nu_0^2}\delta\nu
\label{angularfrequency}
\end{aligned}
\end{align}
with $ \nu_0\equiv \ell_q+n_r+1$ and $\alpha \equiv M\mu$. The corresponding angular frequency expansion $\omega\simeq \omega_0+\delta \omega$ yields
\begin{align}
 \omega &=\sqrt{\mu^2-(k_0+\delta k)^2}\simeq\omega_0-\frac{k_0}{\omega_0}\delta k=\omega_0+\delta\omega.\\
     \quad\omega_0 &= \mu\left[{1-\Big(\frac{\alpha}{\nu_0}\Big)^2}\right]^{1/2},\ 
\delta\omega=\frac{\delta\nu}{M}\left(\frac{\alpha}{\nu_0}\right)^3 \left[1-\Big(\frac{\alpha}{\nu_0}\Big)^2\right]^{1/2}.
\end{align}
Finally, the growth rate of this scalar cloud, determined by the imaginary part of the frequency shift $\delta\omega$, takes the form 
\begin{align}
\omega_{I}= \mathrm{Im}[\delta\omega]=\frac{\delta\nu_I}{M}\left(\frac{{\alpha}}{\nu_0}\right)^3\left[{1-\left(\frac{\alpha}{\nu_0}\right)^2}\right]^{-1/2},
\label{Growth rate}
\end{align}
where we define $\delta\nu_I\equiv{\rm Im}[\delta\nu]$.

\begin{figure}
    \centering
    \includegraphics[width=0.6\linewidth]{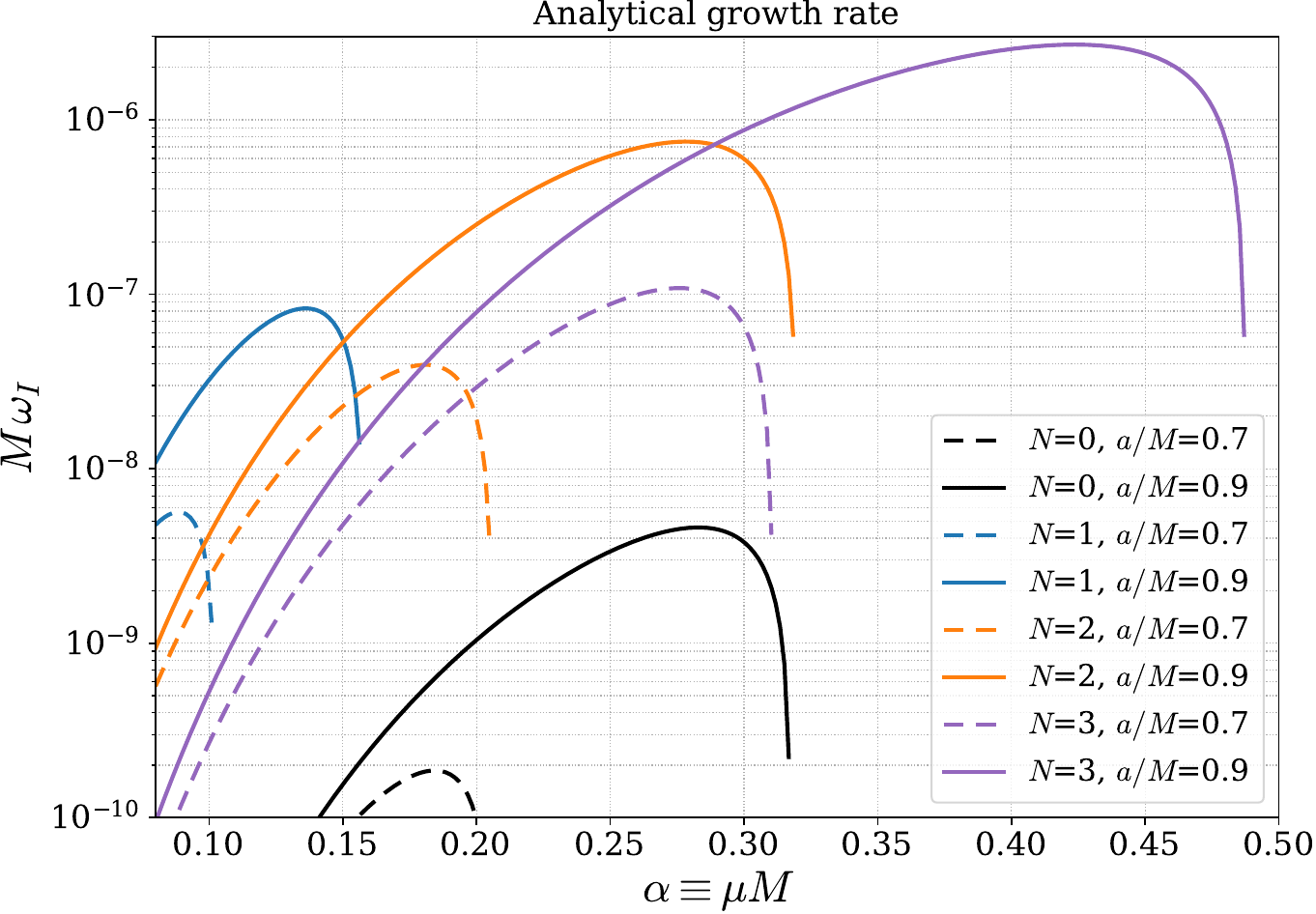}
    \caption{Analytical growth rate $M\omega_I$ of charged scalar clouds around a magnetic BH as a function of $\alpha\equiv\mu M$. Solid (dashed) curves correspond to $a/M=0.9$ ($a/M=0.7$). We take the fundamental radial mode $n_r=0$ and set $P/M\sim10^{-19}$. Colors label the monopole number: black/blue/orange/purple denote $N=0,1,2,3$, respectively. For $N=0$, we take $\ell=m=1$. For $N\ge1$, we take $\ell=m=q=N/2$ with $\ell_q=\sqrt{(\ell+1/2)^2-q^2}-1/2$. The analytical results agree with the numerical values of Ref.~\cite{Pereniguez:2024fkn} to within an order of magnitude.}
    \label{Fig:Analytical growth rate}
\end{figure}

The growth rate enhancement factor $\omega_I/\omega_{I(\text{Kerr})}$ is estimated to be several orders of magnitude by comparing the $\alpha$-scaling in both Kerr and magnetic cases. The radial transmission factor 
\begin{align}
  \left[2 k\left(r_{+}-r_{-}\right)\right]^{2 \ell_q+1}\propto k^{2\ell_q+1}\sim\alpha^{4\ell_q+2}, 
\end{align}
with $k\sim\alpha^2/M$ in $\delta\nu_I$ (Eq.~\eqref{Growth rate delta nu}) and the additional factor $\alpha^3$ in Eq.~\eqref{Growth rate} set the $\alpha$-dependence of the growth rate. For the neutral Kerr case \cite{Starobinskii:1973vzb,Baumann:2019eav,Brito:2015oca}, $\omega_{I(\text{Kerr})}$ scales as
\begin{align}
\omega_{I(\text{Kerr})}=\frac{\delta\nu_I}{M}\left(\frac{{\alpha_{(\rm Kerr)}}}{\ell+n_r+1}\right)^3\propto  \alpha_{(\rm Kerr)}^{4\ell+5}.
\end{align} 
However, for the magnetic case,
\begin{align}
\omega_{I}=\frac{\delta\nu_I}{M}\left(\frac{{\alpha}}{\ell_q+n_r+1}\right)^3\propto  \alpha^{4\ell_q+5}.
\label{Growth rate alpha}
\end{align} 
Here $\ell_q <\ell$. This reduction in the exponent, combined with the small-coupling limit $\alpha,\alpha_{\rm (Kerr)}\ll1$, leads to an enhancement of up to several orders of magnitude.  As a concrete example, the peak growth rate for neutral scalar cloud ($\ell=m=1$) around a Kerr BH at $\alpha_{\rm (Kerr)}\approx0.3$ scales as 
\begin{align}
\begin{aligned}
\omega_{I(\text{Kerr})}&\propto\alpha_{\rm (Kerr)}^{4\ell+5}=\alpha_{\rm (Kerr)}^9\approx0.3^{9}\approx2.0\times10^{-5}.
\end{aligned}
\end{align}
For a charged scalar cloud ($\ell=m=q=N/2=3/2$) around a magnetic BH with $N=3$ magnetic charges at $\alpha\approx0.45$, with $\ell_q\approx 0.82$, the scaling is 
\begin{align}
\omega_{I}\propto\alpha^{4\ell_q+5}=\alpha^{8.3}\approx0.45^{8.3}\approx1.3\times10^{-3},
\end{align}
yielding a growth rate enhancement
\begin{align}
\omega_I/\omega_{I(\text{Kerr})}\approx10^2.
\end{align}
Therefore, the growth rate of the charged scalar cloud around the magnetic BH with $N=3$ magnetic charges can be about two orders of magnitude larger than that of the neutral scalar cloud around the Kerr BH. The above enhancement is illustrated in Fig.~\ref{Fig:Analytical growth rate}.  Furthermore, the numerical results in Fig.~3 of \cite{Pereniguez:2024fkn} are consistent at the order-of-magnitude level with this estimate and our analytical results presented in Fig.~\ref{Fig:Analytical growth rate}.

\subsection{Matched-asymptotic derivation of the growth rate}
\label{Matched-asymptotic derivation of the growth rate}

The analytical calculation of the growth rate for charged scalar clouds around magnetic BHs is performed using a matched-asymptotic expansion. This involves solving the radial equation~\eqref{radial equation} separately in the far region and near region, with the growth rate ultimately extracted via matching of these solutions in the overlap region.

\subsubsection{Far region: hydrogenic solution and the reduced orbital index $\ell_q$}
\label{The far region solution}

In the far region limit (\(r\gg M\) and \(r \sim \ell/\omega\)), the radial equation~\eqref{radial equation} can be approximated to a hydrogen-atom-like equation
\begin{align}
    \frac{d^2(rR)}{d\tilde{r}^2}+\left[\omega^2-\mu^2\big(1-\frac{2M}{r}\big)-\frac{\ell(\ell+1)-q^2}{r^2}\right](rR)=0.
\label{radial hydrogen atom like equation}
\end{align}
Introducing \(\rho=2kr\) transforms the  equation into the Whittaker form \cite{Whittaker_Watson_1996}
\begin{align}\label{rhoR}
    \frac{d^2(\rho R)}{d\rho^2}+\left[-\frac{1}{4}+\frac{\nu}{\rho}-\frac{\ell(\ell+1)-q^2}{\rho^2}\right](\rho R)=0,
\end{align}
where $ k\equiv\sqrt{\mu^2-\omega^2}$ and $\nu\equiv{M\mu^2}/{k}$. Its solution is given by the Whittaker function, expressed in terms of the confluent hypergeometric function 
\begin{align}\rho R(\rho)=e^{-\rho/2}\rho^{\frac{1}{2}+\sqrt{(\ell+\frac{1}{2})^2-q^2}}U(\frac{1}{2}+\sqrt{(\ell+\frac{1}{2})^2-q^2}-\nu,2\sqrt{(\ell+\frac{1}{2})^2-q^2}+1;\rho),
\end{align}
Defining $\ell_q+1\equiv\sqrt{(\ell+\frac{1}{2})^2-q^2}+\frac{1}{2}$, the far-region radial solution can be rewritten as 
\begin{align}
    R(r)=(2kr)^{\ell_q} e^{-kr} U(\ell_q+1-\nu,2\ell_q+2;2kr),
    \label{eqB7}
\end{align}
where \(U(\ell+1-\nu,2\ell+2; 2kr)\) is the confluent hypergeometric function \cite{MacRobert1955HigherTF}. Remarkably, although the centrifugal term $\ell(\ell+1)$ is modified to $\ell(\ell+1)-q^2$ in Eq.~\eqref{radial hydrogen atom like equation}, the solution retains the same confluent hypergeometric functional form as in the neutral case, simply by replacing $\ell$ with $\ell_q$. In addition, the event horizon of the BH induces a complex shift \(\nu=\nu_0+\delta\nu\), with
the correction
\begin{align}
  \nu_0\equiv n_r+\ell_q+1=n_r+\sqrt{(\ell+\frac{1}{2})^2-q^2}+ \frac{1}{2},
    \label{eqB8}
\end{align}
leading to the alternative representation 
\begin{align}
    R(r)=(2kr)^{\ell_q} e^{-kr}U(-n_r-\delta \nu,2\ell_q+2;2kr).
\end{align}

\subsubsection{Near region: hypergeometric solution and horizon boundary condition}
\label{ab2}

In the near region, we introduce the rescaled radial coordinate 
\begin{align}
    z\equiv \frac{r-r_+}{r_+-r_-},\qquad
    \Delta=z(z+1)(r_+-r_-)^2.
\end{align}
In the limit \(r\!\ll\!\ell/\omega\) and \(r\!\ll\!\ell/\mu\), the radial equation in Eq.~(\ref{radial equation}) reduces to
\begin{align}\label{eqB16}
    \begin{aligned}
& z(z+1)\frac{d}{dz}\left[z(z+1)\frac{dR}{dz}\right] 
+\big[{p}^2-(\ell(\ell+1)-q^2)z(z+1)\big]R=0,
    \end{aligned}
\end{align}
with the superradiant parameter 
\begin{align}
p\equiv-\frac{\omega_*}{2\kappa}=-\frac{\omega-m\Omega_H}{2\kappa},
\end{align}
where $\Omega_H=\frac{a}{r_+^2+a^2}$ and $\kappa=\frac{r_+-r_-}{2(r_+^2+a^2)}$ are the angular velocity and surface gravity of the BH respectively. The differential equation (\ref{eqB16}) possesses three regular singularities at \{\(0, -1, \infty\)\}. Its solutions $R(z)$ take the standard form in terms of hypergeometric functions \cite{MacRobert1955HigherTF}, characterized by six indices $(x_0^\pm,x_{-1}^\pm,x_{\infty}^\pm)$ corresponding to these singularities. The indices are determined by the indical equations
\begin{align}
    x_n(x_n-1)+A_n+B_n=0,
    \label{indicial equation}
\end{align}
where \(A_n\) is the first-order characteristic coefficient and \(B_n\) is the second-order characteristic coefficient for $n$-th singularity.
Substituting the specific coefficients
\begin{align}
    \begin{aligned}
&A_0=1,\qquad B_0={p}^2,\\
&A_{-1}=1,\quad~ B_{-1}={p}^2,\\
&A_{\infty}=0,\quad~~ B_{\infty}=-\ell(\ell+1)+q^2,
    \end{aligned}
\end{align}
we obtain the indices
\begin{align}
    \begin{aligned}
& x_0^\pm=\pm i {p},\quad
x_{-1}^\pm=\pm i {p}, \\
& x_{\infty}^+=1/2+\sqrt{(\ell+1/2)^2-q^2},\quad 
x_{\infty}^-=1/2- \sqrt{(\ell+1/2)^2-q^2}.
    \end{aligned}
\end{align}
Consequently, the solutions $R(z)$ in Eq.~(\ref{eqB16}) can be written as the standard form : 
\begin{align}
R=
\left(\frac{z}{z+1}\right)^{i {p}}{_2F_1}(-\ell_q,\ell_q+1,1-2i {p};z+1),
\end{align}
with the hypergeometric function ${_2F_1}(-\ell_q,\ell_q+1,1-2i {p};z+1)$. 

The standard form can be expressed through a set of linearly independent solutions $(R_3,R_4)$, where the hypergeometric function ${_2F_1}(-\ell_q,\ell_q+1,1-2i {p};z+1)$ transforms into two another linearly independent functions $w_3$ and $w_4$:
\begin{align}
    \begin{aligned}
        R_3&\equiv\left(\frac{z}{z+1}\right)^{i {p}}(-z)^{\ell_q}w_3,\  w_3\equiv(-z)^{\ell_q}{_2F_1}(-\ell_q,-\ell_q-2i {p},-2\ell_q,-z^{-1}),\\
        R_4&\equiv\left(\frac{z}{z+1}\right)^{i {p}}(-z)^{-\ell_q-1} w_4,\ w_4\equiv (-z)^{-\ell_q-1}{_2F_1}(\ell_q+1,\ell_q+1-2i {p},2\ell_q+2,-z^{-1}).
    \end{aligned}
    \label{centrifugal hypergeometry function in the overlap region}
\end{align}
Physically, $(R_3,R_4)$ represent the solutions described by the centrifugal potential. These centrifugal solutions can be transformed into the ingoing and outgoing solutions $(R_1,R_2)$ near the horizon:
\begin{align}
    \begin{aligned}
    &R_1(z)=\left(\frac{z}{z+1}\right)^{i {p}}w_1, \ w_1\equiv{}_2{F}_1(-\ell_q,\ell_q+1,1+2i {p};-z),\\
    &R_2(z)=\left(\frac{z}{z+1}\right)^{i {p}}z^{-2ip}w_2, \ w_2\equiv {_2F_1}(-\ell_q-2ip,\ell_q+1-2ip,1-2i {p};z+1).
    \end{aligned}
\end{align}
The relation between these two sets of solutions as $z\to0$ (near horizon) is
\begin{align}
    \begin{aligned}
        R_3=f_{13}R_1+f_{23}R_2,\\
        R_4=f_{14}R_1+f_{24}R_2,
    \end{aligned}
    \label{R1,R2 transformed by R3,R4}
\end{align}
with the transformation coefficients 
\begin{align}
\begin{aligned}
&f_{13}=(-1)^{\ell_q}\frac{\Gamma(-2\ell_q)\Gamma(-2i {p})}{\Gamma(-\ell_q)\Gamma(-2i {p}-\ell_q)},\\
&f_{23}=(-1)^{\ell_q}\frac{\Gamma(-2\ell_q)\Gamma(2i {p})}{\Gamma(-\ell_q+2i {p})\Gamma(-\ell_q)},\\
&f_{14}=(-1)^{1-\ell_q}\frac{\Gamma(2\ell_q+2)\Gamma(-2i {p})}{\Gamma(\ell_q+1)\Gamma(\ell_q-2i {p}+1)},\\
&f_{24}=(-1)^{1-\ell_q}\frac{\Gamma(2\ell_q+2)\Gamma(2i {p})}{\Gamma(\ell_q+1)\Gamma(\ell_q+2i {p}+1)}.
\end{aligned}
\label{the connection coefficients between different hypergeometric function bases}
\end{align}
The general solution $R(z)$ is represented by  \begin{align}
\begin{aligned}
R(z)&=AR_3+BR_4\\
&=(Af_{13}+Bf_{14})R_1+(Af_{23}+Bf_{24})R_2\\
&\simeq (Af_{13}+Bf_{14})z^{ip}+(Af_{23}+Bf_{24})z^{-ip} (z\to 0)
\end{aligned}
\end{align}
where $A$ and $B$ are the linear coefficients in the $(R_3,R_4)$ basis. Imposing the purely ingoing boundary condition at the event horizon (the reflectivity $\mathcal{R}\equiv\frac{Af_{13}+Bf_{14}}{Af_{23}+Bf_{24}}=0$). Then we obtain the relation between \(A\) and \(B\) 
\begin{align}
    Af_{13}+Bf_{14}=0.
    \label{ABr}
\end{align}

\subsubsection{Overlap matching and analytic expression for $\delta\nu$}

In the overlap region, we connect the far-region solutions with the near-region solutions to obtain the complex shift $\delta\nu$. Using the properties of confluent hypergeometric functions 
\begin{align}
\begin{split}
    &{U}(-n_r-\delta\nu,2\ell_q+2;2kr)\\
&=\frac{\Gamma(-2\ell_q-1)}{\Gamma(-n_r-\delta\nu-2\ell_q-1)}{U}(-n_r-\delta\nu,2\ell_q+2;2kr)\\
&+\frac{(2kr)^{-2\ell_q-1}\Gamma(2\ell_q+1)}{\Gamma(-n_r-\delta\nu)}{U}(-n_r-\delta\nu-2\ell_q-1,-2\ell_q;2kr),
 \end{split}
 \label{property confluent hypergeometric functions}
\end{align}
and Gamma functions
\begin{align}
\begin{aligned}
&\frac{\Gamma(-2\ell_q-1)}{\Gamma(-n_r-\delta\nu-2\ell_q-1)}=(-1)^{n_r}\frac{\Gamma(2\ell_q+n_r+2)}{\Gamma(2\ell_q+2)},\\
&\frac{\Gamma(2\ell_q+1)}{\Gamma(-n_r-\delta\nu)}=(-1)^{n_r+1}\delta\nu \Gamma(n_r+1)\Gamma(2\ell_q+1),
\end{aligned}
\label{property gamma functions}
\end{align}
the far-region solution's asymptotic behavior in the overlap region (\(2kr\to 0\)) is
\begin{align}
\begin{split}
    &R_{\rm far}(r)=(2kr)^{\ell_q}e^{-kr}{U}(-n_r-\delta\nu,2\ell_q+2;2kr)
    \\
 &\simeq (-1)^{n_r}\frac{\Gamma(2\ell_q+n_r+2)}{\Gamma(2\ell_q+2)}(2kr)^{\ell_q}+(-1)^{n_r+1}\Gamma(n_r+1){\Gamma(2\ell_q+1)}\delta\nu(2kr)^{-\ell_q-1}.
\end{split}
\label{eqB42}
\end{align}
Meanwhile, the near-region solution's asymptotic behavior for \(r\gg M\) in the overlap region is
\begin{align}
    R_{\rm near}(r) \simeq  A \left(\frac{-r}{r_+-r_-}\right)^{\ell_q}+B \left(\frac{-r}{r_+-r_-}\right)^{-\ell_q-1}.
    \label{eqB45}
\end{align}
where
\begin{align}
\begin{aligned}
    &R_3=(-z)^{\ell_q}w_3 \simeq (-z)^{\ell_q}=\left(\frac{-r}{r_+-r_-}\right)^{\ell_q},\\
    &R_4=(-z)^{-\ell_q-1}w_4 \simeq (-z)^{-\ell_q-1}=\left(\frac{-r}{r_+-r_-}\right)^{-\ell_q-1},
    \end{aligned}
\end{align}
Matching these two solutions (Eq.~(\ref{eqB42}) and Eq.~(\ref{eqB45})) in the overlap region determines the coefficients $A$ and $B$
\begin{align}
    &A= (-1)^{n_r+\ell_q}[2k(r_+-r_-)]^{\ell_q}\frac{\Gamma(2\ell_q+n_r+2)}{\Gamma(2\ell_q+2)},\\&
    B= (-1)^{n_r+\ell_q}[2k(r_+-r_-)]^{-\ell_q-1}\Gamma(n_r+1)\Gamma(2\ell_q+1)\delta\nu.
\end{align}
Considering the relation $Af_{13}+Bf_{14}=0$ from Eq.~(\ref{ABr}), we derive the expression of \(\delta\nu\):
\begin{align}
    \begin{aligned}
    \delta\nu=&2{ip} \left[2 k\left(r_{+}-r_{-}\right)\right]^{2 \ell_q+1}   
    \\&\times\frac{\Gamma(2\ell_q+n_r+2)}{\Gamma(n_r+1)}\left[\frac{\Gamma(\ell_q+1)}{\Gamma(2\ell_q+1)\Gamma(2\ell_q+2)}\right]^2 
    \\&\times\left|\frac{\Gamma(1+\ell_q+2ip)}{\Gamma(1+2ip)}\right|^2.
\end{aligned}
\label{Growthratedeltanu}
\end{align}

\section{Gravitational-wave power from gauge-charged scalar clouds}
\label{gravitational-wave power from gauge-charged scalar clouds}

This section derives the GW source and luminosity for the charged complex scalar configuration used in the main text. Throughout this section, $q\equiv |eP|\geq0$ denotes the magnitude of the monopole number. The north and south cloud sectors are related by charge conjugation and carry gauge charges $+e$ and $-e$, respectively. The superscripts $+$ and $-$ on the monopole harmonics label regular gauge-patch representatives and should not themselves be identified with the sign of the physical gauge charge.

The central distinction from the familiar real-scalar annihilation picture is that the stress tensor of either charged complex sector alone is stationary. The oscillatory source at $|\tilde\omega|\simeq2\omega_R$ is instead generated by the coherent north--south cross term. The positive- and negative-frequency source components are complex conjugates and describe one physical real GW field; only the independent positive-frequency component is retained when defining $P_{\rm GW}$.

\subsection{General sourced-wave setup}

We write the metric perturbation as $\hat g_{\mu\nu}=g_{\mu\nu}+h_{\mu\nu}$. Within the flat-spacetime approximation adopted for the sourced GW
calculation, the trace-reversed perturbation obeys
\begin{equation}
\Box \bar h_{\mu\nu}=-16\pi T_{\mu\nu},
\qquad
\bar h_{\mu\nu}=h_{\mu\nu}-\frac12 g_{\mu\nu}h^\rho{}_{\rho},
\end{equation}
and equivalently
\begin{equation}
\Box h_{\mu\nu}=-16\pi\left(T_{\mu\nu}-\frac12T^\rho{}_{\rho}g_{\mu\nu}\right).
\label{GW equation}
\end{equation}
Expanding the sourced solution in homogeneous outgoing modes,
\begin{equation}
h_{\mu\nu}=\sum_{\tilde j}C_{\tilde j}u^{(\tilde j)}_{\mu\nu},
\qquad
\Box u^{(\tilde j)}_{\mu\nu}=0,
\label{homogeneous out-mode}
\end{equation}
the GW power can be written as
\begin{equation}
P_{\rm GW}=\frac{1}{64\pi}\sum_{\tilde j}\tilde\omega^2|C_{\tilde j}|^2J_{\tilde j},
\qquad
J_{\tilde j}=\int g^{\mu\rho}g^{\nu\sigma}
 u^{(\tilde j){\rm TT}}_{\mu\nu}u^{*(\tilde j){\rm TT}}_{\rho\sigma}\,dS.
\label{homogeneous J}
\end{equation}
Green's identity gives~\cite{Yoshino:2013ofa}
\begin{equation}
C_{\tilde j}=8\pi\frac{\langle u^{(\tilde j){\rm TT}},T\rangle}{\tilde\omega J_{\tilde j}},
\qquad
P_{\rm GW}=\pi\sum_{\tilde j}\frac{|\langle u^{(\tilde j)},T\rangle|^2}{J_{\tilde j}}.
\end{equation}
For the flat-spacetime outgoing mode,
\begin{equation}
J_{\tilde j}\simeq2|Z_{\rm out}|^2\tilde\omega^4,
\qquad
Z_{\rm out}(\tilde\omega)=
\frac{\Gamma(2\tilde\ell+2)}{(2i)^{\tilde\ell+3}\Gamma(\tilde\ell-1)}\tilde\omega^{-5}.
\end{equation}
Using the independent positive-frequency component $\tilde\omega=\Omega>0$, the physical luminosity is therefore
\begin{equation}
P_{\rm GW}=\frac{\pi}{2}
\left|\frac{\langle u,T\rangle_{+\Omega}}{Z_{\rm out}(\Omega)}\right|^2\Omega^{-4}.
\label{eq:SM_PGW_innerproduct}
\end{equation}
The negative-frequency component is fixed by complex conjugation and is not added as a second independent contribution.

\subsection{Charge-conjugate two-sector source}
\label{sec:complex_scalar_cross_source}

The dominant north and south mode functions are
\begin{equation}
\psi_N=e^{-i\omega_R t}Y^+_{qq,q}(\theta,\phi)R_N(r),
\qquad
\psi_S=e^{+i\omega_R t}Y^-_{qq,-q}(\theta,\phi)R_S(r),
\label{eq:SM_psi_NS}
\end{equation}
where
\begin{equation}
Y^+_{qq,q}=A_qe^{iq\phi}\left(\cos\frac{\theta}{2}\right)^{2q},
\qquad
Y^-_{qq,-q}=A_qe^{-iq\phi}\left(\sin\frac{\theta}{2}\right)^{2q},
\qquad
A_q=\sqrt{\frac{2q+1}{4\pi}}.
\end{equation}
The coherent configuration is written schematically as
\begin{equation}
\Psi=c_N\psi_N+c_S\psi_S,
\label{eq:SM_psi_two_sector}
\end{equation}
where the amplitudes $c_N$ and $c_S$ are absorbed below into the sector energies $E_N$ and $E_S$. For the mirror-symmetric configuration the two sectors have the same radial shape; their normalizations need not be equal.

For a charged complex scalar,
\begin{equation}
{\cal L}_{\rm scalar}=-g^{\mu\nu}D^c_\mu\Psi^*D_\nu\Psi-\mu^2\Psi^*\Psi,
\qquad
D_\mu=\nabla_\mu-ieA_\mu,
\qquad
D^c_\mu=\nabla_\mu+ieA_\mu,
\end{equation}
with $D^c_\mu\Psi^*\equiv(D_\mu\Psi)^*$. The stress tensor is
\begin{equation}
T_{\mu\nu}=D^c_\mu\Psi^*D_\nu\Psi+D^c_\nu\Psi^*D_\mu\Psi
-g_{\mu\nu}\left(g^{ab}D^c_a\Psi^*D_b\Psi+\mu^2\Psi^*\Psi\right).
\label{eq:SM_complex_Tmunu}
\end{equation}
For either sector alone, the rapid phase cancels:
\begin{equation}
T^{NN}_{\mu\nu}\propto e^{+i\omega_Rt}e^{-i\omega_Rt},
\qquad
T^{SS}_{\mu\nu}\propto e^{-i\omega_Rt}e^{+i\omega_Rt}.
\label{eq:SM_single_sector_stationary}
\end{equation}
Thus an isolated charged complex sector has no annihilation-type $2\omega_R$ source. The oscillatory terms are
\begin{align}
T^{(+\Omega)}_{\mu\nu}
&\equiv T^{SN,-}_{\mu\nu}
=D^c_\mu\psi_S^*D_\nu\psi_N+D^c_\nu\psi_S^*D_\mu\psi_N
\propto e^{-i\Omega t+2iq\phi},
\\
T^{(-\Omega)}_{\mu\nu}
&\equiv T^{NS,+}_{\mu\nu}
=D^c_\mu\psi_N^*D_\nu\psi_S+D^c_\nu\psi_N^*D_\mu\psi_S
=\left[T^{(+\Omega)}_{\mu\nu}\right]^*
\propto e^{+i\Omega t-2iq\phi},
\label{eq:SM_cross_terms}
\end{align}
where $\Omega\equiv2\omega_R>0$. The labels $SN,-$ and $NS,+$ are retained to match the main text; the physically useful distinction is between the independent $+\Omega$ component and its $-\Omega$ conjugate.

\subsection{Positive-frequency source projection and selection rule}
\label{GW emission in the flat-spacetime approximation}

In the flat approximation the relevant Kinnersley tetrad is
\begin{equation}
\ell^\mu=(1,1,0,0),
\qquad
n^\mu=\left(\frac12,-\frac12,0,0\right),
\qquad
m^\mu=\frac{1}{\sqrt2r}\left(0,0,1,\frac{i}{\sin\theta}\right).
\end{equation}
The trace term in Eq.~\eqref{eq:SM_complex_Tmunu} vanishes after projection onto the null tetrad. Defining
\begin{equation}
{\cal D}_{\pm}=\partial_\theta\pm\frac{i}{\sin\theta}(\partial_\phi-ieA_\phi),
\qquad
{\cal D}^c_{\pm}=\partial_\theta\pm\frac{i}{\sin\theta}(\partial_\phi+ieA_\phi),
\qquad
\tilde r\equiv\Omega r,
\end{equation}
and factoring out the sector-dependent radial normalizations, the nonzero projections of the independent source $T^{(+\Omega)}_{\mu\nu}=T^{SN,-}_{\mu\nu}$ are
\begin{align}
T^{(+\Omega)}_{mm}
&=\frac{\Omega^2}{\tilde r^2}R^2({\cal D}^c_+Y_S^*)({\cal D}_+Y_N)e^{-i\Omega t},
\\
T^{(+\Omega)}_{m^*m^*}
&=\frac{\Omega^2}{\tilde r^2}R^2({\cal D}^c_-Y_S^*)({\cal D}_-Y_N)e^{-i\Omega t},
\\
T^{(+\Omega)}_{nn}
&=\frac{\Omega^2}{2}\left(\frac{iR}{2}+R_{,\tilde r}\right)^2Y_S^*Y_Ne^{-i\Omega t},
\\
T^{(+\Omega)}_{nm}
&=-\frac{\Omega^2}{2\sqrt2\tilde r}\left(\frac{iR}{2}+R_{,\tilde r}\right)R
\left[Y_S^*({\cal D}_+Y_N)+({\cal D}^c_+Y_S^*)Y_N\right]e^{-i\Omega t},
\\
T^{(+\Omega)}_{nm^*}
&=-\frac{\Omega^2}{2\sqrt2\tilde r}\left(\frac{iR}{2}+R_{,\tilde r}\right)R
\left[Y_S^*({\cal D}_-Y_N)+({\cal D}^c_-Y_S^*)Y_N\right]e^{-i\Omega t}.
\label{eq:SM_Tminus_projections}
\end{align}
The negative-frequency projections are the complex conjugates of these expressions. Relative to the real-scalar calculation~\cite{Yoshino:2013ofa}, the useful replacement rules are
\begin{align}
Y^2&\longrightarrow Y_S^*Y_N,
\\
(\eth Y)^2&\longrightarrow({\cal D}^c_+Y_S^*)({\cal D}_+Y_N),
\qquad
(\eth^*Y)^2\longrightarrow({\cal D}^c_-Y_S^*)({\cal D}_-Y_N),
\\
(\eth Y)Y&\longrightarrow\frac12\left[Y_S^*({\cal D}_+Y_N)+({\cal D}^c_+Y_S^*)Y_N\right],
\\
(\eth^*Y)Y&\longrightarrow\frac12\left[Y_S^*({\cal D}_-Y_N)+({\cal D}^c_-Y_S^*)Y_N\right].
\label{eq:SM_replacement_rules}
\end{align}

\subsubsection{North/south selection rules and cross-source emission}
\label{North/south selection rules and cross-source emission}

After inserting the homogeneous Chrzanowski out-mode and integrating over angles,
\begin{equation}
\int \hat u^*_{\mu\nu}T^{\pm,\mu\nu}\,d\Omega
=\frac{e^{i(\tilde\omega\pm\Omega)t}}{4}{\cal F}^{\pm}_{\tilde\ell\tilde m}(\tilde r).
\label{eq:SM_angular_inner_product_cross}
\end{equation}
The monopole-harmonic phases are
\begin{equation}
Y_N^*Y_S=A_q^2e^{-2iq\phi}\left(\sin\frac\theta2\cos\frac\theta2\right)^{2q},
\qquad
Y_S^*Y_N=A_q^2e^{+2iq\phi}\left(\sin\frac\theta2\cos\frac\theta2\right)^{2q}.
\end{equation}
The time and azimuthal integrations therefore select
\begin{equation}
T^{SN,-}_{\mu\nu}\to\quad
(\tilde\ell,\tilde m,\tilde\omega)=(2q,+2q,+2\omega_R),
\qquad
T^{NS,+}_{\mu\nu}\to\quad
(\tilde\ell,\tilde m,\tilde\omega)=(2q,-2q,-2\omega_R).
\label{eq:SM_cross_selection_rule}
\end{equation}
The second mode is the negative-frequency conjugate of the first, rather than a second independent radiation channel. The tensor mode considered here requires $\tilde\ell=2q\geq2$, hence $q\geq1$; the $q=1/2$ case does not furnish this radiative spin-2 multipole.

The cross source is parity even. Hence the odd-parity angular projection vanishes. For the even-parity positive-frequency mode, we define the three real angular integrals
\begin{align}
I_1&=8\pi\tilde A_0A_q^2
\sqrt{(\tilde\ell-1)\tilde\ell(\tilde\ell+1)(\tilde\ell+2)}
B(2q+1,2q+1),
\label{the I1 angular integral}
\\
I_2&=4\pi\tilde A_2A_q^2B(2q+1,2q-1),
\label{the I2 angular integral}
\\
I_3&=16\pi q\tilde A_1A_q^2
\sqrt{(2q+2)(2q-1)}B(2q,2q+2),
\label{the I3 angular integral}
\end{align}
where
\begin{equation}
\tilde A_0=\frac{1}{\sqrt{4\pi B(2q+1,2q+1)}},
\qquad
\tilde A_1=\frac{1}{\sqrt{4\pi B(2q,2q+2)}},
\qquad
\tilde A_2=\frac{1}{\sqrt{4\pi B(2q+3,2q-1)}}.
\end{equation}
Here $B(x,y)$ is the Euler beta function. The apparent relative sign of $I_3$ in the conjugate calculation is accompanied by the corresponding conjugation of the radial coefficient. It should not be interpreted as producing a different physical luminosity.

\subsection{Radial overlap, explicit coefficient, and power scaling}
\label{Radial integral of the Ffunction}

The full positive-frequency source product is
\begin{equation}
\langle u,T\rangle_{+\Omega}
=\frac{1}{4\Omega^3}\int_0^\infty{\cal F}^{-}_{2q,2q}(\tilde r)\tilde r^2\,d\tilde r.
\label{full inner product}
\end{equation}
Let $n\equiv n_r+\ell_q+1$ and $\beta\equiv k/\omega_R\simeq\alpha/n$. The common radial shape has the small-$\beta$ expansion
\begin{equation}
R_A(\tilde r)=A_{R,A}e^{-\beta\tilde r/2}
\left(\frac{\beta}{2}\right)^{\ell_q}
\left[\tilde r^{\ell_q}+\beta\gamma\tilde r^{\ell_q+1}+{\cal O}(\beta^2)\right],
\qquad
\gamma=-\frac{n_r}{2\ell_q+2},
\label{radialexpand}
\end{equation}
where $A=N,S$. To leading nonrelativistic order, the energy normalization of each sector is
\begin{equation}
E_A\simeq2\omega_R^2\int r^2R_A^2\,dr,
\end{equation}
which gives
\begin{equation}
A_{R,A}=(-1)^{n_r}\frac{\sqrt{E_A}}{\omega_R}(2k)^{3/2}
\frac{2^{\ell_q}}{\Gamma(2\ell_q+2)}
\sqrt{\frac{\Gamma(n_r+2\ell_q+2)}{4n\Gamma(n_r+1)}}.
\label{eq:SM_AR_norm}
\end{equation}
Thus the cross-source amplitude is proportional to $A_{R,N}A_{R,S}\propto\sqrt{E_NE_S}$.

The remaining outgoing-mode integrals can be generated from
\begin{equation}
f^{(0)}(\beta)=
\frac{\Gamma(S)}{(\beta-i)^S}
{}_2F_1\!\left(\tilde\ell-1,S;2\tilde\ell+2;
\frac{-2i}{\beta-i}\right),
\qquad
S\equiv2\ell_q+\tilde\ell+1,
\end{equation}
through $f^{(N)}=(-1)^N d^Nf^{(0)}/d\beta^N$. For the independent positive-frequency mode, the argument approaches the hypergeometric branch cut as $2-i0$. We therefore define
\begin{align}
H_0&\equiv{}_2F_1(\tilde\ell-1,S;2\tilde\ell+2;2-i0),
\\
H_1&\equiv{}_2F_1(\tilde\ell,S+1;2\tilde\ell+3;2-i0),
\\
H_2&\equiv{}_2F_1(\tilde\ell+1,S+2;2\tilde\ell+4;2-i0),
\end{align}
and
\begin{align}
P_1&\equiv SH_0+\frac{(\tilde\ell-1)S}{\tilde\ell+1}H_1,
\\
P_2&\equiv-\frac{S(S+1)}{2}H_0
-\frac{(\tilde\ell-1)S(S+1)}{\tilde\ell+1}H_1
-\frac{(\tilde\ell-1)\tilde\ell S(S+1)}{(\tilde\ell+1)(2\tilde\ell+3)}H_2.
\end{align}
Then
\begin{equation}
f^{(0)}(\beta)=\Gamma(S)(-i)^{-S}
\left[H_0-i\beta P_1+\beta^2P_2+{\cal O}(\beta^3)\right].
\end{equation}
Because $z=2$ lies on the standard branch cut, $H_j$ and $P_j$ are not assumed to be real. The negative-frequency calculation uses the $2+i0$ boundary value and is the complex conjugate of the expression above.

Collecting the leading ${\cal O}(\beta^{2\ell_q})$ terms gives
\begin{equation}
\langle u,T\rangle_{+\Omega}\simeq
\frac{A_{R,N}A_{R,S}}{4\Omega^3}
\left(\frac{\beta}{2}\right)^{2\ell_q}
\Gamma(S)(-i)^{-S}\Sigma,
\label{eq:SM_inner_sigma}
\end{equation}
with
\begin{equation}
\Sigma=I_1C_1+I_2C_2+I_3C_3,
\end{equation}
where
\begin{align}
C_1&=\frac12P_2+\ell_qP_1-(3\ell_q^2+2\ell_q)H_0,
\\
C_2&=4P_2+(6+4\ell_q)P_1-(\tilde\ell^2+\tilde\ell-2)H_0,
\\
C_3&=-P_2-(2\ell_q+1)P_1+(2\ell_q^2+\ell_q)H_0.
\label{eq:SM_C123}
\end{align}
The conjugate source satisfies
\begin{equation}
\langle u,T\rangle_{-\Omega}=\langle u,T\rangle_{+\Omega}^*,
\qquad
|\langle u,T\rangle_{-\Omega}|=|\langle u,T\rangle_{+\Omega}|,
\end{equation}
when all angular, radial, and outgoing-wave phase conventions are transformed together. Individual sign changes in $I_3$ are therefore convention-dependent intermediate features, not evidence for unequal physical powers.

Combining Eqs.~\eqref{eq:SM_PGW_innerproduct} and \eqref{eq:SM_inner_sigma}, the physical one-sided GW power is
\begin{equation}
P_{\rm GW}=C_{n\ell\ell_q}
\left(\frac{E_NE_S}{M^2}\right)\alpha^{4\ell_q+8},
\label{eq:SM_PGW_product}
\end{equation}
where
\begin{align}
C_{n\ell\ell_q}
={}&\frac{2^{2\tilde\ell+3}\pi}{n^{4\ell_q+8}}
\frac{\Gamma(\tilde\ell-1)^2}{\Gamma(2\tilde\ell+2)^2}
\frac{\Gamma(2\ell_q+\tilde\ell+1)^2}{\Gamma(2\ell_q+2)^4}
\frac{\Gamma(n+\ell_q+1)^2}{\Gamma(n-\ell_q)^2}
|\Sigma|^2,
\qquad \tilde\ell=2q.
\label{eq:SM_CGW_explicit}
\end{align}
For a balanced configuration,
\begin{equation}
E_N=E_S\equiv E_s,
\qquad
E_c\equiv E_N+E_S=2E_s,
\end{equation}
so that
\begin{equation}
P_{\rm GW}=\bar C_{n\ell\ell_q}
\left(\frac{E_c}{M}\right)^2\alpha^{4\ell_q+8},
\qquad
\bar C_{n\ell\ell_q}\equiv\frac14C_{n\ell\ell_q}.
\label{GW power charged scalar cloud}
\end{equation}
The enhancement relative to the Kerr scaling $P_{\rm GW}\propto\alpha^{4\ell+10}$ has two connected origins. The reduced index $\ell_q<\ell$ weakens the small-$r$ suppression of the bound-state wavefunction and enlarges its support in the effective GW-emitting region. At the same time, the leading charged cross-source projection does not satisfy the Kerr angular identities responsible for the leading-order cancellation, so the additional Kerr factor $\alpha^2$ is absent.

\section{Mass and spin evolution of the two-sector cloud}
\label{sec4}

The dynamical cloud variable is the total scalar energy
\begin{equation}
E_c\equiv E_N+E_S.
\end{equation}
In the nonrelativistic regime this equals the cloud mass to leading order, $M_c\simeq E_c$; the symbol $M_c$ retained in the numerical figure should be understood in this sense. For the mirror-symmetric charge-conjugate pair, both sectors have the same positive growth rate $\omega_I$. Although their mode labels are $(m,\omega_R)$ and $(-m,-\omega_R)$, the ratio $m/\omega_R$ is the same for the two sectors. The BH evolution is therefore
\begin{align}
\frac{dM}{dt}&=-2\omega_I(E_N+E_S)=-2\omega_IE_c,
\label{BHmassevo}
\\
\frac{dJ}{dt}&=-\frac{2m\omega_I}{\omega_R}(E_N+E_S)
=-\frac{2m\omega_I}{\omega_R}E_c.
\label{BHJevo}
\end{align}
Each cross-source emission event removes equal energies from the two sectors because their constituent frequencies have the same magnitude. Hence
\begin{align}
\frac{dE_N}{dt}&=2\omega_IE_N-\frac12P_{\rm GW},
\\
\frac{dE_S}{dt}&=2\omega_IE_S-\frac12P_{\rm GW},
\end{align}
and the total cloud energy obeys
\begin{equation}
\frac{dE_c}{dt}=2\omega_IE_c-P_{\rm GW}(E_N,E_S),
\qquad
P_{\rm GW}\propto E_NE_S.
\label{cloudmassevo}
\end{equation}
A balanced initial state remains balanced under these equations. The factors of $1/2$ prevent double counting: $P_{\rm GW}$ is the total physical luminosity of the one real GW field, not the luminosity of each Fourier component separately.

At early times, when GW losses and the evolution of the BH parameters are negligible,
\begin{equation}
E_A(t)\simeq E_{A,i}e^{2\omega_It},
\qquad A=N,S,
\end{equation}
with e-folding time
\begin{equation}
\frac{\tau_e}{M}\equiv\frac{1}{2M\omega_I}
\sim\alpha^{-(4\ell_q+6)}.
\label{eq:SM_efolding}
\end{equation}
The actual build-up time additionally contains the logarithmic seed factor,
\begin{equation}
t_{\rm grow}\simeq\tau_e\ln\!\left(\frac{E_c}{E_{c,i}}\right).
\end{equation}
Using the representative peak rates $M\omega_I\sim10^{-6}$ for the magnetic $N=3$ mode and $M\omega_I^{\rm Kerr}\sim2\times10^{-8}$ for the neutral 211 mode gives
\begin{align}
\tau_e^{(N=3)}&\simeq2.47\left(\frac{M}{M_\odot}\right){\rm s},
\\
\tau_e^{(211)}&\simeq1.23\times10^2\left(\frac{M}{M_\odot}\right){\rm s}.
\end{align}
The corresponding reference values are listed in Table~\ref{Table: efolding timescale comparison}.

\begin{table}[h]
\centering
\renewcommand{\arraystretch}{1.35}
\begin{tabular}{|c|c|c|}
\hline
BH mass $M$
& Magnetic $N=3$ ($q=1.5$, $M\omega_I\sim10^{-6}$)
& Neutral 211 ($M\omega_I^{\rm Kerr}\sim2\times10^{-8}$)\\
\hline
$10M_\odot$
& $2.47\times10^1\,{\rm s}$ $(0.41\,{\rm min})$
& $1.23\times10^3\,{\rm s}$ $(20.5\,{\rm min})$\\
$10^4M_\odot$
& $2.47\times10^4\,{\rm s}$ $(6.86\,{\rm hr})$
& $1.23\times10^6\,{\rm s}$ $(14.3\,{\rm d})$\\
$10^6M_\odot$
& $2.47\times10^6\,{\rm s}$ $(28.6\,{\rm d})$
& $1.23\times10^8\,{\rm s}$ $(3.9\,{\rm yr})$\\
\hline
\end{tabular}
\caption{Reference e-folding times $\tau_e=(2\omega_I)^{-1}$. The build-up time to a macroscopic cloud is longer by the logarithmic factor $\ln(E_c/E_{c,i})$.}
\label{Table: efolding timescale comparison}
\end{table}

The exponential estimate is only a reference scale. During the full evolution, $M(t)$ and $J(t)$ change, reducing $\Omega_H$ and suppressing $\omega_I\propto m\Omega_H-\omega_R$ as the superradiant threshold is approached. In the magnetic configuration, the enhanced cross-source luminosity can also become important before this threshold is reached. The cloud then enters a GW-limited regime determined by the competition between the linear growth term $2\omega_IE_c$ and the nonlinear loss term $P_{\rm GW}\propto E_NE_S$.

\begin{figure}[h]
\centering
\includegraphics[width=1\textwidth]{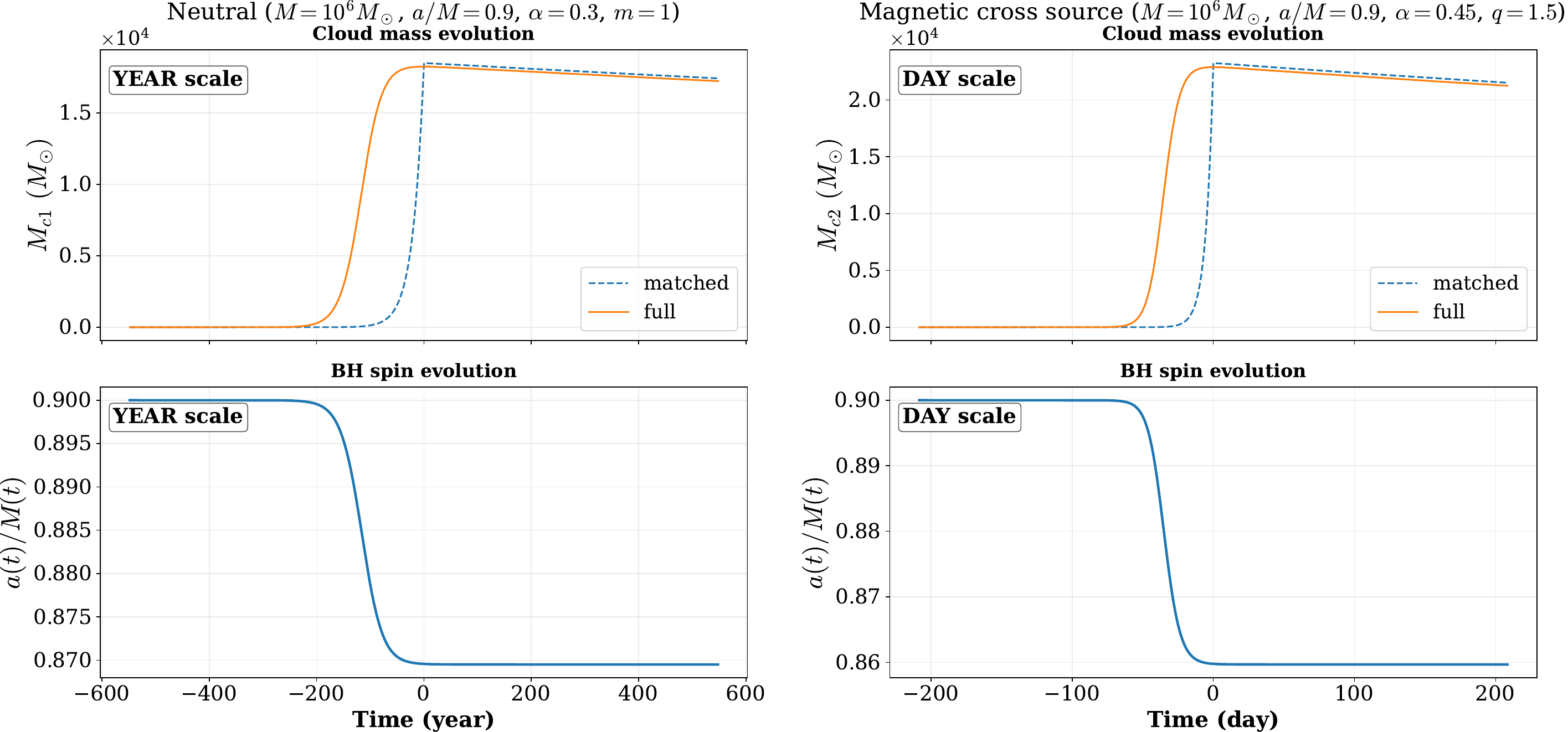}
\caption{Evolution of the total scalar-cloud energy (top, labeled $M_c\simeq E_c$ in the numerical output) and BH spin (bottom) for the neutral (left) and magnetic two-sector (right) configurations around a BH with $M=10^6M_\odot$ and $a/M=0.9$. The neutral case has $\alpha=0.3$, while the magnetic case has $\alpha=0.45$ and $q=1.5$. Dashed curves show the matched growth--depletion approximation, whereas solid curves show the full evolution including GW losses throughout. In the magnetic case, $P_{\rm GW}$ is the physical one-sided luminosity sourced by $E_NE_S$; the negative-frequency conjugate is required for waveform reality but is not an additional energy-loss channel.}
\label{Fig:Cloud evolution}
\end{figure}

We therefore integrate Eqs.~\eqref{BHmassevo}--\eqref{cloudmassevo} while updating $\omega_I[M(t),J(t)]$, $\alpha(t)=\mu M(t)$, and $P_{\rm GW}[E_N(t),E_S(t)]$. Figure~\ref{Fig:Cloud evolution} compares this full evolution with a matched growth--depletion approximation. In the neutral Kerr case, GW losses remain negligible during growth and the two treatments nearly coincide. In the magnetic case, GW quenching reduces and broadens the maximum cloud energy and limits the BH spin-down before the idealized superradiant threshold is reached.

\section{Gravitational-wave waveform of the cross-source channel}
\label{Waveform construction: cross source channel}

The TT waveform is conveniently encoded in the complex strain
\begin{equation}
h(t,\iota,\varphi_o)\equiv h_+(t,\iota,\varphi_o)-ih_\times(t,\iota,\varphi_o),
\end{equation}
with spin-weighted expansion
\begin{equation}
h(t,\iota,\varphi_o)=\frac{1}{D}\sum_{\tilde\ell,\tilde m}
h_{\tilde\ell\tilde m}(t){}_{-2}Y_{\tilde\ell\tilde m}(\iota,\varphi_o),
\label{eq:total_GW_amplitude}
\end{equation}
where $D$ is the luminosity distance and $(\iota,\varphi_o)$ specify the line of sight. The cross-source calculation selects the independent positive-frequency mode
\begin{equation}
(\tilde\ell,\tilde m,\tilde\omega)=(2q,+2q,+2\omega_R),
\end{equation}
together with its negative-frequency conjugate. A convenient analytic-signal representation is
\begin{equation}
h(t,\iota,\varphi_o)=
\frac{{\cal H}(t)}{D}
{}_{-2}Y_{2q,2q}(\iota,\varphi_o)
\exp[-i\Phi_{\rm GW}(t)],
\label{eq:h_minimal_cross}
\end{equation}
where
\begin{equation}
\dot\Phi_{\rm GW}(t)=\Omega(t)=2\omega_R(t),
\qquad
f_{\rm GW}(t)=\frac{\Omega(t)}{2\pi}=\frac{\omega_R(t)}{\pi}.
\label{eq:SM_fGW}
\end{equation}
Equation~\eqref{eq:h_minimal_cross} should not be supplemented by an additional independent power from the negative-frequency mode. The latter is the complex-conjugate component required when reconstructing the real metric perturbation. The two real polarizations are obtained directly from
\begin{equation}
h_+=\operatorname{Re}h,
\qquad
h_\times=-\operatorname{Im}h.
\end{equation}

With the normalization $\int|{}_{-2}Y_{2q,2q}|^2d\Omega=1$ and the standard asymptotic GW flux, the mode amplitude is related to the luminosity by
\begin{equation}
P_{\rm GW}(t)=\frac{\Omega(t)^2}{16\pi}|{\cal H}(t)|^2,
\qquad
|{\cal H}(t)|=\frac{4\sqrt{\pi P_{\rm GW}(t)}}{\Omega(t)}.
\label{eq:SM_H_from_power}
\end{equation}
Thus the waveform envelope is fixed by the two-sector evolution through
\begin{equation}
P_{\rm GW}(t)=C_{n\ell\ell_q}
\left[\frac{E_N(t)E_S(t)}{M(t)^2}\right]
\alpha(t)^{4\ell_q+8},
\end{equation}
while its phase evolution follows $\omega_R[M(t),J(t)]$. For a balanced cloud, $E_N=E_S=E_c/2$, so the strain envelope is proportional to $E_c(t)$ at fixed BH parameters.

\end{document}